\documentclass[12pt,a4paper]{article}

\makeatletter
\newcount\@hour\newcount\@minute\chardef\@x10\chardef\@xv60
\def\tcitime{
\def\@time{%
  \@minute\time\@hour\@minute\divide\@hour\@xv
  \ifnum\@hour<\@x 0\fi\the\@hour:%
  \multiply\@hour\@xv\advance\@minute-\@hour
  \ifnum\@minute<\@x 0\fi\the\@minute
  }}%

\@ifundefined{hyperref}{}{}

\@ifundefined{qExtProgCall}{\def\qExtProgCall#1#2#3#4#5#6{\relax}}{}
\def\QCTOpt[#1]#2{%
  \def\QCTOptB{#1}
  \def\QCTOptA{#2}
}
\def\QCTNOpt#1{%
  \def\QCTOptA{#1}
  \let\QCTOptB\empty
}
\def\Qct{%
  \@ifnextchar[{%
    \QCTOpt}{\QCTNOpt}
}
\def\QCBOpt[#1]#2{%
  \def\QCBOptB{#1}
  \def\QCBOptA{#2}
}
\def\QCBNOpt#1{%
  \def\QCBOptA{#1}
  \let\QCBOptB\empty
}
\def\Qcb{%
  \@ifnextchar[{%
    \QCBOpt}{\QCBNOpt}
}
\def\PrepCapArgs{%
  \ifx\QCBOptA\empty
    \ifx\QCTOptA\empty
      {}%
    \else
      \ifx\QCTOptB\empty
        {\QCTOptA}%
      \else
        [\QCTOptB]{\QCTOptA}%
      \fi
    \fi
  \else
    \ifx\QCBOptA\empty
      {}%
    \else
      \ifx\QCBOptB\empty
        {\QCBOptA}%
      \else
        [\QCBOptB]{\QCBOptA}%
      \fi
    \fi
  \fi
}
\newcount\GRAPHICSTYPE
\GRAPHICSTYPE=\z@
\def\GRAPHICSPS#1{%
 \ifcase\GRAPHICSTYPE
   \special{ps: #1}%
 \or
   \special{language "PS", include "#1"}%
 \fi
}%
\def\graffile#1#2#3#4{%
    \leavevmode
    \raise -#4 \BOXTHEFRAME{%
        \hbox to #2{\raise #3\hbox to #2{\null #1\hfil}}}%
}%
\def\draftbox#1#2#3#4{%
 \leavevmode\raise -#4 \hbox{%
  \frame{\rlap{\protect\tiny #1}\hbox to #2%
   {\vrule height#3 width\z@ depth\z@\hfil}%
  }%
 }%
}%
\newcount\draft
\draft=\z@

\newif\ifwasdraft
\wasdraftfalse

\def\GRAPHIC#1#2#3#4#5{%
 \ifnum\draft=\@ne\draftbox{#2}{#3}{#4}{#5}%
  \else\graffile{#1}{#3}{#4}{#5}%
  \fi
 }%
\def\addtoLaTeXparams#1{%
    \edef\LaTeXparams{\LaTeXparams #1}}%
\newif\ifBoxFrame \BoxFramefalse
\newif\ifOverFrame \OverFramefalse
\newif\ifUnderFrame \UnderFramefalse

\def\BOXTHEFRAME#1{%
   \hbox{%
      \ifBoxFrame
         \frame{#1}%
      \else
         {#1}%
      \fi
   }%
}

\def\doFRAMEparams#1{\BoxFramefalse\OverFramefalse\UnderFramefalse\readFRAMEparams#1\end}%
\def\readFRAMEparams#1{%
 \ifx#1\end%
  \let\next=\relax
  \else
  \ifx#1i\dispkind=\z@\fi
  \ifx#1d\dispkind=\@ne\fi
  \ifx#1f\dispkind=\tw@\fi
  \ifx#1t\addtoLaTeXparams{t}\fi
  \ifx#1b\addtoLaTeXparams{b}\fi
  \ifx#1p\addtoLaTeXparams{p}\fi
  \ifx#1h\addtoLaTeXparams{h}\fi
  \ifx#1X\BoxFrametrue\fi
  \ifx#1O\OverFrametrue\fi
  \ifx#1U\UnderFrametrue\fi
  \ifx#1w
    \ifnum\draft=1\wasdrafttrue\else\wasdraftfalse\fi
    \draft=\@ne
  \fi
  \let\next=\readFRAMEparams
  \fi
 \next
 }%
\def\IFRAME#1#2#3#4#5#6{%
      \bgroup
      \let\QCTOptA\empty
      \let\QCTOptB\empty
      \let\QCBOptA\empty
      \let\QCBOptB\empty
      #6%
      \parindent=0pt%
      \leftskip=0pt
      \rightskip=0pt
      \setbox0 = \hbox{\QCBOptA}%
      \@tempdima = #1\relax
      \ifOverFrame
          \typeout{This is not implemented yet}%
          \show\HELP
      \else
         \ifdim\wd0>\@tempdima
            \advance\@tempdima by \@tempdima
            \ifdim\wd0 >\@tempdima
               \textwidth=\@tempdima
               \setbox1 =\vbox{%
                  \noindent\hbox to \@tempdima{\hfill\GRAPHIC{#5}{#4}{#1}{#2}{#3}\hfill}\\%
                  \noindent\hbox to \@tempdima{\parbox[b]{\@tempdima}{\QCBOptA}}%
               }%
               \wd1=\@tempdima
            \else
               \textwidth=\wd0
               \setbox1 =\vbox{%
                 \noindent\hbox to \wd0{\hfill\GRAPHIC{#5}{#4}{#1}{#2}{#3}\hfill}\\%
                 \noindent\hbox{\QCBOptA}%
               }%
               \wd1=\wd0
            \fi
         \else
            \ifdim\wd0>0pt
              \hsize=\@tempdima
              \setbox1 =\vbox{%
                \unskip\GRAPHIC{#5}{#4}{#1}{#2}{0pt}%
                \break
                \unskip\hbox to \@tempdima{\hfill \QCBOptA\hfill}%
              }%
              \wd1=\@tempdima
           \else
              \hsize=\@tempdima
              \setbox1 =\vbox{%
                \unskip\GRAPHIC{#5}{#4}{#1}{#2}{0pt}%
              }%
              \wd1=\@tempdima
           \fi
         \fi
         \@tempdimb=\ht1
         \advance\@tempdimb by \dp1
         \advance\@tempdimb by -#2%
         \advance\@tempdimb by #3%
         \leavevmode
         \raise -\@tempdimb \hbox{\box1}%
      \fi
      \egroup%
}%
\def\DFRAME#1#2#3#4#5{%
 \begin{center}
     \let\QCTOptA\empty
     \let\QCTOptB\empty
     \let\QCBOptA\empty
     \let\QCBOptB\empty
     \ifOverFrame 
        #5\QCTOptA\par
     \fi
     \GRAPHIC{#4}{#3}{#1}{#2}{\z@}
     \ifUnderFrame 
        \nobreak\par #5\QCBOptA
     \fi
 \end{center}%
 }%
\def\FFRAME#1#2#3#4#5#6#7{%
 \begin{figure}[#1]%
  \let\QCTOptA\empty
  \let\QCTOptB\empty
  \let\QCBOptA\empty
  \let\QCBOptB\empty
  \ifOverFrame
    #4
    \ifx\QCTOptA\empty
    \else
      \ifx\QCTOptB\empty
        \caption{\QCTOptA}%
      \else
        \caption[\QCTOptB]{\QCTOptA}%
      \fi
    \fi
    \ifUnderFrame\else
      \label{#5}%
    \fi
  \else
    \UnderFrametrue%
  \fi
  \begin{center}\GRAPHIC{#7}{#6}{#2}{#3}{\z@}\end{center}%
  \ifUnderFrame
    #4
    \ifx\QCBOptA\empty
      \caption{}%
    \else
      \ifx\QCBOptB\empty
        \caption{\QCBOptA}%
      \else
        \caption[\QCBOptB]{\QCBOptA}%
      \fi
    \fi
    \label{#5}%
  \fi
  \end{figure}%
 }%
\newcount\dispkind%

\def\makeactives{
  \catcode`\"=\active
  \catcode`\;=\active
  \catcode`\:=\active
  \catcode`\'=\active
  \catcode`\~=\active
}
\bgroup
   \makeactives
   \gdef\activesoff{%
      \def"{\string"}
      \def;{\string;}
      \def:{\string:}
      \def'{\string'}
      \def~{\string~}
    }
\egroup

\def\FRAME#1#2#3#4#5#6#7#8{%
 \bgroup
 \@ifundefined{bbl@deactivate}{}{\activesoff}
 \ifnum\draft=\@ne
   \wasdrafttrue
 \else
   \wasdraftfalse%
 \fi
 \def\LaTeXparams{}%
 \dispkind=\z@
 \def\LaTeXparams{}%
 \doFRAMEparams{#1}%
 \ifnum\dispkind=\z@\IFRAME{#2}{#3}{#4}{#7}{#8}{#5}\else
  \ifnum\dispkind=\@ne\DFRAME{#2}{#3}{#7}{#8}{#5}\else
   \ifnum\dispkind=\tw@
    \edef\@tempa{\noexpand\FFRAME{\LaTeXparams}}%
    \@tempa{#2}{#3}{#5}{#6}{#7}{#8}%
    \fi
   \fi
  \fi
  \ifwasdraft\draft=1\else\draft=0\fi{}%
  \egroup
 }%
\def\TEXUX#1{"texux"}

\long\def\QQQ#1#2{%
     \long\expandafter\def\csname#1\endcsname{#2}}%
\@ifundefined{QTP}{\def\QTP#1{}}{}
\@ifundefined{QEXCLUDE}{\def\QEXCLUDE#1{}}{}
\@ifundefined{Qlb}{}{}
\@ifundefined{Qlt}{}{}
\long\def\QQA#1#2{}%
\def\QTR#1#2{{\csname#1\endcsname #2}}
\def\EXPAND#1[#2]#3{}%
\def\NOEXPAND#1[#2]#3{}%
\def\LaTeXparent#1{}%
\def\ChildStyles#1{}%
\def\ChildDefaults#1{}%
\def\QTagDef#1#2#3{}%
\@ifundefined{StyleEditBeginDoc}{}{}
\def\QQfnmark#1{\footnotemark}

\@ifundefined{INDEX}{\def\INDEX#1#2{}{}}{}%
\@ifundefined{SUBINDEX}{\def\SUBINDEX#1#2#3{}{}{}}{}%
\@ifundefined{initial}%
   {\def\initial#1{\bigbreak{\raggedright\large\bf #1}\kern 2\p@\penalty3000}}%
   {}%
\@ifundefined{entry}{}{}%
\@ifundefined{primary}{}{}%
\@ifundefined{secondary}{}{}%
\@ifundefined{ZZZ}{}{\makeatletter\input gnuindex.sty\makeatother\makeindex\makeatletter}%
\@ifundefined{abstract}{%
 \def\abstract{%
  \if@twocolumn
   \section*{Abstract (Not appropriate in this style!)}%
   \else \small 
   \begin{center}{\bf Abstract\vspace{-.5em}\vspace{\z@}}\end{center}%
   \quotation 
   \fi
  }%
 }{%
 }%
\@ifundefined{endabstract}{\def\endabstract
  {\if@twocolumn\else\endquotation\fi}}{}%
\@ifundefined{maketitle}{\def\maketitle#1{}}{}%
\@ifundefined{affiliation}{\def\affiliation#1{}}{}%
\@ifundefined{proof}{}{}%
\@ifundefined{endproof}{}{}%
\@ifundefined{newfield}{\def\newfield#1#2{}}{}%
\@ifundefined{chapter}{\def\chapter#1{\par(Chapter head:)#1\par }%
 \newcount\c@chapter}{}%
\@ifundefined{part}{\def\part#1{\par(Part head:)#1\par }}{}%
\@ifundefined{section}{\def\section#1{\par(Section head:)#1\par }}{}%
\@ifundefined{subsection}{\def\subsection#1%
 {\par(Subsection head:)#1\par }}{}%
\@ifundefined{subsubsection}{\def\subsubsection#1%
 {\par(Subsubsection head:)#1\par }}{}%
\@ifundefined{paragraph}{\def\paragraph#1%
 {\par(Subsubsubsection head:)#1\par }}{}%
\@ifundefined{subparagraph}{\def\subparagraph#1%
 {\par(Subsubsubsubsection head:)#1\par }}{}%
\@ifundefined{therefore}{}{}%
\@ifundefined{backepsilon}{}{}%
\@ifundefined{yen}{}{}%
\@ifundefined{registered}{%
   \def\registered{\relax\ifmmode{}\r@gistered
                    \else$\m@th\r@gistered$\fi}%
 \def\r@gistered{^{\ooalign
  {\hfil\raise.07ex\hbox{$\scriptstyle\rm\text{R}$}\hfil\crcr
  \mathhexbox20D}}}}{}%
\@ifundefined{Eth}{}{}%
\@ifundefined{eth}{}{}%
\@ifundefined{Thorn}{}{}%
\@ifundefined{thorn}{}{}%
\@ifundefined{degree}{}{}%
\newdimen\theight
\def\Column{%
 \vadjust{\setbox\z@=\hbox{\scriptsize\quad\quad tcol}%
  \theight=\ht\z@\advance\theight by \dp\z@\advance\theight by \lineskip
  \kern -\theight \vbox to \theight{%
   \rightline{\rlap{\box\z@}}%
   \vss
   }%
  }%
 }%
\def\qed{%
 \ifhmode\unskip\nobreak\fi\ifmmode\ifinner\else\hskip5\p@\fi\fi
 \hbox{\hskip5\p@\vrule width4\p@ height6\p@ depth1.5\p@\hskip\p@}%
 }%
\def\miss{\hbox{\vrule height2\p@ width 2\p@ depth\z@}}%
\def\tcol#1{{\baselineskip=6\p@ \vcenter{#1}} \Column}  %
\def\newfmtname{LaTeX2e}
\def\chkcompat{%
   \if@compatibility
   \else
     \usepackage{latexsym}
   \fi
}

\ifx\fmtname\newfmtname
  \DeclareOldFontCommand{\rm}{\normalfont\rmfamily}{\mathrm}
  \DeclareOldFontCommand{\sf}{\normalfont\sffamily}{\mathsf}
  \DeclareOldFontCommand{\tt}{\normalfont\ttfamily}{\mathtt}
  \DeclareOldFontCommand{\bf}{\normalfont\bfseries}{\mathbf}
  \DeclareOldFontCommand{\it}{\normalfont\itshape}{\mathit}
  \DeclareOldFontCommand{\sl}{\normalfont\slshape}{\@nomath\sl}
  \DeclareOldFontCommand{\sc}{\normalfont\scshape}{\@nomath\sc}
  \chkcompat
\fi

\def\alpha{{\Greekmath 010B}}%
\def\beta{{\Greekmath 010C}}%
\def\gamma{{\Greekmath 010D}}%
\def\delta{{\Greekmath 010E}}%
\def\epsilon{{\Greekmath 010F}}%
\def\zeta{{\Greekmath 0110}}%
\def\eta{{\Greekmath 0111}}%
\def\theta{{\Greekmath 0112}}%
\def\iota{{\Greekmath 0113}}%
\def\kappa{{\Greekmath 0114}}%
\def\lambda{{\Greekmath 0115}}%
\def\mu{{\Greekmath 0116}}%
\def\nu{{\Greekmath 0117}}%
\def\xi{{\Greekmath 0118}}%
\def\pi{{\Greekmath 0119}}%
\def\rho{{\Greekmath 011A}}%
\def\sigma{{\Greekmath 011B}}%
\def\tau{{\Greekmath 011C}}%
\def\upsilon{{\Greekmath 011D}}%
\def\phi{{\Greekmath 011E}}%
\def\chi{{\Greekmath 011F}}%
\def\psi{{\Greekmath 0120}}%
\def\omega{{\Greekmath 0121}}%
\def\varepsilon{{\Greekmath 0122}}%
\def\vartheta{{\Greekmath 0123}}%
\def\varpi{{\Greekmath 0124}}%
\def\varrho{{\Greekmath 0125}}%
\def\varsigma{{\Greekmath 0126}}%
\def\varphi{{\Greekmath 0127}}%

\def\nabla{{\Greekmath 0272}}
\def\FindBoldGroup{%
   {\setbox0=\hbox{$\mathbf{x\global\edef\theboldgroup{\the\mathgroup}}$}}%
}

\def\Greekmath#1#2#3#4{%
    \if@compatibility
        \ifnum\mathgroup=\symbold
           \mathchoice{\mbox{\boldmath$\displaystyle\mathchar"#1#2#3#4$}}%
                      {\mbox{\boldmath$\textstyle\mathchar"#1#2#3#4$}}%
                      {\mbox{\boldmath$\scriptstyle\mathchar"#1#2#3#4$}}%
                      {\mbox{\boldmath$\scriptscriptstyle\mathchar"#1#2#3#4$}}%
        \else
           \mathchar"#1#2#3#4%
        \fi 
    \else 
        \FindBoldGroup
        \ifnum\mathgroup=\theboldgroup 
           \mathchoice{\mbox{\boldmath$\displaystyle\mathchar"#1#2#3#4$}}%
                      {\mbox{\boldmath$\textstyle\mathchar"#1#2#3#4$}}%
                      {\mbox{\boldmath$\scriptstyle\mathchar"#1#2#3#4$}}%
                      {\mbox{\boldmath$\scriptscriptstyle\mathchar"#1#2#3#4$}}%
        \else
           \mathchar"#1#2#3#4%
        \fi     	    
	  \fi}

\newif\ifGreekBold  \GreekBoldfalse
\let\SAVEPBF=\pbf
\def\pbf{\GreekBoldtrue\SAVEPBF}%

\@ifundefined{theorem}{}{}
\@ifundefined{lemma}{}{}
\@ifundefined{corollary}{}{}
\@ifundefined{conjecture}{}{}
\@ifundefined{proposition}{}{}
\@ifundefined{axiom}{}{}
\@ifundefined{remark}{}{}
\@ifundefined{example}{}{}
\@ifundefined{exercise}{}{}
\@ifundefined{definition}{}{}

\@ifundefined{mathletters}{%
  \newcounter{equationnumber}  
  \def\mathletters{%
     \addtocounter{equation}{1}
     \edef\@currentlabel{\theequation}%
     \setcounter{equationnumber}{\c@equation}
     \setcounter{equation}{0}%
     \edef\theequation{\@currentlabel\noexpand\alph{equation}}%
  }
  
}{}

\@ifundefined{BibTeX}{%
    \def\BibTeX{{\rm B\kern-.05em{\sc i\kern-.025em b}\kern-.08em
                 T\kern-.1667em\lower.7ex\hbox{E}\kern-.125emX}}}{}%
\@ifundefined{AmS}%
    {\def\AmS{{\protect\usefont{OMS}{cmsy}{m}{n}%
                A\kern-.1667em\lower.5ex\hbox{M}\kern-.125emS}}}{}%
\@ifundefined{AmSTeX}{}{}%
\ifx\ds@amstex\relax
\makeatother\endinput
\else
   \@ifpackageloaded{amstex}%
      {\message{amstex already loaded}\makeatother\endinput}
      {}
   \@ifpackageloaded{amsgen}%
      {\message{amsgen already loaded}\makeatother\endinput}
      {}
\fi
\let\DOTSI\relax
\def\RIfM@{\relax\ifmmode}%
\def\FN@{\futurelet\next}%
\newcount\intno@
\def\iint{\DOTSI\intno@\tw@\FN@\ints@}%
\def\iiint{\DOTSI\intno@\thr@@\FN@\ints@}%
\def\iiiint{\DOTSI\intno@4 \FN@\ints@}%
\def\idotsint{\DOTSI\intno@\z@\FN@\ints@}%
\def\ints@{\findlimits@\ints@@}%
\newif\iflimtoken@
\newif\iflimits@
\def\findlimits@{\limtoken@true\ifx\next\limits\limits@true
 \else\ifx\next\nolimits\limits@false\else
 \limtoken@false\ifx\ilimits@\nolimits\limits@false\else
 \ifinner\limits@false\else\limits@true\fi\fi\fi\fi}%
\def\multint@{\int\ifnum\intno@=\z@\intdots@                          
 \else\intkern@\fi                                                    
 \ifnum\intno@>\tw@\int\intkern@\fi                                   
 \ifnum\intno@>\thr@@\int\intkern@\fi                                 
 \int}
\def\multintlimits@{\intop\ifnum\intno@=\z@\intdots@\else\intkern@\fi
 \ifnum\intno@>\tw@\intop\intkern@\fi
 \ifnum\intno@>\thr@@\intop\intkern@\fi\intop}%
\def\intic@{%
    \mathchoice{\hskip.5em}{\hskip.4em}{\hskip.4em}{\hskip.4em}}%
\def\negintic@{\mathchoice
 {\hskip-.5em}{\hskip-.4em}{\hskip-.4em}{\hskip-.4em}}%
\def\ints@@{\iflimtoken@                                              
 \def\ints@@@{\iflimits@\negintic@
   \mathop{\intic@\multintlimits@}\limits                             
  \else\multint@\nolimits\fi                                          
  \eat@}
 \else                                                                
 \def\ints@@@{\iflimits@\negintic@
  \mathop{\intic@\multintlimits@}\limits\else
  \multint@\nolimits\fi}\fi\ints@@@}%
\def\intkern@{\mathchoice{\!\!\!}{\!\!}{\!\!}{\!\!}}%
\def\plaincdots@{\mathinner{\cdotp\cdotp\cdotp}}%
\def\intdots@{\mathchoice{\plaincdots@}%
 {{\cdotp}\mkern1.5mu{\cdotp}\mkern1.5mu{\cdotp}}%
 {{\cdotp}\mkern1mu{\cdotp}\mkern1mu{\cdotp}}%
 {{\cdotp}\mkern1mu{\cdotp}\mkern1mu{\cdotp}}}%
\def\RIfM@{\relax\protect\ifmmode}
\def\text{\RIfM@\expandafter\text@\else\expandafter\mbox\fi}
\let\nfss@text\text
\def\text@#1{\mathchoice
   {\textdef@\displaystyle\f@size{#1}}%
   {\textdef@\textstyle\tf@size{\firstchoice@false #1}}%
   {\textdef@\textstyle\sf@size{\firstchoice@false #1}}%
   {\textdef@\textstyle \ssf@size{\firstchoice@false #1}}%
   \glb@settings}

\def\textdef@#1#2#3{\hbox{{%
                    \everymath{#1}%
                    \let\f@size#2\selectfont
                    #3}}}
\newif\iffirstchoice@
\firstchoice@true
\def\Let@{\relax\iffalse{\fi\let\\=\cr\iffalse}\fi}%
\def\vspace@{\def\vspace##1{\crcr\noalign{\vskip##1\relax}}}%
\def\multilimits@{\bgroup\vspace@\Let@
 \baselineskip\fontdimen10 \scriptfont\tw@
 \advance\baselineskip\fontdimen12 \scriptfont\tw@
 \lineskip\thr@@\fontdimen8 \scriptfont\thr@@
 \lineskiplimit\lineskip
 \vbox\bgroup\ialign\bgroup\hfil$\m@th\scriptstyle{##}$\hfil\crcr}%
\def\Sb{_\multilimits@}%
\def\endSb{\crcr\egroup\egroup\egroup}%
\def\Sp{^\multilimits@}%

\newdimen\ex@
\def\rightarrowfill@#1{$#1\m@th\mathord-\mkern-6mu\cleaders
 \hbox{$#1\mkern-2mu\mathord-\mkern-2mu$}\hfill
 \mkern-6mu\mathord\rightarrow$}%
\def\leftarrowfill@#1{$#1\m@th\mathord\leftarrow\mkern-6mu\cleaders
 \hbox{$#1\mkern-2mu\mathord-\mkern-2mu$}\hfill\mkern-6mu\mathord-$}%
\def\leftrightarrowfill@#1{$#1\m@th\mathord\leftarrow
\mkern-6mu\cleaders
 \hbox{$#1\mkern-2mu\mathord-\mkern-2mu$}\hfill
 \mkern-6mu\mathord\rightarrow$}%
\def\overrightarrow{\mathpalette\overrightarrow@}%
\def\overrightarrow@#1#2{\vbox{\ialign{##\crcr\rightarrowfill@#1\crcr
 \noalign{\kern-\ex@\nointerlineskip}$\m@th\hfil#1#2\hfil$\crcr}}}%

\def\overleftarrow{\mathpalette\overleftarrow@}%
\def\overleftarrow@#1#2{\vbox{\ialign{##\crcr\leftarrowfill@#1\crcr
 \noalign{\kern-\ex@\nointerlineskip}$\m@th\hfil#1#2\hfil$\crcr}}}%
\def\overleftrightarrow{\mathpalette\overleftrightarrow@}%
\def\overleftrightarrow@#1#2{\vbox{\ialign{##\crcr
   \leftrightarrowfill@#1\crcr
 \noalign{\kern-\ex@\nointerlineskip}$\m@th\hfil#1#2\hfil$\crcr}}}%
\def\underrightarrow{\mathpalette\underrightarrow@}%
\def\underrightarrow@#1#2{\vtop{\ialign{##\crcr$\m@th\hfil#1#2\hfil
  $\crcr\noalign{\nointerlineskip}\rightarrowfill@#1\crcr}}}%

\def\underleftarrow{\mathpalette\underleftarrow@}%
\def\underleftarrow@#1#2{\vtop{\ialign{##\crcr$\m@th\hfil#1#2\hfil
  $\crcr\noalign{\nointerlineskip}\leftarrowfill@#1\crcr}}}%
\def\underleftrightarrow{\mathpalette\underleftrightarrow@}%
\def\underleftrightarrow@#1#2{\vtop{\ialign{##\crcr$\m@th
  \hfil#1#2\hfil$\crcr
 \noalign{\nointerlineskip}\leftrightarrowfill@#1\crcr}}}%

\def\qopnamewl@#1{\mathop{\operator@font#1}\nlimits@}
\let\nlimits@\displaylimits
\def\setboxz@h{\setbox\z@\hbox}

\def\varlim@#1#2{\mathop{\vtop{\ialign{##\crcr
 \hfil$#1\m@th\operator@font lim$\hfil\crcr
 \noalign{\nointerlineskip}#2#1\crcr
 \noalign{\nointerlineskip\kern-\ex@}\crcr}}}}

 \def\rightarrowfill@#1{\m@th\setboxz@h{$#1-$}\ht\z@\z@
  $#1\copy\z@\mkern-6mu\cleaders
  \hbox{$#1\mkern-2mu\box\z@\mkern-2mu$}\hfill
  \mkern-6mu\mathord\rightarrow$}
\def\leftarrowfill@#1{\m@th\setboxz@h{$#1-$}\ht\z@\z@
  $#1\mathord\leftarrow\mkern-6mu\cleaders
  \hbox{$#1\mkern-2mu\copy\z@\mkern-2mu$}\hfill
  \mkern-6mu\box\z@$}

\def\projlim{\qopnamewl@{proj\,lim}}
\def\injlim{\qopnamewl@{inj\,lim}}
\def\varinjlim{\mathpalette\varlim@\rightarrowfill@}
\def\varprojlim{\mathpalette\varlim@\leftarrowfill@}
\def\varliminf{\mathpalette\varliminf@{}}
\def\varliminf@#1{\mathop{\underline{\vrule\@depth.2\ex@\@width\z@
   \hbox{$#1\m@th\operator@font lim$}}}}
\def\varlimsup{\mathpalette\varlimsup@{}}
\def\varlimsup@#1{\mathop{\overline
  {\hbox{$#1\m@th\operator@font lim$}}}}

\def\tfrac#1#2{{\textstyle {#1 \over #2}}}%
\def\dfrac#1#2{{\displaystyle {#1 \over #2}}}%
\def\dsum{\mathop{\displaystyle \sum }}%
\def\stackunder#1#2{\mathrel{\mathop{#2}\limits_{#1}}}%
\begingroup \catcode `|=0 \catcode `[= 1
\catcode`]=2 \catcode `\{=12 \catcode `\}=12
\catcode`\\=12 
|gdef|@alignverbatim#1\end{align}[#1|end[align]]
|gdef|@salignverbatim#1\end{align*}[#1|end[align*]]

|gdef|@alignatverbatim#1\end{alignat}[#1|end[alignat]]
|gdef|@salignatverbatim#1\end{alignat*}[#1|end[alignat*]]

|gdef|@xalignatverbatim#1\end{xalignat}[#1|end[xalignat]]
|gdef|@sxalignatverbatim#1\end{xalignat*}[#1|end[xalignat*]]

|gdef|@gatherverbatim#1\end{gather}[#1|end[gather]]
|gdef|@sgatherverbatim#1\end{gather*}[#1|end[gather*]]

|gdef|@gatherverbatim#1\end{gather}[#1|end[gather]]
|gdef|@sgatherverbatim#1\end{gather*}[#1|end[gather*]]

|gdef|@multilineverbatim#1\end{multiline}[#1|end[multiline]]
|gdef|@smultilineverbatim#1\end{multiline*}[#1|end[multiline*]]

|gdef|@arraxverbatim#1\end{arrax}[#1|end[arrax]]
|gdef|@sarraxverbatim#1\end{arrax*}[#1|end[arrax*]]

|gdef|@tabulaxverbatim#1\end{tabulax}[#1|end[tabulax]]
|gdef|@stabulaxverbatim#1\end{tabulax*}[#1|end[tabulax*]]

|endgroup

\def\align{\@verbatim \frenchspacing\@vobeyspaces \@alignverbatim
You are using the "align" environment in a style in which it is not defined.}

\@namedef{align*}{\@verbatim\@salignverbatim
You are using the "align*" environment in a style in which it is not defined.}
\expandafter\let\csname endalign*\endcsname =\endtrivlist

\def\alignat{\@verbatim \frenchspacing\@vobeyspaces \@alignatverbatim
You are using the "alignat" environment in a style in which it is not defined.}

\@namedef{alignat*}{\@verbatim\@salignatverbatim
You are using the "alignat*" environment in a style in which it is not defined.}
\expandafter\let\csname endalignat*\endcsname =\endtrivlist

\def\xalignat{\@verbatim \frenchspacing\@vobeyspaces \@xalignatverbatim
You are using the "xalignat" environment in a style in which it is not defined.}

\@namedef{xalignat*}{\@verbatim\@sxalignatverbatim
You are using the "xalignat*" environment in a style in which it is not defined.}
\expandafter\let\csname endxalignat*\endcsname =\endtrivlist

\def\gather{\@verbatim \frenchspacing\@vobeyspaces \@gatherverbatim
You are using the "gather" environment in a style in which it is not defined.}

\@namedef{gather*}{\@verbatim\@sgatherverbatim
You are using the "gather*" environment in a style in which it is not defined.}
\expandafter\let\csname endgather*\endcsname =\endtrivlist

\def\multiline{\@verbatim \frenchspacing\@vobeyspaces \@multilineverbatim
You are using the "multiline" environment in a style in which it is not defined.}

\@namedef{multiline*}{\@verbatim\@smultilineverbatim
You are using the "multiline*" environment in a style in which it is not defined.}
\expandafter\let\csname endmultiline*\endcsname =\endtrivlist

\def\arrax{\@verbatim \frenchspacing\@vobeyspaces \@arraxverbatim
You are using a type of "array" construct that is only allowed in AmS-LaTeX.}

\def\tabulax{\@verbatim \frenchspacing\@vobeyspaces \@tabulaxverbatim
You are using a type of "tabular" construct that is only allowed in AmS-LaTeX.}

\@namedef{arrax*}{\@verbatim\@sarraxverbatim
You are using a type of "array*" construct that is only allowed in AmS-LaTeX.}
\expandafter\let\csname endarrax*\endcsname =\endtrivlist

\@namedef{tabulax*}{\@verbatim\@stabulaxverbatim
You are using a type of "tabular*" construct that is only allowed in AmS-LaTeX.}
\expandafter\let\csname endtabulax*\endcsname =\endtrivlist

\def\@@eqncr{\let\@tempa\relax
    \ifcase\@eqcnt \def\@tempa{& & &}\or \def\@tempa{& &}%
      \else \def\@tempa{&}\fi
     \@tempa
     \if@eqnsw
        \iftag@
           \@taggnum
        \else
           \@eqnnum\stepcounter{equation}%
        \fi
     \fi
     \global\tag@false
     \global\@eqnswtrue
     \global\@eqcnt\z@\cr}

 \def\endequation{%
     \ifmmode\ifinner 
      \iftag@
        \addtocounter{equation}{-1} 
        $\hfil
           \displaywidth\linewidth\@taggnum\egroup \endtrivlist
        \global\tag@false
        \global\@ignoretrue   
      \else
        $\hfil
           \displaywidth\linewidth\@eqnnum\egroup \endtrivlist
        \global\tag@false
        \global\@ignoretrue 
      \fi
     \else   
      \iftag@
        \addtocounter{equation}{-1} 
        \eqno \hbox{\@taggnum}
        \global\tag@false%
        $$\global\@ignoretrue
      \else
        \eqno \hbox{\@eqnnum}
        $$\global\@ignoretrue
      \fi
     \fi\fi
 } 

 \newif\iftag@ \tag@false
 
 \def\tag{\@ifnextchar*{\@tagstar}{\@tag}}
 \def\@tag#1{%
     \global\tag@true
     \global\def\@taggnum{(#1)}}
 \def\@tagstar*#1{%
     \global\tag@true
     \global\def\@taggnum{#1}%
}

\makeatother

\begin{document}

\author{Christopher Smith\thanks{\protect\bigskip \protect\bigskip \protect\bigskip 
\protect\bigskip e-mail : smith@fyma.fyma.ucl.ac.be} \and \smallskip \\
\textit{Institut de Physique Th\'{e}orique}\\
\textit{Universit\'{e} Catholique de Louvain-la-Neuve}\\
\textit{Chemin du Cyclotron, 2, B-1348 Louvain-la-Neuve, Belgium}}
\title{{\huge SU(N)\ Elastic\ Rescattering\ in\ B\ and\ D\ Decays.}}
\date{February 17, 1999}
\maketitle

\begin{abstract}
The treatment of elastic final state interactions (FSI) under a symmetry
group is presented. The proposed model is based on Watson's theorem, i.e.
on symmetry properties of the \textbf{S}-matrix and on its unitarity. This
theorem provides an easy way to introduce rescattering effects by defining
final state interactions mixing matrices. A symmetry group fixes the
structure of such mixing matrices, and the passage from one group to another
is studied (for example, SU(2) to SU(3)). Mixings among two charmless
pseudoscalar decay product states will be systematically analyzed. Finally,
these mixing matrices will be used on quark diagram parametrizations of B
and D decay amplitudes. This will have some important consequences on the
definition of quark diagrams. It will be argued that these diagrams should
not contain any FSI effects, i.e. they should be real (except for CKM
factors). FSI are then introduced at the hadronic level, by mixing basic
quark diagram topologies.
\end{abstract}

\pagebreak

\section{\protect\smallskip Introduction}

In this work, we will present a method for implementing final state
interactions (FSI or rescattering), i.e. the strong interactions between
weak decay products. These FSI will be treated as elastic, and a special
care will be devoted to define this concept. In particular, elasticity under
a symmetry group will be defined as a special case of a generalized
elasticity concept. This work has to be understood as a first step beyond the
trivial treatment of FSI, where FSI are introduced as elastic under SU(2). 
The decays we have in mind are the B and D decays to two charmless pseudoscalars. 
In these B decays, CP violation is expected to
occur. To be able to extract the values of the relevant parameters from
experiments, in order to compare them to standard model values, we must
dispose of an appropriate parametrization. Quark diagrams are usually
thought to be appropriate for such a goal, but, as we will see, these quark
diagrams should be properly defined in order to be of any use. In all this,
FSI play no fundamental role, they just mix up final states. Therefore, it
is necessary to treat them to reach the underlying dynamics. The interesting
point is that the model we propose to treat elastic FSI, based on the \textbf{S}%
-matrix and Watson's theorem,\ will point towards a specific definition of
the quark diagrams. A lot of papers exist on this subject, some of them are
listed in the bibliography.

The decays we are considering proceed via the weak decay of $b$ or $%
\overline{c}$ quark. We will treat weak interaction at the lowest order. The
strong interactions are involved in the three following processes : they
renormalize the weak interaction, they confine quarks into hadrons and they
determine the asymptotic out states (FSI). Obviously, these two last
manifestations of strong interactions are a priori difficult to distinguish
from each other, because the out states can be considered as completely
hadronized only when they no longer interact. The definition of FSI will be
based on the following consideration : only hadrons, and not quarks enter
the S-matrix. Consequently, FSI will be defined as the (strong) interactions
between hadrons. A typical decay process like $\overline{D^{0}}\rightarrow
K^{+}\pi ^{-}$ is a heavy quark $\overline{c}$ decaying ''quickly'' followed
by the hadronization. This produces an intermediate real hadron state noted
inside accolades: for example $\overline{D^{0}}\rightarrow \left\{
P_{1}P_{2}\right\} $. Then these hadrons interact by the FSI towards the
final state: $\left\{ P_{1}P_{2}\right\} \rightarrow K^{+}\pi ^{-}$. This
picture is quite schematic and we could say as well that we define
intermediate decay amplitudes $\overline{D^{0}}\rightarrow \left\{
P_{1}P_{2}\right\} $ (also qualified as bare) as free of any FSI effects. In
other words, FSI factorize from weak bare decay amplitudes. These bare
amplitudes have no absorptive part since they must be real except for CKM
factors (equivalently, their behaviour under CP is simply CP$\left( 
\overline{D^{0}}\rightarrow \left\{ P_{1}P_{2}\right\} \right) =\left( 
\overline{D^{0}}\rightarrow \left\{ P_{1}P_{2}\right\} \right) ^{*}$).

To summarize, the model will be based on three main points. (1) Unitarity of
the \textbf{S}-matrix for a given set of rescattering channels, (2) the
identification of bare amplitudes (elementary processes) as the part that get 
complex conjugated under CP. These two points will then imply that (3) FSI are 
treated as elastic among the chosen set of rescattering channels. 
Usually, when adding phases to isospin amplitudes to introduce FSI, one is implicitly 
considering that the \textbf{S}-matrix is unitary when restricted to a set of
rescattering channels belonging to the same isomultiplet. We will extend this to more general 
sets of rescattering channels. 
This model can be characterized by the way bare amplitudes are identified. This is a
hypothesis, which is strictly equivalent to the general elasticity hypothesis as soon as 
the \textbf{S}-matrix is unitary. The validity of the present approach is discussed in the 
conclusion.  Note that other propositions exist for the identification of bare amplitudes; see for
instance the K-matrix formalism, which modelize an inelastic approach to the treatment of FSI.

From this picture, we will naturally introduce quark diagrams (QD) at the
bare level. FSI are then viewed as mixings of these bare amplitudes. Since
bare amplitudes must be real, these QD are defined as real. The important
point is that by defining QD at the bare level, they are characterized by
basic topological configurations. This in turn is very important to relate
these QD to elementary dynamical processes. Basic topologies are then mixed
by FSI. We will develop all this further in the text.

For a given final state, we cannot have arbitrary intermediate states. FSI
being strong interactions, these intermediate states must have the same
charge, strangeness,... Also, the available energy will determine the set of
coupled open channels for a given set of quantum numbers. Among these
coupled states, we will only consider two-pseudoscalar states. Thus we are
neglecting transitions between these PP states and many particle states,
vector meson states... This will be used when demonstrating Watson's theorem.

The approximate invariance under flavour exchange of the strong interaction
implies very severe constraints on decay amplitudes and on FSI. At the B or
D mass, SU(2) or SU(3) are expected not to be badly broken. As we will see
extensively, working under a symmetry group fixes the set of coupled states,
this set being bigger under SU(3) than under SU(2). The symmetry group also
fixes the structure of the couplings of these states. These couplings (or
mixings) will be called SU(N) elastic (N=2, 3,...). One immediate question
is to find a link between a SU(2) description and a SU(3) description of
FSI, and this will be thoroughly carried out. Phenomenologically, it is
sometimes questionable to treat SU(3) mixings as elastic, one example
detailed at the end of this paper is the well-known SU(3) prediction $%
\overline{D^{0}}\rightarrow K^{0}\overline{K^{0}}=0$, which can be lifted by
a SU(3) breaking in the FSI. On the other hand the SU(2) restriction may be
too strong, since we neglect many possible rescattering channels. An
intermediate way is proposed in this work, by distinguishing elasticity from
elasticity under a symmetry group.

Let us first recall how FSI are usually treated when working under a flavour
symmetry group.

\subsection{SU(N) analyses of B and D decays}

The B and D decays we wish to describe are those into two charmless
pseudoscalars. We will work under SU(2) or SU(3), at the lowest order in
electroweak interaction.

\subsubsection{Isospin analysis}

Let us analyze the decays $\overline{D^{0}}$ to $K^{+}\pi ^{-}$ and $%
K^{0}\pi ^{0}$ under SU(2). The well-known isospin analysis leads to the
following parametrization of the physical decay amplitudes : 
\begin{equation}
\left\{ 
\begin{array}{l}
\left( \overline{D}\rightarrow K^{+}\pi ^{-}\right) =A^{3/2}+A^{1/2} \\ 
\left( \overline{D}\rightarrow K^{0}\pi ^{0}\right) =\frac{1}{\sqrt{2}}%
\left( 2A^{3/2}-A^{1/2}\right)
\end{array}
\right.  \label{isofull}
\end{equation}

where these isospin amplitudes correspond to $A^{T}=\left\langle T\left|
H_{W}=1\right| 1/2\right\rangle .$ CKM elements are not explicitly written.
These amplitudes contain the weak interaction at the lowest order, and all
the strong interaction, including FSI. The usual procedure to take into
account FSI is to associate phases to the isospin amplitudes as :

\begin{equation}
\left\{ 
\begin{array}{l}
A^{3/2}=e^{i\delta _{3/2}}\ A_{b}^{3/2} \\ 
A^{1/2}=e^{i\delta _{1/2}}\ A_{b}^{1/2}
\end{array}
\right.  \label{isorenorm}
\end{equation}

We can therefore identify the bare amplitudes for these decays :

\begin{equation}
\left\{ 
\begin{array}{l}
\left( \overline{D}\rightarrow \left\{ K^{+}\pi ^{-}\right\} \right)
=A_{b}^{3/2}+A_{b}^{1/2} \\ 
\left( \overline{D}\rightarrow \left\{ K^{0}\pi ^{0}\right\} \right) =\frac{1%
}{\sqrt{2}}\left( 2A_{b}^{3/2}-A_{b}^{1/2}\right)
\end{array}
\right.  \label{isobare}
\end{equation}

From these bare amplitudes, we can reintroduce FSI using a matrix procedure
: 
\begin{equation}
\left( 
\begin{array}{c}
\left( \overline{D^{0}}\rightarrow K^{+}\pi ^{-}\right) \\ 
\left( \overline{D^{0}}\rightarrow K^{0}\pi ^{0}\right)
\end{array}
\right) =M^{SU(2)}\left( 
\begin{array}{c}
\left( \overline{D^{0}}\rightarrow \left\{ K^{+}\pi ^{-}\right\} \right) \\ 
\left( \overline{D^{0}}\rightarrow \left\{ K^{0}\pi ^{0}\right\} \right)
\end{array}
\right)  \label{intro1}
\end{equation}

with $M^{SU(2)}$ given by :

\begin{equation}
M^{SU(2)}=\frac{1}{3}\left( 
\begin{array}{cc}
e^{i\delta _{3/2}}+2e^{i\delta _{1/2}} & \sqrt{2}\left( e^{i\delta
_{3/2}}-e^{i\delta _{1/2}}\right) \\ 
\sqrt{2}\left( e^{i\delta _{3/2}}-e^{i\delta _{1/2}}\right) & 2e^{i\delta
_{3/2}}+e^{i\delta _{1/2}}
\end{array}
\right)  \label{su(2)kpmix}
\end{equation}

This matrix method is strictly equivalent to the usual procedure (\ref
{isorenorm}). However, it is now apparent that FSI are introduced as mixings
between the $\left\{ K^{+}\pi ^{-}\right\} $ and $\left\{ K^{0}\pi
^{0}\right\} $ intermediate states.

\subsubsection{SU(3) analysis}

The SU(3) analyses of B and D decays into two uncharmed pseudoscalars are
given in the appendix. Let us consider the following set of decays :

\begin{equation}
\left\{ 
\begin{array}{l}
\left( \overline{D}^{0}\rightarrow K^{+}\pi ^{-}\right) =\left(
-4A^{27}+4A^{8}-2B^{8}\right) \\ 
\left( \overline{D}^{0}\rightarrow K^{0}\pi ^{0}\right) =\frac{1}{\sqrt{2}}%
\left( -6A^{27}-4A^{8}+2B^{8}\right) \\ 
\left( \overline{D}^{0}\rightarrow K^{0}\eta _{8}\right) =\frac{1}{\sqrt{6}}%
\left( -6A^{27}-4A^{8}+2B^{8}\right)
\end{array}
\right.  \label{exsu(3)}
\end{equation}

again CKM elements are not written explicitly. The usual procedure to take
FSI into account in this SU(3) context is simply :

\begin{equation}
\left\{ 
\begin{array}{l}
A^{27}\rightarrow e^{i\delta _{27}}A_{b}^{27}=A^{27} \\ 
X^{8}\rightarrow e^{i\delta _{8}}X_{b}^{8}=X^{8}\text{ \quad with \quad }%
X=A,\ B
\end{array}
\right.  \label{su(3)usualpre}
\end{equation}

We can therefore identify bare decays as :

\begin{equation}
\left\{ 
\begin{array}{l}
\left( \overline{D}^{0}\rightarrow \left\{ K^{+}\pi ^{-}\right\} \right)
=\left( -4A_{b}^{27}+4A_{b}^{8}-2B_{b}^{8}\right) \\ 
\left( \overline{D}^{0}\rightarrow \left\{ K^{0}\pi ^{0}\right\} \right) =%
\frac{1}{\sqrt{2}}\left( -6A_{b}^{27}-4A_{b}^{8}+2B_{b}^{8}\right) \\ 
\left( \overline{D}^{0}\rightarrow \left\{ K^{0}\eta _{8}\right\} \right) =%
\frac{1}{\sqrt{6}}\left( -6A_{b}^{27}-4A_{b}^{8}+2B_{b}^{8}\right)
\end{array}
\right.  \label{exsu(3)bare}
\end{equation}

And starting with these decompositions, we can reintroduce FSI using a
matrix procedure :

\begin{equation}
\left( 
\begin{array}{c}
\left( \overline{D^{0}}\rightarrow K^{+}\pi ^{-}\right) \\ 
\left( \overline{D^{0}}\rightarrow K^{0}\pi ^{0}\right) \\ 
\left( \overline{D^{0}}\rightarrow K^{0}\eta _{8}\right)
\end{array}
\right) =M^{SU(3)}\left( 
\begin{array}{c}
\left( \overline{D^{0}}\rightarrow \left\{ K^{+}\pi ^{-}\right\} \right) \\ 
\left( \overline{D^{0}}\rightarrow \left\{ K^{0}\pi ^{0}\right\} \right) \\ 
\left( \overline{D^{0}}\rightarrow \left\{ K^{0}\eta _{8}\right\} \right)
\end{array}
\right)  \label{appmex}
\end{equation}

with

\begin{equation}
M^{SU(3)}=\frac{1}{5}\left( 
\begin{array}{ccc}
2e^{i\delta _{27}}+3e^{i\delta _{8}} & \frac{3}{\sqrt{2}}\left( e^{i\delta
_{27}}-e^{i\delta _{8}}\right) & \sqrt{\frac{3}{2}}\left( e^{i\delta
_{27}}-e^{i\delta _{8}}\right) \\ 
\frac{3}{\sqrt{2}}\left( e^{i\delta _{27}}-e^{i\delta _{8}}\right) & \frac{1%
}{2}\left( 7e^{i\delta _{27}}+3e^{i\delta _{8}}\right) & -\frac{\sqrt{3}}{2}%
\left( e^{i\delta _{27}}-e^{i\delta _{8}}\right) \\ 
\sqrt{\frac{3}{2}}\left( e^{i\delta _{27}}-e^{i\delta _{8}}\right) & -\frac{%
\sqrt{3}}{2}\left( e^{i\delta _{27}}-e^{i\delta _{8}}\right) & \frac{1}{2}%
\left( 9e^{i\delta _{27}}+e^{i\delta _{8}}\right)
\end{array}
\right)  \label{su(3)kpmix}
\end{equation}

As for SU(2), we see that FSI effects reduce to some mixings among
intermediate states. But a major difference arises : under SU(3), the $%
\left\{ K^{0}\eta _{8}\right\} $ also mixes with $\left\{ K\pi \right\} $
states. This mixing goes beyond SU(2) since $\left\{ K\pi \right\} $ states
and $\left\{ K^{0}\eta _{8}\right\} $ are in different SU(2) representations.

The same matrix $M^{SU(3)}$ can also be used to introduce FSI in other decay
decompositions into matrix elements. For example, B$^{0}$ bare decays :

\begin{equation}
\left\{ 
\begin{array}{l}
\left( B^{0}\rightarrow \left\{ K^{+}\pi ^{-}\right\} \right)
=V_{ub}^{*}V_{us}\left( -4A_{b}^{27}-A_{b}^{8}-B_{b}^{8}-C_{b}^{8}\right)
+V_{cb}^{*}V_{cs}\left( -C_{b}^{8c}\right) +V_{tb}^{*}V_{ts}\left(
-C_{b}^{8t}\right) \\ 
\left( B^{0}\rightarrow \left\{ K^{0}\pi ^{0}\right\} \right) =\frac{1}{%
\sqrt{2}}\left( V_{ub}^{*}V_{us}\left(
-6A_{b}^{27}+A_{b}^{8}+B_{b}^{8}+C_{b}^{8}\right) +V_{cb}^{*}V_{cs}\left(
C_{b}^{8c}\right) +V_{tb}^{*}V_{ts}\left( C_{b}^{8t}\right) \right) \\ 
\left( B^{0}\rightarrow \left\{ K^{0}\eta _{8}\right\} \right) =\frac{1}{%
\sqrt{6}}\left( V_{ub}^{*}V_{us}\left(
-6A_{b}^{27}+A_{b}^{8}+B_{b}^{8}+C_{b}^{8}\right) +V_{cb}^{*}V_{cs}\left(
C_{b}^{8c}\right) +V_{tb}^{*}V_{ts}\left( C_{b}^{8t}\right) \right)
\end{array}
\right.  \label{su(3)B1}
\end{equation}

And we can see that applying $M^{SU(3)}$ is equivalent to the usual
prescription (\ref{su(3)usualpre}) and $C_{b}^{8}\rightarrow e^{i\delta
_{8}}C_{b}^{8}=C^{8}$.

\subsection{Questions :}

Having written those matrix representations for FSI, the following questions
can be addressed :

(1)\textit{\ What is the underlying theoretical framework ?} We would like
to know precisely the hypotheses concerning this procedure. Also, the
elasticity concept has to be properly defined. Finally, the properties of
these M Matrices like unitarity and symmetry should be explained. This
section is based on Watson's theorem.

(2)\textit{\ SU(N) flavor symmetry implications ?} We would like to find a
systematic way to calculate mixing matrices like $M^{SU(2)}$ and $M^{SU(3)}$%
. The link between these two matrices will also be analyzed. The fact that
the same matrix can be used for different sets of reactions will be
explained.

(3)\textit{\ The use of quark diagrams ?} We will argue that quark diagrams
should be used to parametrize bare decays. In other words, we will
parametrize physical decay amplitudes using FSI mixing matrices for the
rescattering effects, and quark diagrams for the weak decays, their gluonic
corrections (but no absorptive part) and the hadronizations.\medskip

These three points will be considered in the three following sections. In
the last section, we will apply the advocated procedure to analyze
systematically the B and D decays into two uncharmed pseudoscalars.\medskip

Let us summarize the general procedure we suggest in this paper. Watson's
theorem implies that the physical decay amplitudes for a set of processes
can be factorized into a FSI part and a bare part. We can then extract from
the full weak amplitudes the FSI contributions by putting some intermediate
states on-shell, and these states are hadron states entering the S matrix.
Bare amplitudes are then parametrized using quark diagrams, and FSI are
introduced using mixing matrices. Finally, dynamical (e.g. the choice of the
set of coupled states) or symmetry (e.g. SU(2)) considerations will
determine the form of these matrices.

\section{Theoretical framework.}

\subsection{Generalized Watson's theorem}

Part of the following discussion is borrowed from \cite{weinberg} and \cite
{fsiother4}. Watson's theorem will allow us to single out the final state
interaction effects inside the physical weak decay amplitudes of B or D.
Remember that we are working to lowest order in electroweak interaction. Let
us begin by expressing the generalized Watson's theorem. By W we denote the
column vector formed with the weak amplitudes into the possible final states
:

\begin{equation}
W=\left( 
\begin{array}{cccc}
B\rightarrow \pi \pi  & B\rightarrow \pi \pi \pi \pi  & B\rightarrow K%
\overline{K} & \cdots 
\end{array}
\right) ^{t}  \label{weakampl}
\end{equation}

S will be the S-matrix containing the coupling among these final states :

\begin{equation}
S=\left( 
\begin{array}{cccc}
\pi \pi \rightarrow \pi \pi & \pi \pi \rightarrow \pi \pi \pi \pi & \pi \pi
\rightarrow K\overline{K} & \cdots \\ 
\pi \pi \pi \pi \rightarrow \pi \pi & \pi \pi \pi \pi \rightarrow \pi \pi
\pi \pi & \pi \pi \pi \pi \rightarrow K\overline{K} & \cdots \\ 
K\overline{K}\rightarrow \pi \pi & K\overline{K}\rightarrow \pi \pi \pi \pi
& K\overline{K}\rightarrow K\overline{K} & \cdots \\ 
\vdots & \vdots & \vdots & \ddots
\end{array}
\right)  \label{strongcoupl}
\end{equation}

The processes entering this S-matrix proceed dominantly via strong
interaction, the weak contribution being much smaller. Thus this matrix is
block-diagonal, each block representing mixing among states of definite
flavour quantum numbers. The important point is that W and S are built from
hadron states like $K,\pi ,\eta ,D,B$,...; i.e. states which decay via
electroweak interactions only.\smallskip

The \textit{generalized Watson's theorem} then reads:

\begin{eqnarray}
W &=&\sqrt{S}W_{b}  \label{watsontheo} \\
CP(W) &=&\sqrt{S}W_{b}^{*}  \nonumber
\end{eqnarray}

This means that FSI effects contained in $\sqrt{S}$ factorize, leaving bare
amplitudes $W_{b}$, which contain no FSI. These amplitudes will be written
as :

\begin{equation}
W_{b}=\left( 
\begin{array}{cccc}
B\rightarrow \left\{ \pi \pi \right\}  & B\rightarrow \left\{ \pi \pi \pi
\pi \right\}  & B\rightarrow \left\{ K\overline{K}\right\}  & \cdots 
\end{array}
\right) ^{t}  \label{bareampl}
\end{equation}

where the $\left\{ {}\right\} $ denotes intermediate states. These bare
amplitudes contain the weak decay of the heavy quark, with its gluonic
corrections, and the hadronization of the intermediate hadron state, but no
FSI effect. These rescattering effects are introduced as interactions
between these intermediate hadron states using $\sqrt{S}$. We can also say
that this theorem allows one to extract from physical amplitudes the FSI
part from the bare part, and this is done at the hadronic level. In other
words, it is hadron states entering the S matrix that are put on mass-shell
as intermediate states. This can also be interpreted as a renormalization of
bare amplitudes induced by rescattering effects.

The complete demonstration is in the appendix. The main features are :

(1) Watson's theorem follows from the unitarity condition for the complete 
\textbf{S}-matrix built from W and S: 
\begin{equation}
\mathbf{S}^{\dagger }\mathbf{S=SS}^{\dagger }=1\text{ with }\mathbf{S}%
=\left( 
\begin{array}{cc}
1 & iW^{t} \\ 
iCP(W) & S
\end{array}
\right)   \label{unitarity}
\end{equation}

Thus we can say that $W=\sqrt{S}W_{b}$ is a unitarization of weak bare decay
amplitudes, since with the adjunction of the strong phases $\sqrt{S}$, the
full \textbf{S} matrix is unitary.

(2) Bare amplitudes are identified as the part of physical amplitudes that
get complex conjugated under CP. That is the main point, since this identification 
implies elasticity as we will see in the next paragraph. Inverting the argument, 
we want to build a model of FSI based on the elastic hypothesis, we are thus led
to this identification.   

\subsection{Elasticity}

Elasticity is equivalent to the unitarity of the strong S-matrix containing
the coupling. The elasticity hypothesis is then hidden in the feature (1)
above concerning unitarity of \textbf{S}. Indeed, as soon as \textbf{S} is
unitary, $S$ and $\sqrt{S}$ are also unitary (eq(\ref{unitarityimp})). This
implies that we have the conservation of probability among the coupled
channels :

\begin{equation}
W^{\dagger }W=W_{b}^{\dagger }\sqrt{S}^{\dagger }\ \sqrt{S}\
W_{b}=W_{b}^{\dagger }W_{b}  \label{probcons1}
\end{equation}

If we note the intermediate states, i.e. states produced and not yet
rescattered by \{x$_{i}$\} and final out states by x$_{i}$, we can rewrite
the strong S-matrix (\ref{strongcoupl}) as :\medskip \vspace{0in} 
\begin{equation}
S=\left( 
\begin{array}{cccc}
\left\{ x_{1}\right\} \rightarrow x_{1} & \left\{ x_{2}\right\} \rightarrow
x_{1} & \cdots & \left\{ x_{n}\right\} \rightarrow x_{1} \\ 
\left\{ x_{1}\right\} \rightarrow x_{2} & \left\{ x_{2}\right\} \rightarrow
x_{2} & \cdots & \left\{ x_{n}\right\} \rightarrow x_{2} \\ 
\vdots & \vdots & \ddots & \vdots \\ 
\left\{ x_{1}\right\} \rightarrow x_{n} & \left\{ x_{2}\right\} \rightarrow
x_{n} & \cdots & \left\{ x_{n}\right\} \rightarrow x_{n}
\end{array}
\right)  \label{strongcoupl2}
\end{equation}

And probability conservation expresses itself as :

\begin{equation}
\stackunder{i=1}{\stackrel{n}{\dsum }}\left\| \left( B\rightarrow
x_{i}\right) \right\| ^{2}=\stackunder{i=1}{\stackrel{n}{\dsum }}\left\|
\left( B\rightarrow \left\{ x_{i}\right\} \right) \right\| ^{2}
\label{probcons2}
\end{equation}

This is another expression of elasticity. It is clear that if we consider
all the possible final states, \textbf{S }will be unitary. In practice
however, we consider couplings only among a subset of final states (for
example, only $\pi \pi $ states), and thus we neglect many other possible
mixings. We then impose the unitarity of a truncated \textbf{S}-matrix,
limited to this subset of states. This is the elastic hypothesis; it is
characterized by probability conservation among this subset of states.

The most important restriction we will impose on the mixings is to consider
coupling between states of two pseudoscalars (in the s-wave) only. Mixing
with states containing vector mesons, or many particle states are thus
neglected. This restriction is convenient in order to ensure a symmetric
form for S. Indeed, S will be symmetric if the transition amplitudes are
invariant under time-reversal, and since a general state may catch a
different phase than PP states under CP, they will not be considered.

\subsubsection{FSI\ Eigenchannels}

Let us define some technical tools used in the rest of the paper :

(1) The basis of \textit{eigenchannels} $\left| C_{i}\right\rangle $ where S
is diagonal, with matrix elements :

\begin{equation}
\left\| \left\langle C_{i}\left| S\right| C_{j}\right\rangle \right\|
^{2}=\delta _{ij}\quad \Rightarrow \quad \left\langle C_{i}\left| S\right|
C_{j}\right\rangle =\delta _{ij}e^{2i\delta _{C_{i}}}  \label{eigenelast}
\end{equation}

So these states $C_{i}$ do not mix under rescattering. Elasticity is
manifested as the unit norm, which in turn is equivalent to unitarity for S.

(2) These diagonal elements of S are the \textit{strong phases} : 
\begin{equation}
S_{diag}\equiv \left( 
\begin{array}{llll}
e^{2i\delta _{C1}} & 0 & \cdots & 0 \\ 
0 & e^{2i\delta _{C2}} & \cdots & 0 \\ 
\vdots & \vdots & \ddots & \vdots \\ 
0 & 0 & \cdots & e^{2i\delta _{Cn}}
\end{array}
\right)  \label{strongcpldiag}
\end{equation}

(3) and since the S-matrix is symmetric and unitary, we can diagonalize it
using \textit{a real orthogonal transformation} $O$ :

\begin{equation}
S=O^{t}S_{diag}O  \label{diago}
\end{equation}
The orthogonal diagonalizing matrix O also relates the eigenchannel basis to
the physical one :

\begin{equation}
\left| \overrightarrow{C}\right\rangle \equiv \left( 
\begin{array}{c}
\left| C_{1}\right\rangle \\ 
\left| C_{2}\right\rangle \\ 
\vdots \\ 
\left| C_{n}\right\rangle
\end{array}
\right) =O\left( 
\begin{array}{c}
\left\{ x_{1}\right\} \\ 
\left\{ x_{n}\right\} \\ 
\vdots \\ 
\left\{ x_{n}\right\}
\end{array}
\right) \equiv O\left| \overrightarrow{\left\{ x\right\} }\right\rangle
\label{neweigench}
\end{equation}

(4) The \textit{mixing matrix} $M$ is simply the square root of S, which
appears in Watson's theorem, and defined as :

\begin{equation}
\left\{ 
\begin{array}{l}
M\equiv \sqrt{S}=O^{t}\sqrt{S_{diag}}O \\ 
M_{diag}\equiv \sqrt{S_{diag}}
\end{array}
\right.  \label{defMmatrix}
\end{equation}

(5) With these tools, we can give another derivation of Watson's theorem.
Let us define some renormalized out eigenstates as :

\begin{equation}
\left| \overrightarrow{C}\right\rangle \rightarrow \left| \overrightarrow{C}%
_{out}\right\rangle =\sqrt{S_{diag}}\left| \overrightarrow{C}\right\rangle
\label{eigenrenorm}
\end{equation}

and this is equivalent to $\left( \text{\ref{eigenelast}}\right) $ :

\begin{equation}
\left\langle C_{i,out}\mid C_{j,out}\right\rangle =\left\langle C_{i}\left|
S_{diag}\right| C_{j}\right\rangle =\delta _{ij}e^{2i\delta _{i}}
\label{eigenelast2}
\end{equation}

In the physical basis, we have the following situation :

\begin{equation}
\left\{ 
\begin{array}{l}
\left| \overrightarrow{C}\right\rangle =O\left| \overrightarrow{\left\{
x\right\} }\right\rangle \\ 
\left| \overrightarrow{C}_{out}\right\rangle =O\left| \overrightarrow{x}%
\right\rangle
\end{array}
\right.  \label{watsonatstate}
\end{equation}

i.e. the development of the intermediate states in terms of intermediate
eigenchannels is the same as the development of final asymptotic states in
terms of out eigenchannels. Putting all this in equations gives :

\begin{equation}
\left| \overrightarrow{x}\right\rangle =O^{t}\left| \overrightarrow{C}%
_{out}\right\rangle =O^{t}\sqrt{S_{diag}}\left| \overrightarrow{C}%
\right\rangle =O^{t}\sqrt{S_{diag}}O\left| \overrightarrow{\left\{ x\right\} 
}\right\rangle  \label{watsonatstate2}
\end{equation}

and the final result is :

\begin{equation}
\left| \overrightarrow{x}\right\rangle =\sqrt{S}\left| \overrightarrow{%
\left\{ x\right\} }\right\rangle \equiv M\left| \overrightarrow{\left\{
x\right\} }\right\rangle  \label{watsonatsatet3}
\end{equation}

For the decay amplitudes, we recover Watson's theorem :

\begin{equation}
\overrightarrow{\left( B\rightarrow x\right) }=\left( 
\begin{array}{c}
\left( B\rightarrow x_{1}\right) \\ 
\left( B\rightarrow x_{2}\right) \\ 
\vdots \\ 
\left( B\rightarrow x_{n}\right)
\end{array}
\right) =M\ \left( 
\begin{array}{c}
\left( B\rightarrow \left\{ x_{1}\right\} \right) \\ 
\left( B\rightarrow \left\{ x_{2}\right\} \right) \\ 
\vdots \\ 
\left( B\rightarrow \left\{ x_{n}\right\} \right)
\end{array}
\right) =M\ \overrightarrow{\left( B\rightarrow \left\{ x\right\} \right) }
\label{facttoampl}
\end{equation}

\subsubsection{Application to D decays}

The treatment of weak decay amplitudes of D is the same as in the B case. We
will write : 
\begin{equation}
\overrightarrow{\left( D\rightarrow x\right) }=M\ \overrightarrow{\left(
D\rightarrow \left\{ x\right\} \right) }  \label{facttoampl2}
\end{equation}

However, since FSI in D decays proceed at a lower energy than in B decays,
and since the number of open channels in D decays is much smaller than in B
decays, the M matrix structure is not necessarily the same for B and D
decays, and the rescattering phases are different.

\subsubsection{One rescattering channel}

In the case of only one rescattering channel, writing $W_{b}=We^{i\gamma }$
and $\sqrt{S}=e^{i\delta }$, $W=\sqrt{S}W_{b}$ is equivalent to the
well-known result :

\begin{equation}
\quad \left\{ 
\begin{array}{l}
W=We^{i\gamma }e^{i\delta } \\ 
CP\left( W\right) =We^{-i\gamma }e^{i\delta }
\end{array}
\right. \text{ }  \label{onechannelwatson}
\end{equation}

i.e. that under CP the weak phase $\gamma $ is reversed and not the strong
phase $\delta $. The matrix version for many rescattering channels really
appears as a simple generalization.

\section{SU(N) flavour symmetry implications}

\subsection{SU(N) elasticity : Definition}

A mixing will be elastic under SU(N) if

\begin{quote}
(1) the FSI eigenchannels C$_{i}$ are definite states of SU(N).

(2) The orthogonal transformation is the Clebsch-Gordan coefficient matrix
relating the physical states to the SU(N) states.

(3) The phases (M$_{diag}$ matrix elements) are only function of the
representation of the corresponding eigenchannels.
\end{quote}

The point (2) has a direct and important consequence : working under a
symmetry group \textit{fixes the form of mixing matrices} (they are
completely determined by the representation contents of the final states),
and \textit{imposes some restrictions on the mixings}. Indeed, a group of
states coupled together under SU(N) is a group of states containing some
common representations of SU(N). The decaying meson (B, D,...), on the other
hand, fixes the energy scale at which FSI take place, and thus determines
the values of the strong phases.

In general, a symmetry group fixes the set of coupled channels, and this set
increases with N. On the other hand, the symmetry breaking also increases
with N. For example, in B decays, we can treat the mixing K$\pi $, K$\eta $
under SU(4), in order to include charmed meson channels, but the breaking of
SU(4) at the B-mass energy is such that SU(4) elasticity is expected to be
inappropriate.

\paragraph{Terminology :}

To define properly elasticity, we can distinguish the following concepts :

\begin{quotation}
(i) \textit{Pure elastic transitions} like $\left\{ K^{+}\pi ^{-}\right\}
\rightarrow K^{+}\pi ^{-}$ with just a phase as amplitude.

(ii) \textit{SU(N) elasticity}, for which we have elasticity in a basis of
eigenchannels corresponding to SU(N) states.

(iii) \textit{Elasticity}, for which we can define some general
eigenchannels C$_{i}$ by a set of mixing parameters (eq. \ref{generalmix}).
This is sometimes considered as inelastic since we can have, for example,
eigenchannels with no specific isospin.
\end{quotation}

\subsection{SU(2) analysis of $K^{+}\pi ^{-}$, $K^{0}\pi ^{0}$, $K^{0}\eta
_{8}$}

To accomplish the connection with section 1, consider the system $K^{+}\pi
^{-}$, $K^{0}\pi ^{0}$, $K^{0}\eta _{8}$ and suppose we are working under
SU(2). The SU(2) FSI eigenchannels are then

\begin{equation}
\left( 
\begin{array}{c}
\left| C_{1}\right\rangle \\ 
\left| C_{2}\right\rangle \\ 
\left| C_{3}\right\rangle
\end{array}
\right) \equiv \left( 
\begin{array}{c}
\left| 3/2,-1/2\right\rangle \\ 
\left| 1/2^{(1)},-1/2\right\rangle \\ 
\left| 1/2^{(2)},-1/2\right\rangle
\end{array}
\right) =\left( 
\begin{array}{ccc}
-\sqrt{1/3} & -\sqrt{2/3} & 0 \\ 
-\sqrt{2/3} & \sqrt{1/3} & 0 \\ 
0 & 0 & 1
\end{array}
\right) \left( 
\begin{array}{c}
\left\{ K^{+}\pi ^{-}\right\} \\ 
\left\{ K^{0}\pi ^{0}\right\} \\ 
\left\{ K^{0}\eta _{8}\right\}
\end{array}
\right)  \label{su(2)eigench}
\end{equation}

Which means that $K^{+}\pi ^{-}$, $K^{0}\pi ^{0}$ are mixed under SU(2),
since they contain the same representations of isospin 3/2 and 1/2$^{(1)}$,
but $K^{0}\eta _{8}$, being in a different 1/2 representation, stays alone.
The $O_{SU(2)}$ matrix (Clebsch-Gordan coefficients) is block-diagonal, and
so is $M^{SU(2)}$. The phases entering $M_{diag}^{SU(2)}$, depending only on
the eigenchannels SU(2) representations, are:

\begin{equation}
M_{diag}^{SU(2)}=\left( 
\begin{array}{ccc}
e^{i\delta _{3/2}} & 0 & 0 \\ 
0 & e^{i\delta _{1/2}^{(1)}} & 0 \\ 
0 & 0 & e^{i\delta _{1/2}^{(2)}}
\end{array}
\right)  \label{su(2)diagM}
\end{equation}

The SU(2) elasticity is expressed in these eigenchannels as :

\begin{equation}
\left\{ 
\begin{array}{c}
\left\| \left\langle 3/2,-1/2\left| S\right| 3/2,-1/2\right\rangle \right\|
^{2}=1 \\ 
\left\| \left\langle 1/2^{(1)},-1/2\left| S\right|
1/2^{(1)},-1/2\right\rangle \right\| ^{2}=1 \\ 
\left\| \left\langle 1/2^{(2)},-1/2\left| S\right|
1/2^{(2)},-1/2\right\rangle \right\| ^{2}=1
\end{array}
\right. \Rightarrow \left\{ 
\begin{array}{c}
\left\langle 3/2,-1/2\left| S\right| 3/2,-1/2\right\rangle =e^{2i\delta
_{3/2}} \\ 
\left\langle 1/2^{(1)},-1/2\left| S\right| 1/2^{(1)},-1/2\right\rangle
=e^{2i\delta _{1/2}^{(1)}} \\ 
\left\langle 1/2^{(2)},-1/2\left| S\right| 1/2^{(2)},-1/2\right\rangle
=e^{2i\delta _{1/2}^{(2)}}
\end{array}
\right.  \label{isoeigench2}
\end{equation}

The M matrix is then calculated as $M^{SU(2)}=O_{SU(2)}^{t}$ $%
M_{diag}^{SU(2)}O_{SU(2)}$, and we finally have :

\begin{equation}
M^{SU(2)}=\left( 
\begin{array}{ccc}
\frac{1}{3}\left( e^{i\delta _{3/2}}+2e^{i\delta _{1/2}^{(1)}}\right) & 
\frac{\sqrt{2}}{3}\left( e^{i\delta _{3/2}}-e^{i\delta _{1/2}^{(1)}}\right)
& 0 \\ 
\frac{\sqrt{2}}{3}\left( e^{i\delta _{3/2}}-e^{i\delta _{1/2}^{(1)}}\right)
& \frac{1}{3}\left( 2e^{i\delta _{3/2}}+e^{i\delta _{1/2}^{(1)}}\right) & 0
\\ 
0 & 0 & e^{i\delta _{1/2}^{(2)}}
\end{array}
\right)  \label{isoMill2}
\end{equation}

which is the same matrix (for the $K\pi $ sector) as in section 1 (eq. \ref
{su(2)kpmix}).

\subsubsection{Intermediate and asymptotic states}

At this point, we can repeat the discussion of the preceding section and
define out isospin eigenchannels :

\begin{equation}
\left( 
\begin{array}{l}
\left| C_{1}\right\rangle =\left| 3/2,-1/2\right\rangle \\ 
\left| C_{2}\right\rangle =\left| 1/2,-1/2\right\rangle
\end{array}
\right) \rightarrow \left( 
\begin{array}{l}
\left| C_{1,out}\right\rangle =\left| 3/2,-1/2\right\rangle _{out} \\ 
\left| C_{2,out}\right\rangle =\left| 1/2,-1/2\right\rangle _{out}
\end{array}
\right) =\left( 
\begin{array}{l}
e^{i\delta _{3/2}}\left| 3/2,-1/2\right\rangle \\ 
e^{i\delta _{1/2}}\left| 1/2,-1/2\right\rangle
\end{array}
\right)  \label{renormeigeniso}
\end{equation}

and for the physical states, we write :

\begin{eqnarray}
\left( 
\begin{array}{l}
\left\{ K^{+}\pi ^{-}\right\} \\ 
\left\{ K^{0}\pi ^{0}\right\}
\end{array}
\right) &=&O_{SU(2)}^{t}\left( 
\begin{array}{l}
\left| C_{1}\right\rangle =\left| 3/2,-1/2\right\rangle \\ 
\left| C_{2}\right\rangle =\left| 1/2,-1/2\right\rangle
\end{array}
\right)  \label{renormeigeniso2} \\
\left( 
\begin{array}{c}
K^{+}\pi ^{-} \\ 
K^{0}\pi ^{0}
\end{array}
\right) &=&O_{SU(2)}^{t}\left( 
\begin{array}{l}
\left| C_{1,out}\right\rangle =\left| 3/2,-1/2\right\rangle _{out} \\ 
\left| C_{2,out}\right\rangle =\left| 1/2,-1/2\right\rangle _{out}
\end{array}
\right)  \nonumber
\end{eqnarray}

So the link between intermediate and asymptotic states is given by

\begin{equation}
\left( 
\begin{array}{c}
K^{+}\pi ^{-} \\ 
K^{0}\pi ^{0}
\end{array}
\right) =M^{SU(2)}\left( 
\begin{array}{l}
\left\{ K^{+}\pi ^{-}\right\} \\ 
\left\{ K^{0}\pi ^{0}\right\}
\end{array}
\right)  \label{renormeigeniso3}
\end{equation}

This shows once again that SU(2) fixes the structure of $M^{SU(2)}$ and that
only the representation contents of the final states is relevant. This in
turn implies that the same matrix is appropriate for B and D decays. Of
course, the phases $\delta _{3/2}$ and $\delta _{1/2}$ can be different
since their specific values is a dynamical question (they depend on the
energy available, i.e. the mass of the decaying meson). So, for example, we
can write :

\begin{equation}
\left( 
\begin{array}{c}
\left( B^{0}\rightarrow K^{+}\pi ^{-}\right) \\ 
\left( B^{0}\rightarrow K^{0}\pi ^{0}\right)
\end{array}
\right) =M^{SU(2)}\left( 
\begin{array}{c}
\left( B^{0}\rightarrow \left\{ K^{+}\pi ^{-}\right\} \right) \\ 
\left( B^{0}\rightarrow \left\{ K^{0}\pi ^{0}\right\} \right)
\end{array}
\right)  \label{isofullB}
\end{equation}

\subsubsection{Decay amplitudes}

As we have shown in section 1, by applying this $M^{SU(2)}$ matrix on the
isospin decompositions into bare amplitudes (\ref{isobare}), we find again (%
\ref{isofull}). This is a general principle. The usual procedure to take
into account SU(N) FSI in a SU(N) bare amplitude decomposition is to add
phases to the SU(N) bare amplitudes according to their SU(N)
representations. As we have said, applying the SU(N) mixing matrix on SU(N)
bare decompositions is equivalent, i.e. the usual prescription is equivalent
to a mixing of states, and the M matrices provide a clear representation of
these mixings.

Let us illustrate this fact in the example of $K^{+}\pi ^{-}$, $K^{0}\pi
^{0} $ under SU(2). From

\begin{equation}
\left| H_{W}\quad \overline{D^{0}}\right\rangle =\sqrt{\frac{1}{3}}\left|
3/2\right\rangle -\sqrt{\frac{2}{3}}\left| 1/2\right\rangle
\label{isosphamilt}
\end{equation}

and eq.(\ref{su(2)eigench}), we found the isospin decomposition eq.(\ref
{isobare}) with $A_{b}^{T}\sim \left\langle T\mid T\right\rangle $. The
point is to note that the same orthogonal transformation is used in the
calculation of decomposition and of mixing matrices (this remains valid
under any SU(N)). To find the decompositions in terms of full amplitudes,
just replace the intermediate states $\left| T\right\rangle $ by out states $%
e^{i\delta _{T}}\left| T\right\rangle $, and this is strictly equivalent to
renormalize $A_{b}^{T}\rightarrow A^{T}\sim e^{i\delta _{T}}\left\langle
T\mid T\right\rangle .$

\subsection{SU(3) analysis of $K^{+}\pi ^{-}$, $K^{0}\pi ^{0}$, $K^{0}\eta
_{8}$}

Under SU(3), the same group of states mixes completely because they all
contain the same 27 and 8 :

\begin{equation}
\left( 
\begin{array}{c}
\left| C_{1}\right\rangle \\ 
\left| C_{2}\right\rangle \\ 
\left| C_{3}\right\rangle
\end{array}
\right) \equiv \left( 
\begin{array}{c}
\left| 27,3/2,-1/2,1\right\rangle \\ 
\left| 27,1/2,-1/2,1\right\rangle \\ 
\left| 8_{S},1/2,-1/2,1\right\rangle
\end{array}
\right) =\left( 
\begin{array}{ccc}
-\sqrt{1/3} & -\sqrt{2/3} & 0 \\ 
-\sqrt{1/15} & \sqrt{1/30} & -\sqrt{9/10} \\ 
-\sqrt{3/5} & \sqrt{3/10} & \sqrt{1/10}
\end{array}
\right) \left( 
\begin{array}{c}
\left\{ K^{+}\pi ^{-}\right\} \\ 
\left\{ K^{0}\pi ^{0}\right\} \\ 
\left\{ K^{0}\eta _{8}\right\}
\end{array}
\right)  \label{su(3)eigench}
\end{equation}

Where SU(3) states are specified as $\left| rep.,T,T_{3},Y\right\rangle $. M$%
_{diag}^{SU(3)}$ is given by :

\begin{equation}
M_{diag}^{SU(3)}=\left( 
\begin{array}{ccc}
e^{i\delta _{27}} & 0 & 0 \\ 
0 & e^{i\delta _{27}} & 0 \\ 
0 & 0 & e^{i\delta _{8}}
\end{array}
\right)  \label{su(3)diagM}
\end{equation}

From this orthogonal transformation and M$_{diag}^{SU(3)}$, we immediately
recover eq. (\ref{su(3)kpmix}) :

\[
M^{SU(3)}=\frac{1}{5}\left( 
\begin{array}{ccc}
2e^{i\delta _{27}}+3e^{i\delta _{8}} & \frac{3}{\sqrt{2}}\left( e^{i\delta
_{27}}-e^{i\delta _{8}}\right) & \sqrt{\frac{3}{2}}\left( e^{i\delta
_{27}}-e^{i\delta _{8}}\right) \\ 
\frac{3}{\sqrt{2}}\left( e^{i\delta _{27}}-e^{i\delta _{8}}\right) & \frac{1%
}{2}\left( 7e^{i\delta _{27}}+3e^{i\delta _{8}}\right) & -\frac{\sqrt{3}}{2}%
\left( e^{i\delta _{27}}-e^{i\delta _{8}}\right) \\ 
\sqrt{\frac{3}{2}}\left( e^{i\delta _{27}}-e^{i\delta _{8}}\right) & -\frac{%
\sqrt{3}}{2}\left( e^{i\delta _{27}}-e^{i\delta _{8}}\right) & \frac{1}{2}%
\left( 9e^{i\delta _{27}}+e^{i\delta _{8}}\right)
\end{array}
\right) 
\]

The same matrix $M^{SU(3)}$ can also be used for other decays, since it is
determined by the representation contents of the final states only. For
example :

\begin{equation}
\left\{ 
\begin{array}{l}
\left( B_{s}\rightarrow \left\{ K^{-}\pi ^{+}\right\} \right) \\ 
\left( B_{s}\rightarrow \left\{ \overline{K}^{0}\pi ^{0}\right\} \right) \\ 
\left( B_{s}\rightarrow \left\{ \overline{K}^{0}\eta _{8}\right\} \right)
\end{array}
\right. \qquad \text{or}\qquad \left\{ 
\begin{array}{l}
\left( \overline{D}^{0}\rightarrow \left\{ K^{+}\pi ^{-}\right\} \right) \\ 
\left( \overline{D}^{0}\rightarrow \left\{ K^{0}\pi ^{0}\right\} \right) \\ 
\left( \overline{D}^{0}\rightarrow \left\{ K^{0}\eta _{8}\right\} \right)
\end{array}
\right.  \label{su(3)D}
\end{equation}

And we have thus recovered and explained the results of section 1.2.

\subsection{Probability conservation}

We can use probability conservation to characterize the difference between
SU(2) and SU(3) elasticity (see eq(\ref{probconstwoblocwat})). Probability
conservation expresses itself under SU(2) as :

\begin{equation}
SU(2)\left\{ 
\begin{array}{c}
\left\| \left( \overline{D}\rightarrow \left\{ K^{+}\pi ^{-}\right\} \right)
\right\| ^{2}+\left\| \left( \overline{D}\rightarrow \left\{ K^{0}\pi
^{0}\right\} \right) \right\| ^{2}=\left\| \left( \overline{D}\rightarrow
K^{+}\pi ^{-}\right) \right\| ^{2}+\left\| \left( \overline{D}\rightarrow
K^{0}\pi ^{0}\right) \right\| ^{2} \\ 
\left\| (\overline{D}\rightarrow \left\{ K^{0}\eta _{8}\right\} )\right\|
^{2}=\left\| (\overline{D}\rightarrow K^{0}\eta _{8})\right\| ^{2}
\end{array}
\right.  \label{prbcvsu(2)}
\end{equation}

and under SU(3), in a less restrictive way as :

\begin{equation}
SU(3)\left\{ 
\begin{array}{l}
\left\| \left( \overline{D}\rightarrow \left\{ K^{+}\pi ^{-}\right\} \right)
\right\| ^{2}+\left\| \left( \overline{D}\rightarrow \left\{ K^{0}\pi
^{0}\right\} \right) \right\| ^{2}+\left\| \left( \overline{D}\rightarrow
\left\{ K^{0}\eta _{8}\right\} \right) \right\| ^{2} \\ 
\qquad \qquad =\left\| \left( \overline{D}\rightarrow K^{+}\pi ^{-}\right)
\right\| ^{2}+\left\| \left( \overline{D}\rightarrow K^{0}\pi ^{0}\right)
\right\| ^{2}+\left\| \left( \overline{D}\rightarrow K^{0}\eta _{8}\right)
\right\| ^{2}
\end{array}
\right.  \label{prbcvsu(3)}
\end{equation}

\subsection{Links between SU(N) and SU(N$\pm $1)}

\subsubsection{Principle}

As we have seen above, an elastic mixing under SU(N) is not in general
elastic under SU(N-1). Suppose we have the matrices M$^{SU(N)}$ and M$%
^{SU(N-1)}$. This last matrix is block-diagonal, since it mixes only some
subsets of states. Therefore, to go from SU(N-1) towards SU(N) elastic
mixings, we will have to add to M$^{SU(N-1)}$ some extra mixing between
different sets of SU(N-1) coupled states. These new mixings are not
completely arbitrary : they must be compatible with the SU(N-1) included in
SU(N). Such extra mixings will be parametrized by mixing parameters $\alpha
,\beta ,...$, and we will obtain a generalized mixing matrix M$^{gen.}$($%
\alpha ,\beta ,...$). Finally, for a specific value of $\alpha ,\beta ,...$,
this matrix will correspond to M$^{SU(N)}.$

We will not describe the most general case, but we will take again the
system $K^{+}\pi ^{-}$, $K^{0}\pi ^{0}$, $K^{0}\eta _{8}$ and carry the
transition from a SU(2) description to a SU(3) description. The discussion
in the general case is then straightforward.

\subsubsection{From SU(2) to SU(3)}

\paragraph{Step 1 :}

\textit{Building of the most general mixing among }$K^{+}\pi ^{-}$\textit{, }%
$K^{0}\pi ^{0}$\textit{, }$K^{0}\eta _{8}$\textit{\ compatible with isospin.}

As we have seen, (\ref{su(2)eigench}) defines isospin eigenchannels for
SU(2) FSI. The only possible extra mixing is between the two 1/2
eigenchannels, since we want to keep isospin as a good quantum number for
FSI\ eigenchannels. This extra mixing can be parametrized by a general 2 x 2
orthogonal matrix (\ref{generalmix}) :

\begin{equation}
\left( 
\begin{array}{c}
\left| C_{1}\right\rangle \\ 
\left| C_{2}\right\rangle \\ 
\left| C_{3}\right\rangle
\end{array}
\right) \equiv \left( 
\begin{array}{ccc}
1 & 0 & 0 \\ 
0 & \cos \alpha & -\sin \alpha \\ 
0 & \sin \alpha & \cos \alpha
\end{array}
\right) \left( 
\begin{array}{c}
\left| 3/2,-1/2\right\rangle \\ 
\left| 1/2^{(1)},-1/2\right\rangle \\ 
\left| 1/2^{(2)},-1/2\right\rangle
\end{array}
\right)  \label{su(2)tosu(3)1}
\end{equation}

in terms of physical intermediate states, this gives:

\begin{equation}
\left( 
\begin{array}{c}
\left| C_{1}\right\rangle \\ 
\left| C_{2}\right\rangle \\ 
\left| C_{3}\right\rangle
\end{array}
\right) \equiv \left( 
\begin{array}{ccc}
-\sqrt{1/3} & -\sqrt{2/3} & 0 \\ 
-\sqrt{2/3}\cos \alpha & \sqrt{1/3}\cos \alpha & -\sin \alpha \\ 
-\sqrt{2/3}\sin \alpha & \sqrt{1/3}\sin \alpha & \cos \alpha
\end{array}
\right) \left( 
\begin{array}{c}
\left\{ K^{+}\pi ^{-}\right\} \\ 
\left\{ K^{0}\pi ^{0}\right\} \\ 
\left\{ K^{0}\eta _{8}\right\}
\end{array}
\right)  \label{su(2)tosu(3)2}
\end{equation}

This equation defines new eigenchannels, and the corresponding orthogonal
transformation O$\left( \alpha \right) .$ Remark that these new
eigenchannels keep isospin as a good quantum number (C$_{1}$: 3/2, C$_{2}$
and C$_{3}$: 1/2). M$_{diag}^{general}$ in this basis is given by :

\begin{equation}
M_{diag}^{general}=\left( 
\begin{array}{ccc}
e^{i\delta _{C1}} & 0 & 0 \\ 
0 & e^{i\delta _{C2}} & 0 \\ 
0 & 0 & e^{i\delta _{C3}}
\end{array}
\right)  \label{su(2)tosu(3)3}
\end{equation}

And we can calculate M$^{general}\left( \alpha \right) =O(\alpha
)^{t}M_{diag}^{general}O(\alpha ):$ 
\begin{equation}
\frac{1}{3}\left( 
\begin{array}{ccc}
e^{i\delta _{C1}}+2c^{2}e^{i\delta _{C2}}+2s^{2}e^{i\delta _{C3}} & \sqrt{2}%
\left( e^{i\delta _{C1}}-c^{2}e^{i\delta _{C2}}-s^{2}e^{i\delta
_{C3}}\right)  & \sqrt{6}sc\left( e^{i\delta _{C2}}-e^{i\delta _{C3}}\right) 
\\ 
\sqrt{2}\left( e^{i\delta _{C1}}-c^{2}e^{i\delta _{C2}}-s^{2}e^{i\delta
_{C3}}\right)  & 2e^{i\delta _{C1}}+c^{2}e^{i\delta _{C2}}+s^{2}e^{i\delta
_{C3}} & -\sqrt{3}sc\left( e^{i\delta _{C2}}-e^{i\delta _{C3}}\right)  \\ 
\sqrt{6}sc\left( e^{i\delta _{C2}}-e^{i\delta _{C3}}\right)  & -\sqrt{3}%
sc\left( e^{i\delta _{C2}}-e^{i\delta _{C3}}\right)  & 3(s^{2}e^{i\delta
_{C2}}+c^{2}e^{i\delta _{C3}})
\end{array}
\right)   \label{su(2)tosu(3)4}
\end{equation}

where $s=\sin \alpha $ and $c=\cos \alpha $.

For each value of the mixing parameter $\alpha $, M$^{general}\left( \alpha
\right) $ parametrizes an elastic mixing among $K^{+}\pi ^{-}$, $K^{0}\pi
^{0}$, $K^{0}\eta _{8}$ compatible with isospin (but inelastic for SU(2)).

\paragraph{Step 2 :}

\textit{SU(3) mixing appears as a special case of the general mixing M}$%
^{general}\left( \alpha \right) .$

Indeed, SU(3) mixing are obtained from the SU(2) ones by introducing an
extra mixing compatible with SU(2), since isospin group is a subgroup of
SU(3) (a SU(3) elastic eigenchannel could not be a mixture of 3/2 and 1/2
isospin states). The SU(3) limit is now easy to get : for $\cos \alpha =%
\sqrt{1/10}$ and $\sin \alpha =\sqrt{9/10}$, O$(\alpha )$ equals the
orthogonal matrix of SU(3) Clebsch-Gordan (\ref{su(3)eigench}). This means
that the three channels C$_{i}$ tend towards SU(3) states $\left|
27,3/2,-1/2,1\right\rangle $, $\left| 27,1/2,-1/2,1\right\rangle $ and $%
\left| 8_{S},1/2,-1/2,1\right\rangle $. Then the corresponding phases tend
towards SU(3) phases : $e^{i\delta _{C1}},e^{i\delta _{C2}}\rightarrow
e^{i\delta _{27}}$ and $e^{i\delta _{C3}}\rightarrow e^{i\delta _{8}}$. We
have thus completed the passage from SU(2) to SU(3).

\subparagraph{Remarks :}

(1) Note that it is sometimes necessary to introduce a P $=diag(\pm
1,...,\pm 1)$ matrix in order to have $P.O(\alpha )\rightarrow O^{SU(3)}$,
since the later depends on phase conventions. On the other hand, the M
matrix, being physical, is always phase convention independent. In other
words, orthogonal transformations differing by a P matrix give the same M
matrix.

(2) In SU(2), the mixing of $K\pi $ states with state $K^{0}\eta _{8}$ is
neglected. We have shown in this section that the extra mixing needed to
treat the mixings among $K^{+}\pi ^{-}$, $K^{0}\pi ^{0}$, $K^{0}\eta _{8}$
under SU(3) elasticity is quite big (the two channels 1/2 get nearly
inverted). At the cost of one unknown mixing parameter $\alpha $, we can use 
$\left( \text{\ref{su(2)tosu(3)4}}\right) ,$ which is compatible with both
SU(2) and SU(3) to introduce a ''small'' mixing between $K\pi $ and $%
K^{0}\eta _{8}$. This will be done in section 5.

\subsubsection{SU(2) in SU(3)}

Still in the same example, we will illustrate another link between SU(3) and
SU(2). A very interesting form for the mixing matrix is built from the
following diagonal form :

\begin{equation}
M_{diag}^{SU(2\text{ }in\text{ }3)}=\left( 
\begin{array}{ccc}
e^{i\delta _{27}^{3/2}} & 0 & 0 \\ 
0 & e^{i\delta _{27}^{1/2}} & 0 \\ 
0 & 0 & e^{i\delta _{8}^{1/2}}
\end{array}
\right)  \label{su(2)tosu(3)5}
\end{equation}

i.e. we distinguish isospin in SU(3), the three phases are different. From
this matrix, by applying the orthogonal SU(3) transformation, we obtain $%
M^{SU(2\text{ }in\text{ }3)}=$%
\begin{equation}
\left( 
\begin{array}{ccc}
\frac{e^{i\delta _{27}^{3/2}}}{3}+\frac{e^{i\delta _{27}^{1/2}}}{15}+\frac{%
3e^{i\delta _{8}^{1/2}}}{5} & \frac{1}{\sqrt{2}}\left( \frac{2e^{i\delta
_{27}^{3/2}}}{3}-\frac{e^{i\delta _{27}^{1/2}}}{15}-\frac{3e^{i\delta
_{8}^{1/2}}}{5}\right) & \frac{\sqrt{6}}{10}\left( e^{i\delta
_{27}^{1/2}}-e^{i\delta _{8}^{1/2}}\right) \\ 
\frac{1}{\sqrt{2}}\left( \frac{2e^{i\delta _{27}^{3/2}}}{3}-\frac{e^{i\delta
_{27}^{1/2}}}{15}-\frac{3e^{i\delta _{8}^{1/2}}}{5}\right) & \frac{%
2e^{i\delta _{27}^{3/2}}}{3}+\frac{e^{i\delta _{27}^{1/2}}}{30}+\frac{%
3e^{i\delta _{8}^{1/2}}}{10} & \frac{\sqrt{3}}{10}\left( e^{i\delta
_{8}^{1/2}}-e^{i\delta _{27}^{1/2}}\right) \\ 
\frac{\sqrt{6}}{10}\left( e^{i\delta _{27}^{1/2}}-e^{i\delta
_{8}^{1/2}}\right) & \frac{\sqrt{3}}{10}\left( e^{i\delta
_{8}^{1/2}}-e^{i\delta _{27}^{1/2}}\right) & \frac{1}{10}\left( 9e^{i\delta
_{27}^{1/2}}+e^{i\delta _{8}^{1/2}}\right)
\end{array}
\right)  \label{su(2)tosu(3)6}
\end{equation}

This form can be very useful in phenomenological analyses, since it is a
easy way to implement SU(3) breaking in the FSI . We can recover the SU(2)
and SU(3) limits straightforwardly. If we identify $e^{i\delta _{27}^{3/2}}$
and $e^{i\delta _{27}^{1/2}}$, we find again M$^{SU(3)}$. On the other hand,
if we identify $e^{i\delta _{27}^{3/2}}$ with $e^{i\delta _{3/2}}$, $%
e^{i\delta _{27}^{1/2}}$ and $e^{i\delta _{8}^{1/2}}$ with $e^{i\delta
_{1/2}}$, we find the following mixing :

\begin{equation}
M_{SU(2)}^{modified}=\left( 
\begin{array}{ccc}
\frac{1}{3}\left( e^{i\delta _{3/2}}+2e^{i\delta _{1/2}}\right) & \frac{%
\sqrt{2}}{3}\left( e^{i\delta _{3/2}}-e^{i\delta _{1/2}}\right) & 0 \\ 
\frac{\sqrt{2}}{3}\left( e^{i\delta _{3/2}}-e^{i\delta _{1/2}}\right) & 
\frac{1}{3}\left( 2e^{i\delta _{3/2}}+e^{i\delta _{1/2}}\right) & 0 \\ 
0 & 0 & e^{i\delta _{1/2}}
\end{array}
\right)  \label{su(2)tosu(3)7}
\end{equation}

which is built from the SU(2) orthogonal transformation (\ref{su(2)eigench})
with a modified M$_{diag}^{SU(2)}$ (eq. (\ref{su(2)diagM})). This modified
form is obtained from the identification : $\delta _{1/2}^{(1)}=\delta
_{1/2}^{(2)}=\delta _{1/2}$. We can explain this easily : the SU(3)
orthogonal transformation has the structure $O(\alpha )=O^{extra}O^{SU(2)}$
(see eq(\ref{su(2)tosu(3)2})). Then $%
M_{SU(2)}^{modified}=O^{SU(2),t}O^{extra,t}M_{diag}^{SU(2)}O^{extra}O^{SU(2)}
$and with $\delta _{1/2}^{(1)}=\delta _{1/2}^{(2)}$, $O^{extra}$ simplifies,
leaving the SU(2) transformation.

\subsubsection{Concluding Remarks}

This section is of theoretical and practical importance. On the theoretical
side, we have defined SU(N) elasticity as a special case of the general
elasticity concept.

On the practical side, we have shown how to build mixing matrices
explicitly. We have obtained four different mixing matrices for the system $%
\left\{ K^{+}\pi ^{-},K^{0}\pi ^{0},K^{0}\eta _{8}\right\} $ : $%
M^{general}\left( \alpha \right) ,M^{SU(3)},M^{SU(2)}$ and $M^{SU(2\text{ }in%
\text{ }3)}$ with its limit $M_{SU(2)}^{modified}$. We can now choose to use
any form, independently of the parametrization chosen for bare decay
amplitudes. This illustrates the power of this matrix method for treating
FSI. Since we have factorized FSI from bare amplitudes inside physical
amplitudes, these two aspects can be analyzed independently.

We stress again that all this can be repeated for other mixings, other
flavor symmetry groups and other meson decays.

\section{Quark diagrams}

The B and D decays can be parametrized with quark diagrams (QD) amplitudes.
The ultimate goal of QD is to test the standard model. To achieve this, we
must compare a calculated value of an amplitude to its measured value. We
will show that \textit{by defining quark diagrams as free of any FSI effect,
they are well-defined in terms of basic topologies}, and thus allow in
principle to reach such a goal.

Using Watson' theorem, physical amplitudes can be decomposed into bare
amplitudes and FSI matrices. We are thus naturally led to the following
parametrization :

\[
\text{Physical decay amplitudes}\stackrel{\text{Watson}^{\prime }\text{s
theorem}}{\rightarrow }\left\{ 
\begin{array}{l}
\text{Bare amplitudes : Quark diagrams} \\ 
\text{FSI effects : Mixing matrices}
\end{array}
\right. 
\]

We will now analyze the consequences.

\subsection{Quark diagrams as bare amplitudes}

Since quark diagrams are free of any FSI effects, they are real, except for
CKM phases. In other words, QD are defined at the level of bare amplitudes.
They contain the weak decay of the heavy quark, the hadronization and some
gluonic renormalizations of the weak current (without absorptive part).

FSI are introduced as interactions between hadrons, using FSI phases. In
this way, we avoid the difficulties (if not the inconsistencies), of
on-shell quarks since we are always working with on-shell hadrons, i.e.
states entering the \textbf{S}-matrix. The treatment of FSI at the hadronic
level is ultimately justified by the \textbf{S}-matrix hadronic structure.

\subsection{Quark diagram topologies and FSI topology mixings}

Quark diagrams are built from quark lines and W-lines. The resulting
topologies (or ''shapes'') of these diagrams are the usual tree T,
colour-suppressed C, annihilation A, exchange E, penguin P and penguin
annihilation PA diagrams (see ref. \cite{su(3)bdec1} to \cite{su(3)ddec3}).
These can also be defined from two basic topologies (a bubble, representing
the quark-gluon sea, with a W inside and two bubbles connected by a W), from
which we extract real hadrons (Figure 1).

The assertion that QD allow dynamical considerations when defined at the
bare level comes from the well-known fact that final state interactions mix
the different topologies. In particular, any scheme like factorization
should be carried at the level of these ''bare'' QD. Furthermore, helicity
suppression of A is valid only if A is a bare amplitude, since otherwise it
could contain other basic topologies than A.

The establishment of the QD parametrizations of bare decay amplitudes is
explained in the appendix 2, where the link between SU(3) bare amplitudes
and QD is also written. The QD\ parametrizations of B and D decays are given
in appendix 3.

\paragraph{Discussion :}

In figure 2, we have drawn two possible diagrams ((a) and (b)) contributing
to $\left( \overline{D}^{0}\rightarrow K^{+}\pi ^{-}\right) $. The scheme of
introducing QD at the bare level and FSI as hadronic mixing matrices is
depicted in the passage from (a) and (b) to (c) and (d), i.e. by the
identification of the relevant hadronic intermediate states. These hadrons
then interact, and this interaction is the FSI. Figure (c) and (d) also show
that basic quark diagrams automatically occur as bare topologies, and the
resulting physical amplitude $\left( \overline{D}^{0}\rightarrow K^{+}\pi
^{-}\right) $ receives contributions from its QD topologies $\overline{D}%
^{0}\rightarrow \left\{ K^{+}\pi ^{-}\right\} =T+E,$ and also from some
extra topologies (here, some C) coming from $\overline{D}^{0}\rightarrow
\left\{ K^{0}\pi ^{0}\right\} $.

If we had introduced quark diagrams at the physical level, we would have
obtained for the physical amplitudes $\left( \overline{D}^{0}\rightarrow
K^{+}\pi ^{-}\right) =T^{full}+E^{full}$. Now looking at quark lines in fig.
2(b), we see that this diagram is topologically equivalent to the E
topology, i.e. that it contribute to $E^{full}$. So this $E^{full}$
amplitude, containing FSI effects, contain some C topology. Doing the same
analysis with fig. 2(a), we can see that this diagram is a non-factorizable
contribution to $T^{full}$. When introduced with FSI effects, T looses its
factorization properties. By analyzing some other decays, one can easily see
that an amplitude $A^{full}$ receives contributions from other topologies
than A. Thus $A^{full}$ is not helicity suppressed.

In conclusion, this discussion shows that in order to have well-defined
quark diagrams in terms of basic topologies, they should be defined at the
bare level, without rescattering effects. We have also shown how some
considerations like factorizability or helicity suppression collapse when QD
contain FSI\ effects.

\paragraph{Remark :}

The elastic hypothesis is to introduce these quark diagrams at the level of
bare amplitudes. Other propositions exist, for example to introduce QD at
the level of different amplitudes in the K matrix formalism (for this K
matrix formalism, see for example \cite{fsikamal1} to \cite{fsikamal6}).

\subsection{Use of mixing matrices on QD decompositions}

By using Watson's theorem, we have shown that FSI and bare processes
separate. We can thus analyze each part independently, for example :

\[
\left\{ 
\begin{array}{l}
\text{\textbf{Bare amplitudes} : }\left\{ 
\begin{array}{l}
SU(2),SU(3),...\text{ amplitudes} \\ 
\text{Quark diagrams}\rightarrow \text{under }SU(2),SU(3),...
\end{array}
\right. \\ 
\begin{array}{c}
\text{\textbf{FSI effects : }} \\ 
\text{(\textbf{Mixing matrices)}}
\end{array}
\left\{ 
\begin{array}{l}
\text{under SU(2) with }M^{SU(2)} \\ 
\text{under SU(3) with }M^{SU(3)} \\ 
\text{as a general mixing with }M^{general}(\alpha ,\beta ,...) \\ 
\text{under SU(3) with SU(2) specified with }M^{SU(2\text{ }in\text{ }3)} \\ 
...
\end{array}
\right.
\end{array}
\right. 
\]

And the physical decay amplitudes are obtained by applying the chosen mixing
matrix on the chosen bare decay parametrizations. The next section will
illustrate extensively this procedure.

\section{Applications to B and D decays}

The two-pseudoscalar final states can be grouped into sets of coupled states
under SU(3) (noted inside $\left\{ {}\right\} $) by considering conserved
quantum numbers : Isospin T and hypercharge Y. In fact, these sets
correspond to the sets of definite T$_{3}$ and Y, since they completely mix.
The only exception is $\left\{ \text{ }\pi ^{-}\pi ^{0}\right\} $ states
(pure isospin 2) which do not mix with $\left\{ K^{-}K^{0},\pi ^{-}\eta
_{8}\right\} $(pure isospin 1). Repeating the same analysis, we can also
find sets of coupled states under SU(2). The results, with on the left side
SU(3) mixing and on the right side SU(2) mixings are :

\paragraph{(a) $K\pi $ and $\overline{K}\pi $ Sets $\left( Y=\pm 1,T_{3}=\pm
1/2\right) $ :}

\begin{eqnarray*}
A &:&\left\{ K^{+}\pi ^{-},K^{0}\pi ^{0},K^{0}\eta _{8}\right\} \quad
^{SU(3)}\longleftrightarrow ^{SU(2)}\quad \left\{ K^{+}\pi ^{-},K^{0}\pi
^{0}\right\} \left\{ K^{0}\eta _{8}\right\} \\
B &:&\left\{ K^{0}\pi ^{+},K^{+}\pi ^{0},K^{+}\eta _{8}\right\} \quad
^{SU(3)}\longleftrightarrow ^{SU(2)}\quad \left\{ K^{0}\pi ^{+},K^{+}\pi
^{0}\right\} \left\{ K^{+}\eta _{8}\right\} \\
C &:&\left\{ \overline{K^{0}}\pi ^{-},K^{-}\pi ^{0},K^{-}\eta _{8}\right\}
\quad ^{SU(3)}\longleftrightarrow ^{SU(2)}\quad \left\{ \overline{K^{0}}\pi
^{-},K^{-}\pi ^{0}\right\} \left\{ K^{-}\eta _{8}\right\} \\
D &:&\left\{ K^{-}\pi ^{+},\overline{K^{0}}\pi ^{0},\overline{K^{0}}\eta
_{8}\right\} \quad ^{SU(3)}\longleftrightarrow ^{SU(2)}\quad \left\{
K^{-}\pi ^{+},\overline{K^{0}}\pi ^{0}\right\} \left\{ \overline{K^{0}}\eta
_{8}\right\}
\end{eqnarray*}

\paragraph{(b) $K\overline{K},\pi \eta $ sets $(Y=0,T_{3}=\pm 1)$ :}

\begin{eqnarray*}
E &:&\left\{ K^{-}K^{0},\pi ^{-}\eta _{8}\right\} \left\{ \text{ }\pi
^{-}\pi ^{0}\right\} \quad ^{SU(3)}\longleftrightarrow ^{SU(2)}\quad \left\{
K^{-}K^{0}\right\} \left\{ \pi ^{-}\eta _{8}\right\} \left\{ \text{ }\pi
^{-}\pi ^{0}\right\} \\
F &:&\left\{ K^{+}\overline{K^{0}},\pi ^{+}\eta _{8}\right\} \left\{ \text{ }%
\pi ^{+}\pi ^{0}\right\} \quad ^{SU(3)}\longleftrightarrow ^{SU(2)}\quad
\left\{ K^{+}\overline{K^{0}}\right\} \left\{ \pi ^{+}\eta _{8}\right\}
\left\{ \text{ }\pi ^{+}\pi ^{0}\right\}
\end{eqnarray*}

\paragraph{(c) $K\overline{K},\pi \eta ,\pi \pi ,\eta \eta $ set $%
(Y=0,T_{3}=0)$ :}

\[
G: 
\begin{array}{c}
\left\{ K^{-}K^{+},K^{0}\overline{K^{0}},\eta _{8}\eta _{8},\pi ^{+}\pi
^{-},\pi ^{0}\pi ^{0},\pi ^{0}\eta _{8}\right\} \\ 
\updownarrow _{SU(2)}^{SU(3)} \\ 
\left\{ K^{-}K^{+},K^{0}\overline{K^{0}}\right\} \left\{ \eta _{8}\eta
_{8}\right\} \left\{ \pi ^{+}\pi ^{-},\pi ^{0}\pi ^{0}\right\} \left\{ \pi
^{0}\eta _{8}\right\}
\end{array}
\]

\paragraph{(d) And also some isolated states}

(i.e. which do not mix) like for example $K^{0}\pi ^{-},(Y=1,T_{3}=-3/2)$%
.\bigskip

We see from this analysis that we have to consider two-channel,
three-channel and six-channel mixings under SU(3), and only two-channel
mixings under SU(2).

\subsection{Two-channel mixings}

\subsubsection{Different parametrizations of M}

In this section, we will develop quite extensively the general two-channel
mixing parametrizations. We start from a general orthogonal transformation
(eq $\left( \text{\ref{generalmix}}\right) $) and M$_{diag}$:

\begin{equation}
\left( 
\begin{array}{c}
\left| C_{1}\right\rangle \\ 
\left| C_{2}\right\rangle
\end{array}
\right) =P\left( 
\begin{array}{cc}
\cos \alpha & -\sin \alpha \\ 
\sin \alpha & \cos \alpha
\end{array}
\right) \left( 
\begin{array}{c}
\left\{ x_{1}\right\} \\ 
\left\{ x_{2}\right\}
\end{array}
\right) \qquad M_{diag}=\left( 
\begin{array}{cc}
e^{i\delta _{1}} & 0 \\ 
0 & e^{i\delta _{2}}
\end{array}
\right)  \label{2chmix1}
\end{equation}

By defining $\varepsilon =\cos \alpha \sin \alpha ,\ \lambda =2\varepsilon
\sin \left( \frac{\delta _{2}-\delta _{1}}{2}\right) $, we obtain the
following forms :

\begin{equation}
M^{(1)}=\left( 
\begin{array}{cc}
\cos ^{2}\alpha \ e^{i\delta _{1}}+\sin ^{2}\alpha \ e^{i\delta _{2}} & \cos
\alpha \sin \alpha \left( e^{i\delta _{2}}-e^{i\delta _{1}}\right) \\ 
\cos \alpha \sin \alpha \left( e^{i\delta _{2}}-e^{i\delta _{1}}\right) & 
\sin ^{2}\alpha \ e^{i\delta _{1}}+\cos ^{2}\alpha \ e^{i\delta _{2}}
\end{array}
\right)  \label{2chform1}
\end{equation}

\begin{equation}
M^{(2)}=\frac{1}{2}\left( 
\begin{array}{cc}
\left( {\small 1+}\sqrt{{\small 1-4\varepsilon }^{2}}\right) e^{i\delta
_{1}}+\left( {\small 1-}\sqrt{{\small 1-4\varepsilon }^{2}}\right)
e^{i\delta _{2}} & 2\varepsilon \left( e^{i\delta _{2}}-e^{i\delta
_{1}}\right) \\ 
2\varepsilon \left( e^{i\delta _{2}}-e^{i\delta _{1}}\right) & \left( 
{\small 1-}\sqrt{{\small 1-4\varepsilon }^{2}}\right) e^{i\delta
_{1}}+\left( {\small 1+}\sqrt{{\small 1-4\varepsilon }^{2}}\right)
e^{i\delta _{2}}
\end{array}
\right)  \label{2chform2}
\end{equation}

\begin{equation}
M^{(3)}=\left( 
\begin{array}{cc}
\sqrt{1-\lambda ^{2}}e^{i\beta _{1}} & i\lambda e^{i\left( \frac{\beta
_{2}+\beta _{1}}{2}\right) } \\ 
i\lambda e^{i\left( \frac{\beta _{2}+\beta _{1}}{2}\right) } & \sqrt{%
1-\lambda ^{2}}e^{i\beta _{2}}
\end{array}
\right) \qquad \text{ with }\left\{ 
\begin{array}{c}
\beta _{1}=\arg \left( \cos ^{2}\alpha \ e^{i\delta _{1}}+\sin ^{2}\alpha \
e^{i\delta _{2}}\right) \\ 
\beta _{2}=\arg \left( \sin ^{2}\alpha \ e^{i\delta _{1}}+\cos ^{2}\alpha \
e^{i\delta _{2}}\right)
\end{array}
\right.  \label{2chform3}
\end{equation}

\begin{equation}
M^{(4)}=e^{i\delta _{1}}\left[ \left( 
\begin{array}{cc}
1 & 0 \\ 
0 & 1
\end{array}
\right) +\left( e^{i\left( \delta _{2}-\delta _{1}\right) }-1\right) \left( 
\begin{array}{cc}
\sin ^{2}\alpha & \cos \alpha \sin \alpha \\ 
\cos \alpha \sin \alpha & \cos ^{2}\alpha
\end{array}
\right) \right]  \label{2chform4}
\end{equation}

Where $\beta _{1}+\beta _{2}=\delta _{1}+\delta _{2}$ and the last form is
obtained from eq. (\ref{termdecomp2}).

\subsubsection{Parametrization of $M^{2}=S$ : The elasticity parameters.}

In this section, we will define the elasticity parameter. Let us consider a
general coupled system of two states $X_{1}$ and $X_{2}$. We can describe
this system in three different bases :

The eigenchannel basis $(C_{1},C_{2})$, with

\begin{equation}
S_{eigen}=\left( 
\begin{array}{cc}
e^{2i\delta _{1}} & 0 \\ 
0 & e^{2i\delta _{2}}
\end{array}
\right)  \label{Seigenbasis}
\end{equation}

The isospin basis $(T_{1},T_{2})$, with

\begin{equation}
S_{isospin}=\left( 
\begin{array}{cc}
\eta _{w}e^{2iw_{1}} & i\sqrt{1-\eta _{w}^{2}}e^{i\left( w_{1}+w_{2}\right) }
\\ 
i\sqrt{1-\eta _{w}^{2}}e^{i\left( w_{1}+w_{2}\right) } & \eta _{w}e^{2iw_{2}}
\end{array}
\right)  \label{Sisobasis}
\end{equation}

The physical basis $(X_{1},X_{2})$, with

\begin{equation}
S_{physical}=\left( 
\begin{array}{cc}
\eta e^{2i\alpha _{1}} & i\sqrt{1-\eta ^{2}}e^{i\left( \alpha _{1}+\alpha
_{2}\right) } \\ 
i\sqrt{1-\eta ^{2}}e^{i\left( \alpha _{1}+\alpha _{2}\right) } & \eta
e^{2i\alpha _{2}}
\end{array}
\right)  \label{Sphysicbasis}
\end{equation}

Thus, we have two different possible definitions of the elasticity parameter
:

(1) The parameter $\eta _{w}$ quantifies the deviation of S in the isospin
basis from its diagonal form in the eigenchannel basis. SU(2) elasticity
implies $\eta _{w}=1$ since when $\eta _{w}=1$, $S_{eigen}=S_{isospin}$. The
phases $w_{1}$ and $w_{2}$ are then eigenphases, sometimes called Watson
phases (hence the subscript w to $\eta _{w}$).

(2) The parameter $\eta $ quantifies the deviation of S in the physical
basis from its diagonal form in the eigenchannel basis. This is the way the
elasticity parameter will be defined in this paper. This definition allows
one to define a elasticity parameter for every mixings, including SU(2)
elastic mixings.

This $\eta $ elasticity parameter is defined in terms of mixing parameter in
the following way (the complete discussion is in the appendix) : $%
S_{physical}$ is built in the standard way, as a general coupled channel
mixing :

\begin{equation}
S_{physical}(\delta _{1},\delta _{2},\beta )=O^{t}(\beta )\ S_{eigen}(\delta
_{1},\delta _{2})\ O(\beta )  \label{physicaltoeigen}
\end{equation}

with $O(\beta )=\left( 
\begin{array}{cc}
\cos \beta & -\sin \beta \\ 
\sin \beta & \cos \beta
\end{array}
\right) $. We can change the parameter basis from $\delta _{1},\delta
_{2},\beta $ to $\alpha _{1},\alpha _{2},\eta $. The elasticity parameter is
then given in terms of mixing parameter as :

\begin{equation}
\eta =\sqrt{1-4\varepsilon ^{2}\sin ^{2}(\delta _{2}-\delta _{1})};\quad
\varepsilon =\cos \beta \sin \beta  \label{elasticitydefmixing}
\end{equation}

This formula is quite interesting. We can distinguish two factor
contributing to $\eta :$

- The $\varepsilon $ parameter quantifies the non-diagonal trend of $O(\beta
)$, i.e. the distance between the physical basis and the eingenchannel basis.

- The elasticity parameter $\eta $ quantifies the non-diagonal trend of S,
which is \textit{function of both }$\varepsilon $\textit{\ and the
eigenphase difference}, since if these phases are equal, the mixings
disappear.

\paragraph{Remarks :}

It is now clear that the form (\ref{Sisobasis}) used in some other papers to
introduce inelasticity is equivalent to a general two-channel \textit{elastic%
} mixing in the context of Watson's theorem. Note however that this form (%
\ref{Sisobasis}) is also used as a general parametrization for a 2x2 unitary
symmetric matrix in the K matrix formalism, and it no longer reduces to an
elastic parametrization there.

\subsubsection{SU(2 or 3) two-channel mixings in B and D decays}

In this section, we will give the parameters defined above for the different
two-channel mixings among two pseudoscalar states, in the framework of SU(2)
and SU(3). From these parameters, one can rebuild easily the mixing matrix
using one of the forms $M^{(1)}$ to $M^{(4)}$ (eq. \ref{2chform1} to \ref
{2chform4}).

\begin{equation}
\begin{tabular}{|l|l|l|l|l|l|l|}
\hline
$\text{{\small Flavour}}$ & {\small Coupled states} & \multicolumn{2}{|l}{$
\begin{array}{c}
\text{{\small Strong}} \\ 
\text{{\small phases}}
\end{array}
$} & \multicolumn{2}{|l|}{$
\begin{array}{c}
\text{{\small Mixing}} \\ 
\text{{\small parameters}}
\end{array}
$} & {\small Elasticity Parameters} \\ \cline{3-6}
$\text{{\small groups}}$ &  & $\delta _{1}$ & $\delta _{2}$ & $\cos \beta $
& $\sin \beta $ &  \\ \hline
{\small SU(3)} & $
\begin{array}{c}
\left\{ K^{-}K^{0},\pi ^{-}\eta _{8}\right\} \\ 
\left\{ K^{+}\overline{K^{0}},\pi ^{+}\eta _{8}\right\}
\end{array}
$ & $\delta _{27}$ & $\delta _{8}$ & $\sqrt{2/5}$ & $\sqrt{3/5}$ & $
\begin{array}{c}
\varepsilon ^{2}=6/25=0.24 \\ 
\eta =\sqrt{1-\frac{24}{25}\sin ^{2}(\delta _{27}-\delta _{8})}
\end{array}
$ \\ \hline
& $
\begin{array}{c}
\left\{ K^{+}\pi ^{-},K^{0}\pi ^{0}\right\} \\ 
\left\{ K^{-}\pi ^{+},\overline{K^{0}}\pi ^{0}\right\}
\end{array}
$ & $\delta _{3/2}$ & $\delta _{1/2}$ & $-\sqrt{1/3}$ & $\sqrt{2/3}$ & $
\begin{array}{c}
\varepsilon ^{2}=2/9\approx 0.22 \\ 
\eta =\sqrt{1-\frac{8}{9}\sin ^{2}(\delta _{3/2}-\delta _{1/2})}
\end{array}
$ \\ \cline{2-7}\cline{3-7}
{\small SU(2)} & $
\begin{array}{c}
\left\{ \overline{K^{0}}\pi ^{-},K^{-}\pi ^{0}\right\} \\ 
\left\{ K^{0}\pi ^{+},K^{+}\pi ^{0}\right\}
\end{array}
$ & $\delta _{3/2}$ & $\delta _{1/2}$ & $\sqrt{1/3}$ & $\sqrt{2/3}$ & $
\begin{array}{c}
\varepsilon ^{2}=2/9\approx 0.22 \\ 
\eta =\sqrt{1-\frac{8}{9}\sin ^{2}(\delta _{3/2}-\delta _{1/2})}
\end{array}
$ \\ \cline{2-7}\cline{3-7}
& $\left\{ \pi ^{+}\pi ^{-},\pi ^{0}\pi ^{0}\right\} $ & $\delta _{2}$ & $%
\delta _{0}$ & $\sqrt{1/3}$ & $\sqrt{2/3}$ & $
\begin{array}{c}
\varepsilon ^{2}=2/9\approx 0.22 \\ 
\eta =\sqrt{1-\frac{8}{9}\sin ^{2}(\delta _{2}-\delta _{0})}
\end{array}
$ \\ \cline{2-7}
& $\left\{ K^{+}K^{-},K^{0}\overline{K}^{0}\right\} $ & $\delta _{1}$ & $%
\delta _{0}$ & $\sqrt{1/2}$ & $\sqrt{1/2}$ & $
\begin{array}{c}
\varepsilon ^{2}=1/4\text{ \ (maximal)} \\ 
\eta =\cos (\delta _{1}-\delta _{0})
\end{array}
$ \\ \hline
\end{tabular}
\label{Table2ch}
\end{equation}

Note that in all the mixings, we have $\eta $ very close to $\cos (\delta
_{1}-\delta _{0})$, i.e. maximal mixings.

\subsubsection{Quark diagrams}

We will illustrate the application of the preceding mixing matrices on QD
parametrizations of $\left\{ K^{-}K^{0},\pi ^{-}\eta _{8}\right\} $ only.
The other channels will be treated when dealing with three and six-channel
mixings.

The SU(3) mixings among the $E:\left\{ K^{-}K^{0},\pi ^{-}\eta _{8}\right\} $
in D decays (Cabibbo approximation for CKM) is simply given by $%
M_{E}^{SU(3)} $:

\begin{equation}
\left( 
\begin{array}{c}
D^{-}\rightarrow K^{-}K^{0} \\ 
D^{-}\rightarrow \pi ^{-}\eta _{8}
\end{array}
\right) =V_{cd}^{*}V_{ud}\left( 
\begin{array}{cc}
\frac{1}{5}\left( 2e^{i\delta _{27}}+3e^{i\delta _{8}}\right) & \frac{\sqrt{6%
}}{5}\left( e^{i\delta _{8}}-e^{i\delta _{27}}\right) \\ 
\frac{\sqrt{6}}{5}\left( e^{i\delta _{8}}-e^{i\delta _{27}}\right) & \frac{1%
}{5}\left( 3e^{i\delta _{27}}+2e^{i\delta _{8}}\right)
\end{array}
\right) \left( 
\begin{array}{c}
-T+A \\ 
\frac{1}{\sqrt{6}}\left( T+3C+2A\right)
\end{array}
\right)  \label{mixE}
\end{equation}

We can proceed similarly in B decays (mixing $F:\left\{ K^{+}\overline{K^{0}}%
,\pi ^{+}\eta _{8}\right\} $):

\begin{equation}
\left( 
\begin{array}{c}
B^{+}\rightarrow K^{+}\overline{K^{0}} \\ 
B^{+}\rightarrow \pi ^{+}\eta _{8}
\end{array}
\right) =M_{F}^{SU(3)}\left( 
\begin{array}{c}
V_{ub}^{*}V_{ud}\left( A+P\right) +V_{cb}^{*}V_{cd}\left( P^{c}\right)
+V_{tb}^{*}V_{td}\left( P^{t}\right) \\ 
\frac{1}{\sqrt{6}}\left[ V_{ub}^{*}V_{ud}\left( T+C+2A+2P\right)
+V_{cb}^{*}V_{cd}\left( 2P^{c}\right) +V_{tb}^{*}V_{td}\left( 2P^{t}\right)
\right]
\end{array}
\right)  \label{mixF}
\end{equation}

\subsection{Three-channel mixings}

\subsubsection{Bare amplitudes}

Consider for definiteness the set of decays $\overline{D}^{0}$ to $K^{+}\pi
^{-}$, $K^{0}\pi ^{0}$, $K^{0}\eta _{8}$. The SU(3) QD decompositions are,
omitting CKM elements :

\begin{equation}
\left\{ 
\begin{array}{l}
\left( \overline{D}^{0}\rightarrow \left\{ K^{+}\pi ^{-}\right\} \right) =T+E
\\ 
\left( \overline{D}^{0}\rightarrow \left\{ K^{0}\pi ^{0}\right\} \right) =%
\frac{1}{\sqrt{2}}\left( C-E\right) \\ 
\left( \overline{D}^{0}\rightarrow \left\{ K^{0}\eta _{8}\right\} \right) =%
\frac{1}{\sqrt{6}}\left( C-E\right)
\end{array}
\right.  \label{udq3ch}
\end{equation}

\subsubsection{Physical amplitudes}

We can now apply FSI mixing matrices on these bare amplitudes to obtain a
parametrization of physical decay amplitudes.

\paragraph{SU(3) mixings :}

Applying the SU(3) elastic mixing matrix (\ref{su(3)kpmix}), we find the
full amplitudes :

\begin{equation}
\left\{ 
\begin{array}{l}
\left( \overline{D}^{0}\rightarrow K^{+}\pi ^{-}\right) =T\ \left( \dfrac{%
2e^{i\delta _{27}}+3e^{i\delta _{8}}}{5}\right) +C\ \dfrac{2}{5}\ \left(
e^{i\delta _{27}}-e^{i\delta _{8}}\right) +Ee^{i\delta _{8}} \\ 
\left( \overline{D}^{0}\rightarrow K^{0}\pi ^{0}\right) =\frac{1}{\sqrt{2}}%
\left( T\ \dfrac{3}{5}\ \left( e^{i\delta _{27}}-e^{i\delta _{8}}\right) +C\
\left( \dfrac{3e^{i\delta _{27}}+2e^{i\delta _{8}}}{5}\right) -Ee^{i\delta
_{8}}\right) \\ 
\left( \overline{D}^{0}\rightarrow K^{0}\eta _{8}\right) =\frac{1}{\sqrt{6}}%
\left( T\ \dfrac{3}{5}\ \left( e^{i\delta _{27}}-e^{i\delta _{8}}\right) +C\
\left( \dfrac{3e^{i\delta _{27}}+2e^{i\delta _{8}}}{5}\right) -Ee^{i\delta
_{8}}\right)
\end{array}
\right.  \label{udq3ch2}
\end{equation}

\paragraph{SU(2) mixings :}

If we choose to apply the SU(2) matrix (\ref{isoMill2}), we get :

\begin{equation}
\left\{ 
\begin{array}{l}
\left( \overline{D}^{0}\rightarrow K^{+}\pi ^{-}\right) =T\ \left( \dfrac{%
e^{i\delta _{3/2}}+2e^{i\delta _{1/2}^{(1)}}}{3}\right) +C\ \left( \dfrac{%
e^{i\delta _{3/2}}-e^{i\delta _{1/2}^{(1)}}}{3}\right) +Ee^{i\delta
_{1/2}^{(1)}} \\ 
\left( \overline{D}^{0}\rightarrow K^{0}\pi ^{0}\right) =\frac{1}{\sqrt{2}}%
\left( T\ \left( \dfrac{2\left( e^{i\delta _{3/2}}-e^{i\delta
_{1/2}^{(1)}}\right) }{3}\right) +C\ \left( \dfrac{2e^{i\delta
_{3/2}}+e^{i\delta _{1/2}^{(1)}}}{3}\right) -Ee^{i\delta _{1/2}^{(1)}}\right)
\\ 
\left( \overline{D}^{0}\rightarrow K^{0}\eta _{8}\right) =\frac{1}{\sqrt{6}}%
\left( C-E\right) e^{i\delta _{1/2}^{(2)}}
\end{array}
\right.  \label{udq3ch3}
\end{equation}

\paragraph{General mixings :}

Instead of applying directly the $M^{general}\left( \alpha \right) $ (\ref
{su(2)tosu(3)4}), let us rewrite it in an interesting way. Using a
decomposition like (\ref{termdecomp2}), we write :

\begin{equation}
M^{general}\left( \alpha \right) =\left( 
\begin{array}{c}
\frac{1}{3}\left( 
\begin{array}{ccc}
e^{i\delta _{C1}}+2e^{i\delta _{C2}} & \sqrt{2}\left( e^{i\delta
_{C1}}-e^{i\delta _{C2}}\right) & 0 \\ 
\sqrt{2}\left( e^{i\delta _{C1}}-e^{i\delta _{C2}}\right) & 2e^{i\delta
_{C1}}+e^{i\delta _{C2}} & 0 \\ 
0 & 0 & 3e^{i\delta _{C2}}
\end{array}
\right) \\ 
+\frac{\left( e^{i\delta _{C3}}-e^{i\delta _{C2}}\right) }{3}\left( 
\begin{array}{ccc}
2\sin ^{2}\alpha & -\sqrt{2}\sin ^{2}\alpha & -\sqrt{6}\cos \alpha \sin
\alpha \\ 
-\sqrt{2}\sin ^{2}\alpha & \sin ^{2}\alpha & \sqrt{3}\cos \alpha \sin \alpha
\\ 
-\sqrt{6}\cos \alpha \sin \alpha & \sqrt{3}\cos \alpha \sin \alpha & 3\cos
^{2}\alpha
\end{array}
\right)
\end{array}
\right)  \label{3chtermdcp1}
\end{equation}

We see that the first term correspond to the SU(2) mixing (see eq(\ref
{isoMill2})), and the second one is the perturbation due to the mixing with $%
K^{0}\eta _{8}$. This equation shows that the mixing $K\pi $ with $K^{0}\eta
_{8}$ is function of both the mixing parameter $\alpha $ and the eigenphase
difference $\left( e^{i\delta _{C3}}-e^{i\delta _{C2}}\right) $ (exactly
like the elasticity parameter $\eta $ in two-channel mixing). If $\alpha $
is small, we write :

\begin{equation}
M=\left( 
\begin{array}{c}
\frac{1}{3}\left( 
\begin{array}{ccc}
e^{i\delta _{C1}}+2e^{i\delta _{C2}} & \sqrt{2}\left( e^{i\delta
_{C1}}-e^{i\delta _{C2}}\right) & 0 \\ 
\sqrt{2}\left( e^{i\delta _{C1}}-e^{i\delta _{C2}}\right) & 2e^{i\delta
_{C1}}+e^{i\delta _{C2}} & 0 \\ 
0 & 0 & 3e^{i\delta _{C2}}
\end{array}
\right) \\ 
+\frac{\left( e^{i\delta _{C3}}-e^{i\delta _{C2}}\right) }{3}\left( 
\begin{array}{ccc}
0 & 0 & -\sqrt{6}\alpha \\ 
0 & 0 & \sqrt{3}\alpha \\ 
-\sqrt{6}\alpha & \sqrt{3}\alpha & 3
\end{array}
\right)
\end{array}
\right)  \label{3chtermdcp2}
\end{equation}

This form can be applied to QD parametrizations. From (\ref{udq3ch}), we
have for $K\pi $ decays (omitting CKM factors):

\begin{equation}
\left\{ 
\begin{array}{l}
\left( \overline{D}^{0}\rightarrow K^{+}\pi ^{-}\right) = \\ 
\quad T\ \left( \dfrac{e^{i\delta _{C1}}+2e^{i\delta _{C2}}}{3}\right) +C\
\left( \dfrac{e^{i\delta _{C1}}-e^{i\delta _{C2}}}{3}\right) +Ee^{i\delta
_{C2}}-\dfrac{\alpha \left( e^{i\delta _{C3}}-e^{i\delta _{C2}}\right) }{3}%
\left( C-E\right) \\ 
\left( \overline{D}^{0}\rightarrow K^{0}\pi ^{0}\right) = \\ 
\quad \frac{1}{\sqrt{2}}\left( T\ \left( \dfrac{2\left( e^{i\delta
_{C1}}-e^{i\delta _{C2}}\right) }{3}\right) +C\ \left( \dfrac{2e^{i\delta
_{C1}}+e^{i\delta _{C2}}}{3}\right) -Ee^{i\delta _{C2}}+\dfrac{\alpha \left(
e^{i\delta _{C3}}-e^{i\delta _{C2}}\right) }{3}\left( C-E\right) \right)
\end{array}
\right.  \label{3chperturb}
\end{equation}

to be compared with (\ref{udq3ch3}).

\subsubsection{Other three-channel mixings}

The forms (\ref{su(3)kpmix}, \ref{3chtermdcp1},...) for the mixing matrix is
valid for the mixings in the sets A and D. For the mixings in B and C, the $%
M^{SU(3)}$ matrix is :

\begin{equation}
M^{SU(3)}=\frac{1}{5}\left( 
\begin{array}{ccc}
2e^{i\delta _{27}}+3e^{i\delta _{8}} & -\frac{3}{\sqrt{2}}\left( e^{i\delta
_{27}}-e^{i\delta _{8}}\right) & \sqrt{\frac{3}{2}}\left( e^{i\delta
_{27}}-e^{i\delta _{8}}\right) \\ 
-\frac{3}{\sqrt{2}}\left( e^{i\delta _{27}}-e^{i\delta _{8}}\right) & \frac{1%
}{2}\left( 7e^{i\delta _{27}}+3e^{i\delta _{8}}\right) & \frac{\sqrt{3}}{2}%
\left( e^{i\delta _{27}}-e^{i\delta _{8}}\right) \\ 
\sqrt{\frac{3}{2}}\left( e^{i\delta _{27}}-e^{i\delta _{8}}\right) & \frac{%
\sqrt{3}}{2}\left( e^{i\delta _{27}}-e^{i\delta _{8}}\right) & \frac{1}{2}%
\left( 9e^{i\delta _{27}}+e^{i\delta _{8}}\right)
\end{array}
\right)  \label{su(3)3ch2}
\end{equation}

which differs from eq(\ref{su(3)kpmix}) by some signs only. Note that these
signs are not SU(3) phase conventions dependent, since phase conventions
always disappear when calculating M matrices.

\subsection{Six-channel mixings}

\paragraph{SU(2) analysis of set of states G :}

For the last group of coupled states (G), we can make the following isospin
analysis :

\begin{equation}
\left( 
\begin{array}{l}
\left\{ K\overline{K}\right\} :\text{ I = 1} \\ 
\left\{ K\overline{K}\right\} :\text{I = 0} \\ 
\left\{ \eta _{8}\eta _{8}\right\} :\text{I = 0} \\ 
\left\{ \pi \pi \right\} :\text{I = 2} \\ 
\left\{ \pi \pi \right\} :\text{I = 0} \\ 
\left\{ \pi ^{0}\eta _{8}\right\} :\text{I = 1}
\end{array}
\right) =\left( 
\begin{array}{cccccc}
{\small -}\sqrt{\frac{1}{2}} & \sqrt{\frac{1}{2}} & 0 & 0 & 0 & 0 \\ 
\sqrt{\frac{1}{2}} & \sqrt{\frac{1}{2}} & 0 & 0 & 0 & 0 \\ 
0 & 0 & 1 & 0 & 0 & 0 \\ 
0 & 0 & 0 & {\small -}\sqrt{\frac{1}{3}} & \sqrt{\frac{2}{3}} & 0 \\ 
0 & 0 & 0 & {\small -}\sqrt{\frac{2}{3}} & {\small -}\sqrt{\frac{1}{3}} & 0
\\ 
0 & 0 & 0 & 0 & 0 & 1
\end{array}
\right) \left( 
\begin{array}{l}
\left\{ K^{-}K^{+}\right\} :\text{ I = 1,0} \\ 
\left\{ K^{0}\overline{K^{0}}\right\} :\text{ I = 1,0} \\ 
\left\{ \eta _{8}\eta _{8}\right\} :\text{ I = 0} \\ 
\left\{ \pi ^{+}\pi ^{-}\right\} :\text{ I = 2,0} \\ 
\left\{ \pi ^{0}\pi ^{0}\right\} :\text{ I = 2,0} \\ 
\left\{ \pi ^{0}\eta _{8}\right\} :\text{ I = 1}
\end{array}
\right)  \label{stateyequal0}
\end{equation}

We obtain the already introduced SU(2) mixing matrix (see table (\ref
{Table2ch})): $M^{SU(2)}=$

\begin{equation}
\left( 
\begin{array}{cccccc}
\frac{1}{2}\left( e^{i\delta _{0}^{K\overline{K}}}{\small +}e^{i\delta
_{1}^{K\overline{K}}}\right) & \frac{1}{2}\left( e^{i\delta _{0}^{K\overline{%
K}}}{\small -}e^{i\delta _{1}^{K\overline{K}}}\right) & 0 & 0 & 0 & 0 \\ 
\frac{1}{2}\left( e^{i\delta _{0}^{K\overline{K}}}{\small -}e^{i\delta
_{1}^{K\overline{K}}}\right) & \frac{1}{2}\left( e^{i\delta _{0}^{K\overline{%
K}}}{\small +}e^{i\delta _{1}^{K\overline{K}}}\right) & 0 & 0 & 0 & 0 \\ 
0 & 0 & e^{i\delta _{0}^{\eta \eta }} & 0 & 0 & 0 \\ 
0 & 0 & 0 & \frac{1}{3}\left( e^{i\delta _{2}^{\pi \pi }}{\small +}%
2e^{i\delta _{0}^{\pi \pi }}\right) & \frac{\sqrt{2}}{3}\left( e^{i\delta
_{0}^{\pi \pi }}{\small -}e^{i\delta _{2}^{\pi \pi }}\right) & 0 \\ 
0 & 0 & 0 & \frac{\sqrt{2}}{3}\left( e^{i\delta _{0}^{\pi \pi }}{\small -}%
e^{i\delta _{2}^{\pi \pi }}\right) & \frac{1}{3}\left( 2e^{i\delta _{2}^{\pi
\pi }}{\small +}e^{i\delta _{0}^{\pi \pi }}\right) & 0 \\ 
0 & 0 & 0 & 0 & 0 & e^{i\delta _{1}^{\pi \eta }}
\end{array}
\right)  \label{6chsu(2)mix}
\end{equation}

\paragraph{Most general mixing among the states of set G compatible with
isospin :}

From SU(2), to go towards SU(3), we must introduce an extra mixing between
the two isospin 1 states (one mixing parameter $\alpha _{1}$) and extra
mixings between the three isospin 0 states (three mixing parameters $\alpha
_{2},\alpha _{3}$ and $\alpha _{4}$) :\smallskip

$O(\alpha _{1},\alpha _{2},\alpha _{3},\alpha _{4})\left( \text{Intermediate
states}\right) =$

\begin{equation}
\left( 
\begin{array}{cccccc}
1 & 0 & 0 & 0 & 0 & 0 \\ 
0 & x & x & 0 & x & 0 \\ 
0 & x & x & 0 & x & 0 \\ 
0 & 0 & 0 & 1 & 0 & 0 \\ 
0 & x & x & 0 & x & 0 \\ 
0 & 0 & 0 & 0 & 0 & 1
\end{array}
\right) \left( 
\begin{array}{cccccc}
y & 0 & 0 & 0 & 0 & y \\ 
0 & 1 & 0 & 0 & 0 & 0 \\ 
0 & 0 & 1 & 0 & 0 & 0 \\ 
0 & 0 & 0 & 1 & 0 & 0 \\ 
0 & 0 & 0 & 0 & 1 & 0 \\ 
y & 0 & 0 & 0 & 0 & y
\end{array}
\right) \left( 
\begin{array}{l}
\left\{ K\overline{K}\right\} :\text{ Isospin 1} \\ 
\left\{ K\overline{K}\right\} :\text{Isospin 0} \\ 
\left\{ \eta _{8}\eta _{8}\right\} :\text{Isospin 0} \\ 
\left\{ \pi \pi \right\} :\text{Isospin 2} \\ 
\left\{ \pi \pi \right\} :\text{Isospin 0} \\ 
\left\{ \pi ^{0}\eta _{8}\right\} :\text{Isospin 1}
\end{array}
\right)  \label{generalmixG}
\end{equation}

with x the entries of a 3 by 3 orthogonal matrix (three angles $\alpha
_{2},\alpha _{3},\alpha _{4}$) and y the entries of a 2 by 2 orthogonal
matrix (one angle $\alpha _{1}$). By using $O(\alpha _{1},\alpha _{2},\alpha
_{3},\alpha _{4})$, we can build the most general mixing $M^{general}(\alpha
_{1},\alpha _{2},\alpha _{3},\alpha _{4})$ among this six state set
compatible with isospin.

\paragraph{SU(3) analysis of set of states G :}

Finally, we can find a value for each four parameters $\alpha
_{1},...,\alpha _{4}$ such that $M^{general}(\alpha _{1},\alpha _{2},\alpha
_{3},\alpha _{4})\rightarrow M^{SU(3)}$ $=$

\begin{eqnarray}
&&\left( 
\begin{array}{ccc}
\frac{7a}{20}{\small +}\frac{2b}{5}{\small +}\frac{c}{4} & {\small -}\frac{a%
}{20}{\small -}\frac{b}{5}{\small +}\frac{c}{4} & \frac{1}{\sqrt{2}}\left( 
{\small -}\frac{9a}{20}{\small +}\frac{b}{5}{\small +}\frac{c}{4}\right) \\ 
-\frac{a}{20}{\small -}\frac{b}{5}{\small +}\frac{c}{4} & \frac{7a}{20}%
{\small +}\frac{2b}{5}{\small +}\frac{c}{4} & \frac{1}{\sqrt{2}}\left( -%
\frac{9a}{20}{\small +}\frac{b}{5}{\small +}\frac{c}{4}\right) \\ 
\frac{1}{\sqrt{2}}\left( {\small -}\frac{9a}{20}{\small +}\frac{b}{5}{\small %
+}\frac{c}{4}\right) & \frac{1}{\sqrt{2}}\left( {\small -}\frac{9a}{20}%
{\small +}\frac{b}{5}{\small +}\frac{c}{4}\right) & \frac{27a}{40}{\small +}%
\frac{b}{5}{\small +}\frac{c}{8} \\ 
{\small -}\frac{a}{20}{\small -}\frac{b}{5}{\small +}\frac{c}{4} & {\small -}%
\frac{a}{20}{\small -}\frac{b}{5}{\small +}\frac{c}{4} & \frac{1}{\sqrt{2}}%
\left( \frac{3a}{20}{\small -}\frac{2b}{5}{\small +}\frac{c}{4}\right) \\ 
\frac{1}{\sqrt{2}}\left( {\small -}\frac{a}{20}{\small -}\frac{b}{5}{\small +%
}\frac{c}{4}\right) & \frac{1}{\sqrt{2}}\left( -\frac{a}{20}-\frac{b}{5}+%
\frac{c}{4}\right) & \frac{3a}{40}{\small -}\frac{b}{5}{\small +}\frac{c}{8}
\\ 
\frac{\sqrt{3}}{5}\left( b{\small -}a\right) & -\frac{\sqrt{3}}{5}\left( b%
{\small -}a\right) & 0
\end{array}
\right. \cdots  \label{6chsu(3)mix} \\
&&\cdots \left. 
\begin{array}{ccc}
{\small -}\frac{a}{20}{\small -}\frac{b}{5}{\small +}\frac{c}{4} & \frac{1}{%
\sqrt{2}}\left( {\small -}\frac{a}{20}{\small -}\frac{b}{5}{\small +}\frac{c%
}{4}\right) & \frac{\sqrt{3}}{5}\left( b{\small -}a\right) \\ 
{\small -}\frac{a}{20}{\small -}\frac{b}{5}{\small +}\frac{c}{4} & \frac{1}{%
\sqrt{2}}\left( {\small -}\frac{a}{20}{\small -}\frac{b}{5}{\small +}\frac{c%
}{4}\right) & {\small -}\frac{\sqrt{3}}{5}\left( b{\small -}a\right) \\ 
\frac{1}{\sqrt{2}}\left( \frac{3a}{20}{\small -}\frac{2b}{5}{\small +}\frac{c%
}{4}\right) & \frac{3a}{40}{\small -}\frac{b}{5}{\small +}\frac{c}{8} & 0 \\ 
\frac{7a}{20}{\small +}\frac{2b}{5}{\small +}\frac{c}{4} & \frac{1}{\sqrt{2}}%
\left( {\small -}\frac{13a}{20}{\small +}\frac{2b}{5}{\small +}\frac{c}{4}%
\right) & 0 \\ 
\frac{1}{\sqrt{2}}\left( {\small -}\frac{13a}{20}{\small +}\frac{2b}{5}%
{\small +}\frac{c}{4}\right) & \frac{27a}{40}{\small +}\frac{b}{5}{\small +}%
\frac{c}{8} & 0 \\ 
0 & 0 & \frac{1}{5}\left( 2b{\small +}3a\right)
\end{array}
\right) \qquad \text{with }\left\{ 
\begin{array}{l}
a=e^{i\delta _{27}} \\ 
b=e^{i\delta _{8}} \\ 
c=e^{i\delta _{1}}
\end{array}
\right.  \nonumber
\end{eqnarray}

\paragraph{SU(2) specification in the SU(3) analysis :}

In order to introduce SU(3) breaking in the FSI, we can calculate a form
like (\ref{su(2)tosu(3)6}) for this set of states. We obtain the following
matrix $M^{SU(2\text{ }in\text{ }3)}$:

\begin{equation}
\left( 
\begin{array}{ccc}
\frac{b}{5}+\frac{3c}{20}+\frac{3d}{10}+\frac{e}{10}+\frac{f}{4} & -\frac{b}{%
5}+\frac{3c}{20}-\frac{3d}{10}+\frac{e}{10}+\frac{f}{4} & \frac{1}{\sqrt{2}}%
\left( -\frac{9c}{20}+\frac{e}{5}+\frac{f}{4}\right) \\ 
-\frac{b}{5}+\frac{3c}{20}-\frac{3d}{10}+\frac{e}{10}+\frac{f}{4} & \frac{b}{%
5}+\frac{3c}{20}+\frac{3d}{10}+\frac{e}{10}+\frac{f}{4} & \frac{1}{\sqrt{2}}%
\left( -\frac{9c}{20}+\frac{e}{5}+\frac{f}{4}\right) \\ 
\frac{1}{\sqrt{2}}\left( -\frac{9c}{20}+\frac{e}{5}+\frac{f}{4}\right) & 
\frac{1}{\sqrt{2}}\left( -\frac{9c}{20}+\frac{e}{5}+\frac{f}{4}\right) & 
\frac{27c}{40}+\frac{e}{5}+\frac{f}{8} \\ 
-\frac{c}{20}-\frac{e}{5}+\frac{f}{4} & -\frac{c}{20}-\frac{e}{5}+\frac{f}{4}
& \frac{1}{\sqrt{2}}\left( \frac{3c}{20}-\frac{2e}{5}+\frac{f}{4}\right) \\ 
\frac{1}{\sqrt{2}}\left( -\frac{c}{20}-\frac{e}{5}+\frac{f}{4}\right) & 
\frac{1}{\sqrt{2}}\left( -\frac{c}{20}-\frac{e}{5}+\frac{f}{4}\right) & 
\frac{3c}{40}-\frac{e}{5}+\frac{f}{8} \\ 
\frac{\sqrt{3}}{5}\left( d-b\right) & -\frac{\sqrt{3}}{5}\left( d-b\right) & 
0
\end{array}
\right. \cdots  \label{6chsu(3-2)mix}
\end{equation}

\[
\cdots \left. 
\begin{array}{ccc}
-\frac{c}{20}-\frac{e}{5}+\frac{f}{4} & \frac{1}{\sqrt{2}}\left( -\frac{c}{20%
}-\frac{e}{5}+\frac{f}{4}\right) & \frac{\sqrt{3}}{5}\left( d-b\right) \\ 
-\frac{c}{20}-\frac{e}{5}+\frac{f}{4} & \frac{1}{\sqrt{2}}\left( -\frac{c}{20%
}-\frac{e}{5}+\frac{f}{4}\right) & -\frac{\sqrt{3}}{5}\left( d-b\right) \\ 
\frac{1}{\sqrt{2}}\left( \frac{3c}{20}-\frac{2e}{5}+\frac{f}{4}\right) & 
\frac{3c}{40}-\frac{e}{5}+\frac{f}{8} & 0 \\ 
\frac{a}{3}+\frac{c}{60}+\frac{2e}{5}+\frac{f}{4} & \frac{1}{\sqrt{2}}\left(
-\frac{2a}{3}+\frac{c}{60}+\frac{2e}{5}+\frac{f}{4}\right) & 0 \\ 
\frac{1}{\sqrt{2}}\left( -\frac{2a}{3}+\frac{c}{60}+\frac{2e}{5}+\frac{f}{4}%
\right) & \frac{2a}{3}+\frac{c}{120}+\frac{e}{5}+\frac{f}{8} & 0 \\ 
0 & 0 & \frac{1}{5}\left( 2d+3b\right)
\end{array}
\right) \qquad 
\]

with $a=e^{i\delta _{27}^{2}},\quad b=e^{i\delta _{27}^{1}},\quad
c=e^{i\delta _{27}^{0}},\quad d=e^{i\delta _{8}^{1}},\quad e=e^{i\delta
_{8}^{0}},\quad f=e^{i\delta _{1}^{0}}$ (the notation is $\delta _{SU(3)\
rep.}^{isospin}$), where we have distinguished SU(3) phases according to
isospin. The SU(3) limit can be obtained by identifying $a=b=c$, $d=e$ and
the modified SU(2) limit by identifying $b=d$, $c=e=f$.

\paragraph{Other possibilities :}

We can of course also limit ourself to some intermediate mixings. For
example, forgetting states containing $\eta _{8}$, we can mix $\left\{ K%
\overline{K}\right\} _{T=0}$ with $\left\{ \pi \pi \right\} _{T=0}$ with the
orthogonal matrix :

\begin{equation}
O^{\left\{ K\overline{K},\pi \pi \right\} }=\left( 
\begin{array}{cccc}
1 & 0 & 0 & 0 \\ 
0 & \cos \alpha & 0 & -\sin \alpha \\ 
0 & 0 & 1 & 0 \\ 
0 & \sin \alpha & 0 & \cos \alpha
\end{array}
\right) \left( 
\begin{array}{cccc}
{\small -}\sqrt{\frac{1}{2}} & \sqrt{\frac{1}{2}} & 0 & 0 \\ 
\sqrt{\frac{1}{2}} & \sqrt{\frac{1}{2}} & 0 & 0 \\ 
0 & 0 & {\small -}\sqrt{\frac{1}{3}} & \sqrt{\frac{2}{3}} \\ 
0 & 0 & {\small -}\sqrt{\frac{2}{3}} & {\small -}\sqrt{\frac{1}{3}}
\end{array}
\right)  \label{4chmix}
\end{equation}

and build a mixing matrix $M^{\left\{ K\overline{K},\pi \pi \right\}
}=\left( O^{\left\{ K\overline{K},\pi \pi \right\} }\right)
^{t}M_{diag}^{\left\{ K\overline{K},\pi \pi \right\} }O^{\left\{ K\overline{K%
},\pi \pi \right\} }$, which can be used for example in D decays as :

\begin{equation}
\left( 
\begin{array}{c}
\overline{D^{0}}\rightarrow K^{+}K^{-} \\ 
\overline{D^{0}}\rightarrow K^{0}\overline{K^{0}} \\ 
\overline{D^{0}}\rightarrow \pi ^{+}\pi ^{-} \\ 
\overline{D^{0}}\rightarrow \pi ^{0}\pi ^{0}
\end{array}
\right) =M^{\left\{ K\overline{K},\pi \pi \right\} }\left( 
\begin{array}{c}
\overline{D^{0}}\rightarrow \left\{ K^{+}K^{-}\right\} \\ 
\overline{D^{0}}\rightarrow \left\{ K^{0}\overline{K^{0}}\right\} \\ 
\overline{D^{0}}\rightarrow \left\{ \pi ^{+}\pi ^{-}\right\} \\ 
\overline{D^{0}}\rightarrow \left\{ \pi ^{0}\pi ^{0}\right\}
\end{array}
\right)  \label{4chmix2}
\end{equation}

\subsubsection{Application to QD decompositions}

All these matrices can now be applied on quark diagram decompositions. For D
decays, using Cabibbo approximation (and omitting CKM) :

\begin{equation}
\left( 
\begin{array}{c}
\overline{D^{0}}\rightarrow K^{+}K^{-} \\ 
\overline{D^{0}}\rightarrow K^{0}\overline{K^{0}} \\ 
\overline{D^{0}}\rightarrow \eta _{8}\eta _{8} \\ 
\overline{D^{0}}\rightarrow \pi ^{+}\pi ^{-} \\ 
\overline{D^{0}}\rightarrow \pi ^{0}\pi ^{0} \\ 
\overline{D^{0}}\rightarrow \pi ^{0}\eta _{8}
\end{array}
\right) =M\left( 
\begin{array}{l}
\overline{D^{0}}\rightarrow \left\{ K^{+}K^{-}\right\} =-T-E \\ 
\overline{D^{0}}\rightarrow \left\{ K^{0}\overline{K^{0}}\right\} =0 \\ 
\overline{D^{0}}\rightarrow \left\{ \eta _{8}\eta _{8}\right\} =\frac{1}{%
\sqrt{2}}\left( C-E\right) \\ 
\overline{D^{0}}\rightarrow \left\{ \pi ^{+}\pi ^{-}\right\} =T+E \\ 
\overline{D^{0}}\rightarrow \left\{ \pi ^{0}\pi ^{0}\right\} =\frac{1}{\sqrt{%
2}}\left( -C+E\right) \\ 
\overline{D^{0}}\rightarrow \left\{ \pi ^{0}\eta _{8}\right\} =\frac{1}{%
\sqrt{3}}\left( C-E\right)
\end{array}
\right)  \label{6chdqparam}
\end{equation}

For $B^{0}$ and $B_{s}$ decays into these channels, the QD parametrizations
of decay amplitudes are given in the appendix.

\paragraph{Example : D decays to $K\overline{K}$.}

As is well-known, the amplitudes for the decay $\overline{D^{0}}\rightarrow
K^{0}\overline{K^{0}}$ is identically zero under SU(3). Here we can see that
this decay is zero at the level of bare amplitude. Of course, if we apply
the SU(3) FSI matrix, the full amplitude remains zero. Under SU(2), we find
as usual :

\begin{equation}
\left( 
\begin{array}{c}
\overline{D^{0}}\rightarrow K^{+}K^{-} \\ 
\overline{D^{0}}\rightarrow K^{0}\overline{K^{0}}
\end{array}
\right) =\left( 
\begin{array}{c}
-\frac{1}{2}\left( e^{i\delta _{0}^{K\overline{K}}}+e^{i\delta _{1}^{K%
\overline{K}}}\right) \left( T+E\right) \\ 
-\frac{1}{2}\left( e^{i\delta _{0}^{K\overline{K}}}-e^{i\delta _{1}^{K%
\overline{K}}}\right) \left( T+E\right)
\end{array}
\right)  \label{KKsu(2)mix}
\end{equation}

If we use the $M^{SU(2\text{ }in\text{ }3)}$ form $\left( \text{\ref
{6chsu(3-2)mix}}\right) $, we find :

\begin{equation}
\left\{ 
\begin{array}{l}
\overline{D^{0}}\rightarrow K^{+}K^{-}= \\ 
-T\left( \dfrac{2(e^{i\delta _{27}^{0}}+e^{i\delta _{27}^{1}})+3(e^{i\delta
_{8}^{0}}+e^{i\delta _{8}^{1}})}{10}\right) +C\left( \dfrac{(e^{i\delta
_{8}^{0}}+e^{i\delta _{8}^{1}})-(e^{i\delta _{27}^{0}}+e^{i\delta _{27}^{1}})%
}{5}\right) -E\left( \dfrac{e^{i\delta _{8}^{0}}+e^{i\delta _{8}^{1}}}{2}%
\right) \\ 
\overline{D^{0}}\rightarrow K^{0}\overline{K^{0}}= \\ 
-T\left( \dfrac{2(e^{i\delta _{27}^{0}}-e^{i\delta _{27}^{1}})+3(e^{i\delta
_{8}^{0}}-e^{i\delta _{8}^{1}})}{10}\right) +C\left( \dfrac{(e^{i\delta
_{8}^{0}}-e^{i\delta _{8}^{1}})-(e^{i\delta _{27}^{0}}-e^{i\delta _{27}^{1}})%
}{5}\right) -E\left( \dfrac{e^{i\delta _{8}^{0}}-e^{i\delta _{8}^{1}}}{2}%
\right)
\end{array}
\right.  \label{KKsu(3-2)mix}
\end{equation}

Where we can see that the non-zero $\overline{D^{0}}\rightarrow K^{0}%
\overline{K^{0}}$ is generated by SU(3) breaking in the FSI phases. If we
identify $\delta _{27}^{T}=\delta _{8}^{T}=\delta _{T}^{K\overline{K}}$, we
recover the SU(2) limit (\ref{KKsu(2)mix}) and if we identify $\delta
_{R}^{1}=\delta _{R}^{0}=\delta _{R}$, we recover the SU(3) limit $\left( 
\overline{D^{0}}\rightarrow K^{0}\overline{K^{0}}\right) =0$ .
Experimentally, the amplitude for $\overline{D^{0}}\rightarrow K^{0}%
\overline{K^{0}}$ is non-negligible compared to $\overline{D^{0}}\rightarrow
K^{+}K^{-}$, the SU(3) breaking is therefore quite important. Note that this
interpretation of the non-zero amplitude $\left( \overline{D^{0}}\rightarrow
K^{0}\overline{K^{0}}\right) $ given here is not new, but it shows the
simplicity of the proposed matrix method.

\section{Conclusion}

The main motivation of our work is to obtain a parametrization of B and D
decays which can be used to extract some theoretically interesting
quantities from experimental measurements.

The first step towards this parametrization is the generalized Watson's
theorem $W=\sqrt{S}W_{b}$ applied to a set of decay channels. This theorem
shows that physical decay amplitudes can be factorized into a bare part and
a FSI part. The model character of our procedure enters precisely when identifying 
those bare amplitudes to elementary processes free of FSI effect, this 
identification being strictly equivalent to the elastic hypothesis (as soon as the
 \textbf{S}-matrix\ is unitary). Elementary processes and FSI effect 
can then be analyzed separately. Quark diagrams are used at the bare level, 
and this ensures that they are well-defined in terms of elementary processes. 
For the FSI part, we introduce unitary mixing matrices. Unitarity of these mixing matrices is
equivalent to probability conservation among the set of decay channels, i.e.
to elasticity. The important point is that FSI are treated at the hadronic
level, since our S matrix was build from hadron states. We have then shown
how to build mixing matrices, using symmetry group or introducing selected
mixings among hadron states.

The next step should be to simplify the parametrizations obtained. Indeed,
if we introduce FSI as some general mixings, we introduce many mixing
parameters and many strong phases, and since we have only a limited number
of possible decays, we have to reduce the number of parameters. Dynamical
considerations can lead to the neglect of some quark diagrams (usually, A
and PA), and also to the neglect of some mixings among possible final
states. As we have repeatedly emphasized, it is also possible to use flavour
symmetry to fix some mixings. Finally, Regge phenomenology may be useful to
calculate some strong phases.

The final step is of course comparison with experimental data. Within our 
model framework, one can build simple parametrizations. Wether the various hypotheses, 
emphasized in this work, are valid or not can then be tested, especially the elasticity
 of FSI, and the limited extend of the set of rescattering channels. For example, 
in B decays, treating FSI as elastic under SU(2) may be a sufficient approximation. 
But it could also happen that mixing with $\eta ,  \eta ' $ or charmed meson states are important 
(the present approach is straightforwardly extended to these mixings), or even mixings with multibody states.
Finally, the present elastic approach for FSI could be inappropriate. However, in that case, it 
will have shown where and how severely inelasticity comes into play. 

In our model framework, all these considerations are possible because of the
factorization of physical amplitudes and because of the identification of
quark diagrams with elementary processes. In conclusion, the framework we
propose may lead to a simplified parametrization (using relevant symmetry
and dynamical arguments) that can be used to analyze experimental data.

\section{Acknowledgment.}

Many thanks are due to J.-M. G\'{e}rard, J. Pestieau and J. Weyers for the
numerous discussions, comments and encouragements. This work is supported in
part by the Fonds National de la Recherche Scientifique (Belgique).

\pagebreak

\begin{description}
\item  \appendix 

{\LARGE Appendices}
\end{description}

\section{SU(3) analysis of B and D decay amplitudes}

Let us describe briefly the SU(3) analysis of decay amplitudes (see ref. 
\cite{su(3)fond} and \cite{su(3)bdec1}). We work with the conventions that $%
(u,d,s)$ transform as $3$ and $(\overline{s},\overline{d},-\overline{u})$
transform as $\overline{3}$.

\subsection{Initial States :}

For the decaying mesons, we are considering $\left( B^{+},B^{0},B_{s}\right) 
$ and $(\overline{D}^{0},D^{-},D_{s}^{-})$ because they both transform as $3$
under SU(3).

\subsection{Weak Hamiltonians :}

We give here the weak Hamiltonians at lowest order in electroweak for $%
\Delta $C=0 B decays and for D decays. These hamiltonians are written as a
Fermi current-current interaction and the V-A structure is omitted.

\paragraph{(i) For B decays,}

The weak Hamiltonian for $\Delta $C=0 transitions is at lowest order

\begin{equation}
\left\{ 
\begin{array}{l}
H_{W}^{\Delta S=0}=V_{ub}^{*}V_{ud}\overline{b}u.\overline{u}%
d+V_{cb}^{*}V_{cd}\overline{b}c.\overline{c}d+V_{tb}^{*}V_{td}\overline{b}t.%
\overline{t}d \\ 
H_{W}^{\Delta S=1}=V_{ub}^{*}V_{us}\overline{b}u.\overline{u}%
s+V_{cb}^{*}V_{cs}\overline{b}c.\overline{c}s+V_{tb}^{*}V_{ts}\overline{b}t.%
\overline{t}s
\end{array}
\right.  \label{ham1}
\end{equation}

This is written in terms of SU(3) representations as :

\begin{equation}
\begin{array}{l}
H_{W}^{\Delta S=0}=\left\{ 
\begin{array}{c}
V_{ub}^{*}V_{ud}\left( \sqrt{8}\left| \overline{15},\frac{3}{2},\frac{1}{2},-%
\frac{1}{3}\right\rangle +\left| \overline{15},\frac{1}{2},\frac{1}{2},-%
\frac{1}{3}\right\rangle +\left| 6,\frac{1}{2},\frac{1}{2},-\frac{1}{3}%
\right\rangle +\left| \overline{3},\frac{1}{2},\frac{1}{2},-\frac{1}{3}%
\right\rangle \right) \\ 
+V_{cb}^{*}V_{cd}\left| \overline{3_{c}},\frac{1}{2},\frac{1}{2},-\frac{1}{3}%
\right\rangle +V_{tb}^{*}V_{td}\left| \overline{3_{t}},\frac{1}{2},\frac{1}{2%
},-\frac{1}{3}\right\rangle
\end{array}
\right. \\ 
H_{W}^{\Delta S=1}=\left\{ 
\begin{array}{c}
+V_{ub}^{*}V_{us}\left( \sqrt{6}\left| \overline{15},1,0,\frac{2}{3}%
\right\rangle +\sqrt{3}\left| \overline{15},0,0,\frac{2}{3}\right\rangle
-\left| 6,1,0,\frac{2}{3}\right\rangle +\left| \overline{3},0,0,\frac{2}{3}%
\right\rangle \right) \\ 
+V_{cb}^{*}V_{cs}\left| \overline{3_{c}},0,0,\frac{2}{3}\right\rangle
+V_{tb}^{*}V_{ts}\left| \overline{3_{t}},0,0,\frac{2}{3}\right\rangle
\end{array}
\right.
\end{array}
\label{ham2}
\end{equation}

\paragraph{(ii) For D decays,}

we have : 
\begin{equation}
\left\{ 
\begin{array}{l}
H_{W}^{\Delta S=-1}=V_{cd}^{*}V_{us}\overline{c}d.\overline{s}u \\ 
H_{W}^{\Delta S=0}=V_{cd}^{*}V_{ud}\overline{c}d.\overline{d}%
u+V_{cs}^{*}V_{us}\overline{c}s.\overline{s}u+V_{cb}^{*}V_{ub}\overline{c}b.%
\overline{b}u \\ 
H_{W}^{\Delta S=+1}=V_{cs}^{*}V_{ud}\overline{c}s.\overline{d}u
\end{array}
\right.  \label{ham3}
\end{equation}

and in representations :

\begin{equation}
\begin{array}{l}
H_{W}^{\Delta S=-1}=V_{cd}^{*}V_{us}\left( -\sqrt{12}\left| \overline{15}%
,1,0,-\tfrac{4}{3}\right\rangle +\sqrt{2}\left| 6,0,0,-\tfrac{4}{3}%
\right\rangle \right) \\ 
H_{W}^{\Delta S=0}=\left\{ 
\begin{array}{l}
V_{cd}^{*}V_{ud}\left( {\small -}\sqrt{8}\left| \overline{15},\frac{3}{2},%
\frac{-1}{2},\frac{-1}{3}\right\rangle {\small +}\left| \overline{15},\frac{1%
}{2},\frac{-1}{2},\frac{-1}{3}\right\rangle {\small +}\left| 6,\frac{1}{2},%
\frac{-1}{2},\frac{-1}{3}\right\rangle {\small +}\left| \overline{3},\frac{1%
}{2},\frac{-1}{2},\frac{-1}{3}\right\rangle \right) \\ 
+V_{cs}^{*}V_{us}\left( {\small -}\sqrt{9}\left| \overline{15},\frac{1}{2},%
\frac{-1}{2},\frac{-1}{3}\right\rangle {\small -}\left| 6,\frac{1}{2},\frac{%
-1}{2},\frac{-1}{3}\right\rangle {\small +}\left| \overline{3},\frac{1}{2},%
\frac{-1}{2},\frac{-1}{3}\right\rangle \right) +V_{cb}^{*}V_{ub}\left| 
\overline{3_{b}},\frac{1}{2},\frac{-1}{2},\frac{-1}{3}\right\rangle
\end{array}
\right. \\ 
H_{W}^{\Delta S=+1}=V_{cs}^{*}V_{ud}\left( -\sqrt{12}\left| \overline{15}%
,1,-1,\tfrac{2}{3}\right\rangle -\sqrt{2}\left| 6,1,-1,\tfrac{2}{3}%
\right\rangle \right)
\end{array}
\label{ham4}
\end{equation}

And with the Cabibbo approximation ($V_{cd}^{*}V_{ud}=-V_{cs}^{*}V_{us}=%
\lambda ,V_{cb}^{*}V_{ub}=0):$

\begin{equation}
H_{W}^{\Delta S=0}=\lambda \left( -\sqrt{8}\left| \overline{15},\tfrac{3}{2},%
\tfrac{-1}{2},\tfrac{-1}{3}\right\rangle +4\left| \overline{15},\tfrac{1}{2},%
\tfrac{-1}{2},\tfrac{-1}{3}\right\rangle +2\left| 6,\tfrac{1}{2},\tfrac{-1}{2%
},\tfrac{-1}{3}\right\rangle \right)  \label{ham5}
\end{equation}

\subsection{Final States :}

The pseudoscalars transform as the octet 8, therefore all the final states
of two charmless pseudoscalars can be obtained from the symmetric part (Bose
statistics) of the tensor product $(8\otimes 8)_{S}=27,8_{S},1.$ Note that
under our conventions, the pseudoscalar octet is :

\begin{equation}
\left( K^{+},K^{0},\pi ^{+},-\pi ^{0},-\pi ^{-},-\eta _{8},\overline{K^{0}}%
,-K^{-}\right)  \label{octetconv}
\end{equation}

\subsection{Decay amplitudes}

The complete set of decays can be parametrized with SU(3) amplitudes using
the standard Wigner-Eckart theorem. The results are well-known (see ref. 
\cite{su(3)bdec1} to \cite{su(3)ddec3}), but we give them again for
convenience, and because we have renormalized the SU(3) amplitudes
coherently in B and D decays. This implies that we have a nice
correspondence between SU(3) amplitudes and quark diagrams, valid in both B
and D decays. The following table summarizes the notations for matrix
elements :

\begin{equation}
\ 
\begin{tabular}{|l|l|l|l|l|}
\hline
SU(3) Amplitudes & $3\otimes \overline{15}\rightarrow 8,27$ & $3\otimes
6\rightarrow 8$ & $3\otimes \overline{3}\rightarrow 1,8$ & $3\otimes 
\overline{3_{q}}\rightarrow 1,8$ \\ \hline
Initial = 3 & $A^{27}=\left\langle 27\left| \overline{15}\right|
3\right\rangle $ & $B^{8}=\left\langle 8\left| 6\right| 3\right\rangle $ & $%
C^{8}=\left\langle 8\left| \overline{3}\right| 3\right\rangle $ & $%
C^{8q}=\left\langle 8\left| \overline{3_{q}}\right| 3\right\rangle $ \\ 
Final = 27,8,1 & $A^{8}=\left\langle 8\left| \overline{15}\right|
3\right\rangle $ &  & $C^{1}=\left\langle 1\left| \overline{3}\right|
3\right\rangle $ & $C^{1q}=\left\langle 1\left| \overline{3_{q}}\right|
3\right\rangle $ \\ \hline
\end{tabular}
\label{su(3)ampl}
\end{equation}

Of course, SU(3) amplitudes have different values in B and D decays, but the
SU(3) structure is similar. For D decays, we can use the Cabibbo
approximation for CKM, under which only three SU(3) amplitudes survive : A$%
^{27}$, A$^{8}$ and B$^{8}$.

\paragraph{Remark :}

The procedure just described can be applied to find SU(3) decompositions of
bare or of full decay amplitudes in terms of bare or full SU(3) amplitudes
respectively, since FSI proceed only by strong interactions (to translate
from full to bare, just replace $A^{27}$ by $A_{b}^{27}$ and so on, see eq(%
\ref{exsu(3)}) and (\ref{exsu(3)bare})). If we work at the level of full
amplitudes under SU(3), this means that we are imposing SU(3) invariance on
FSI.

\section{Quark diagram analysis}

The decompositions of bare decay amplitudes in terms of QD are calculated as
usual :

\begin{quote}
(i) For a given initial state, and a given type of QD, write all the
possible flavour ''final'' states (note that since we are working at the
bare level, ''final'' means here intermediate).

It is at this step that we ensure SU(3) symmetry. By identifying diagrams
which correspond under the exchange of $u$, $d$ or $s$ (and $\overline{u}$, $%
\overline{d}$, $\overline{s}$), we are left with the six diagrams T, C, E,
A, P, PA, and some P$^{q}$ and PA$^{q}$ (proceeding via a heavy quark q in
the loop). We can also implement SU(2) by considering exchange of $u$, $d$
(and $\overline{u}$, $\overline{d}$), but we are left with a huge number of
different diagrams.

(ii) Contract these flavour ''final'' states with every hadron states
according to the conventions : 
\begin{eqnarray}
&& 
\begin{array}{c}
K^{+}=u\overline{s} \\ 
K^{0}=d\overline{s}
\end{array}
\quad \quad 
\begin{array}{c}
\overline{K^{0}}=s\overline{u} \\ 
K^{-}=s\overline{d}
\end{array}
\qquad 
\begin{array}{c}
\pi ^{+}=u\overline{d} \\ 
\pi ^{-}=d\overline{u}
\end{array}
\quad  \label{stphys} \\
\pi ^{0} &=&\tfrac{{\small 1}}{\sqrt{{\small 2}}}\left( u\overline{u}-d%
\overline{d}\right) \qquad \eta _{8}=\tfrac{{\small 1}}{\sqrt{{\small 6}}}%
\left( u\overline{u}+d\overline{d}-2s\overline{s}\right)  \nonumber
\end{eqnarray}

This is necessary in order to render QD decompositions compatible with SU(3)
decompositions (with the phase conventions (\ref{octetconv})).

(ii) If the ''final'' state contains identical hadrons, divide by $\sqrt{2}$%
. This is compulsory in order to compare with SU(3) amplitudes, where final
states are symmetric under exchange (Bose statistics). This implies that
when calculating decay widths for identical particle ''final'' states, we
should not divide by 2.

(iv) Finally, add the required CKM elements.
\end{quote}

Note that even if the analysis is the same in B and D, the specific values
of QD are of course different for B and D decays.

\subsection{Link between SU(3) bare amplitudes and QD amplitudes}

Since we have two parametrizations : QD and SU(3) amplitudes (bare), we can
find relations between them. The expressions of SU(3) amplitudes in terms of
QD are :

\begin{equation}
\left\{ 
\begin{array}{l}
A_{b}^{27}=-\frac{1}{10}\left( T+C\right) \\ 
A_{b}^{8}=\frac{1}{40}\left( T+C\right) +\frac{1}{8}\left( E+A\right) \\ 
B_{b}^{8}=\frac{1}{4}\left( -T+C-E+A\right) \\ 
C_{b}^{8}=\frac{1}{8}\left( -3T+C+E-3A\right) -P \\ 
C_{b}^{1}=\frac{1}{12}\left( 3T-C\right) +\frac{2}{3}\left( E+P\right) +PA
\end{array}
\right.  \label{linksu(3)dq}
\end{equation}

These relations are valid for B and D decays. There are also relations
linking $\overline{3^{q}}$ with P$^{q}$ and PA$^{q}$ ($q=c,b$ or $t$) :

\begin{equation}
\left\{ 
\begin{array}{l}
C_{b}^{8q}=-P^{q} \\ 
C_{b}^{1q}=\frac{2}{3}P^{q}+PA^{q}
\end{array}
\right.  \label{linksu(3)dq2}
\end{equation}

Since there are more QD amplitudes than SU(3) amplitudes, there is a
combination of QD that never appear in decay amplitudes. This relation is :

\begin{equation}
T-C-E+A-P+PA=0  \label{dqrelation}
\end{equation}

This relation is to be interpreted as a relation for the corresponding
coefficients in decay amplitude decompositions. For example, $%
B^{+}\rightarrow \left\{ K^{+}\pi ^{0}\right\} =\frac{1}{\sqrt{2}}\left(
T+C+A+P\right) $ and the relation is verified : $\frac{1}{\sqrt{2}}-\frac{1}{%
\sqrt{2}}+\frac{1}{\sqrt{2}}-\frac{1}{\sqrt{2}}=0$. Due to this relation,
the expressions of QD in terms of SU(3) amplitudes are not uniquely defined.
Anyway, one can use the following simple set of relations to translate QD
decompositions into SU(3) amplitude decompositions :

\begin{equation}
\left\{ 
\begin{array}{l}
T=-6A_{b}^{27} \\ 
C=-4A_{b}^{27} \\ 
E=2A_{b}^{27}+4A_{b}^{8}-2B_{b}^{8} \\ 
A=4A_{b}^{8}+2B_{b}^{8} \\ 
P=2A_{b}^{27}-A_{b}^{8}-B_{b}^{8}-C_{b}^{8} \\ 
PA=-\frac{3}{2}A_{b}^{27}-2A_{b}^{8}+2B_{b}^{8}+\frac{2}{3}%
C_{b}^{8}+C_{b}^{1}
\end{array}
\right.  \label{linkdqsu(3)}
\end{equation}

\begin{equation}
\left\{ 
\begin{array}{l}
P^{q}=-C_{b}^{8q} \\ 
PA^{q}=\frac{2}{3}C_{b}^{8q}+C_{b}^{1q}
\end{array}
\right.  \label{linkdqsu(3)2}
\end{equation}

But it should be clear that care is needed when dealing with these relations.

\section{B decays decompositions}

\[
\begin{tabular}{|cc|c|ccc|c|c|}
\hline
$\Delta S=0$ &  &  & \multicolumn{3}{|c|}{$V_{ub}^{*}V_{ud}$} & 
\multicolumn{1}{|c|}{$V_{ub}^{*}V_{ud}C^{8}$} & $V_{ub}^{*}V_{ud}C^{1}$ \\ 
\multicolumn{2}{|c|}{$B\rightarrow PP$} & $Prefactors$ &  &  & 
\multicolumn{1}{c|}{} & $+V_{cb}^{*}V_{cd}C^{8c}$ & $+V_{cb}^{*}V_{cd}C^{1c}$
\\ \cline{4-6}\cline{4-6}
&  &  & $A^{27}$ & \multicolumn{1}{|c}{$A^{8}$} & \multicolumn{1}{|c|}{$%
B^{8} $} & $+V_{tb}^{*}V_{td}C^{8t}$ & $+V_{tb}^{*}V_{td}C^{1t}$ \\ 
\hline\hline
& \multicolumn{1}{|c|}{$\overline{K^{0}}\eta _{8}$} & $\frac{1}{\sqrt{6}}$ & 
$-6$ & \multicolumn{1}{|c}{$1$} & \multicolumn{1}{|c|}{$1$} & $1$ & $0$ \\ 
\cline{2-8}
$B_{S}$ & \multicolumn{1}{|c|}{$\overline{K^{0}}\pi ^{0}$} & $\frac{1}{\sqrt{%
2}}$ & $-6$ & \multicolumn{1}{|c}{$1$} & \multicolumn{1}{|c|}{$1$} & $1$ & $%
0 $ \\ \cline{2-8}
& \multicolumn{1}{|c|}{$K^{-}\pi ^{+}$} & $1$ & $-4$ & \multicolumn{1}{|c}{$%
-1$} & \multicolumn{1}{|c|}{$-1$} & $-1$ & $0$ \\ \hline\hline
& \multicolumn{1}{|c|}{$\pi ^{+}\eta _{8}$} & $\frac{1}{\sqrt{6}}$ & $-6$ & 
\multicolumn{1}{|c}{$6$} & \multicolumn{1}{|c|}{$2$} & $-2$ & $0$ \\ 
\cline{2-8}
$B^{+}$ & \multicolumn{1}{|c|}{$\overline{K^{0}}K^{+}$} & $1$ & $2$ & 
\multicolumn{1}{|c}{$3$} & \multicolumn{1}{|c|}{$1$} & $-1$ & $0$ \\ 
\cline{2-8}
& \multicolumn{1}{|c|}{$\pi ^{+}\pi ^{0}$} & $\frac{1}{\sqrt{2}}$ & $-10$ & 
\multicolumn{1}{|c}{$0$} & \multicolumn{1}{|c|}{$0$} & $0$ & $0$ \\ 
\hline\hline
& \multicolumn{1}{|c|}{$K^{+}K^{-}$} & $1$ & $1/2$ & \multicolumn{1}{|c}{$2$}
& \multicolumn{1}{|c|}{$0$} & $2/3$ & $1$ \\ \cline{2-8}
& \multicolumn{1}{|c|}{$K^{0}\overline{K^{0}}$} & $1$ & $1/2$ & 
\multicolumn{1}{|c}{$-3$} & \multicolumn{1}{|c|}{$1$} & $-1/3$ & $1$ \\ 
\cline{2-8}
$B^{0}$ & \multicolumn{1}{|c|}{$\eta _{8}\eta _{8}$} & $\frac{1}{\sqrt{2}}$
& $-3/2$ & \multicolumn{1}{|c}{$-1$} & \multicolumn{1}{|c|}{$1$} & $1/3$ & $%
1 $ \\ \cline{2-8}
& \multicolumn{1}{|c|}{$\pi ^{0}\eta _{8}$} & $\frac{1}{\sqrt{3}}$ & $0$ & 
\multicolumn{1}{|c}{$5$} & \multicolumn{1}{|c|}{$-1$} & $1$ & $0$ \\ 
\cline{2-8}
& \multicolumn{1}{|c|}{$\pi ^{+}\pi ^{-}$} & $1$ & $-7/2$ & 
\multicolumn{1}{|c}{$1$} & \multicolumn{1}{|c|}{$-1$} & $-1/3$ & $1$ \\ 
\cline{2-8}
& \multicolumn{1}{|c|}{$\pi ^{0}\pi ^{0}$} & $\frac{1}{\sqrt{2}}$ & $13/2$ & 
\multicolumn{1}{|c}{$1$} & \multicolumn{1}{|c|}{$-1$} & $-1/3$ & $1$ \\ 
\hline\hline
\end{tabular}
\]

\[
\begin{tabular}{|cc|c|ccc|c|c|}
\hline
$\Delta S=1$ &  &  & \multicolumn{3}{|c|}{$V_{ub}^{*}V_{us}$} & $%
V_{ub}^{*}V_{us}C^{8}$ & $V_{ub}^{*}V_{us}C^{1}$ \\ 
\multicolumn{2}{|c|}{$B\rightarrow PP$} & $Prefactors$ &  &  &  & $%
+V_{cb}^{*}V_{cs}C^{8c}$ & $+V_{cb}^{*}V_{cs}C^{1c}$ \\ 
\cline{4-6}\cline{4-6}
&  &  & $A^{27}$ & \multicolumn{1}{|c}{$A^{8}$} & \multicolumn{1}{|c|}{$%
B^{8} $} & $+V_{tb}^{*}V_{ts}C^{8t}$ & $+V_{tb}^{*}V_{ts}C^{1t}$ \\ 
\hline\hline
& \multicolumn{1}{|c|}{$K^{0}\eta _{8}$} & $\frac{1}{\sqrt{6}}$ & $-6$ & 
\multicolumn{1}{|c}{$1$} & \multicolumn{1}{|c|}{$1$} & $1$ & $0$ \\ 
\cline{2-8}
$B^{0}$ & \multicolumn{1}{|c|}{$K^{0}\pi ^{0}$} & $\frac{1}{\sqrt{2}}$ & $-6$
& \multicolumn{1}{|c}{$1$} & \multicolumn{1}{|c|}{$1$} & $1$ & $0$ \\ 
\cline{2-8}
& \multicolumn{1}{|c|}{$K^{+}\pi ^{-}$} & $1$ & $-4$ & \multicolumn{1}{|c}{$%
-1$} & \multicolumn{1}{|c|}{$-1$} & $-1$ & $0$ \\ \hline\hline
& \multicolumn{1}{|c|}{$K^{+}\eta _{8}$} & $\frac{1}{\sqrt{6}}$ & $-12$ & 
\multicolumn{1}{|c}{$-3$} & \multicolumn{1}{|c|}{$-1$} & $1$ & $0$ \\ 
\cline{2-8}
$B^{+}$ & \multicolumn{1}{|c|}{$K^{+}\pi ^{0}$} & $\frac{1}{\sqrt{2}}$ & $-8$
& \multicolumn{1}{|c}{$3$} & \multicolumn{1}{|c|}{$1$} & $-1$ & $0$ \\ 
\cline{2-8}
& \multicolumn{1}{|c|}{$K^{0}\pi ^{+}$} & $1$ & $2$ & \multicolumn{1}{|c}{$3$%
} & \multicolumn{1}{|c|}{$1$} & $-1$ & $0$ \\ \hline\hline
& \multicolumn{1}{|c|}{$K^{+}K^{-}$} & $1$ & $-7/2$ & \multicolumn{1}{|c}{$1$%
} & \multicolumn{1}{|c|}{$-1$} & $-1/3$ & $1$ \\ \cline{2-8}
& \multicolumn{1}{|c|}{$K^{0}\overline{K^{0}}$} & $1$ & $1/2$ & 
\multicolumn{1}{|c}{$-3$} & \multicolumn{1}{|c|}{$1$} & $-1/3$ & $1$ \\ 
\cline{2-8}
$B_{S}$ & \multicolumn{1}{|c|}{$\eta _{8}\eta _{8}$} & $\frac{1}{\sqrt{2}}$
& $9/2$ & \multicolumn{1}{|c}{$-2$} & \multicolumn{1}{|c|}{$0$} & $-2/3$ & $%
1 $ \\ \cline{2-8}
& \multicolumn{1}{|c|}{$\pi ^{0}\eta _{8}$} & $\frac{1}{\sqrt{3}}$ & $6$ & 
\multicolumn{1}{|c}{$4$} & \multicolumn{1}{|c|}{$-2$} & $0$ & $0$ \\ 
\cline{2-8}
& \multicolumn{1}{|c|}{$\pi ^{+}\pi ^{-}$} & $1$ & $1/2$ & 
\multicolumn{1}{|c}{$2$} & \multicolumn{1}{|c|}{$0$} & $2/3$ & $1$ \\ 
\cline{2-8}
& \multicolumn{1}{|c|}{$\pi ^{0}\pi ^{0}$} & $\frac{1}{\sqrt{2}}$ & $1/2$ & 
\multicolumn{1}{|c}{$2$} & \multicolumn{1}{|c|}{$0$} & $2/3$ & $1$ \\ 
\hline\hline
\end{tabular}
\]

\[
\begin{tabular}{|cc|c|c|c|c|c|c|c|}
\hline
$\Delta S=0$ &  &  & \multicolumn{4}{|c|}{$V_{ub}^{*}V_{ud}$} & $%
V_{ub}^{*}V_{ud}P$ & $V_{ub}^{*}V_{ud}PA$ \\ 
\multicolumn{2}{|c|}{$B\rightarrow PP$} & $Prefactors$ & 
\multicolumn{4}{|c|}{} & $+V_{cb}^{*}V_{cd}P^{c}$ & $+V_{cb}^{*}V_{cd}PA^{c}$
\\ \cline{4-7}\cline{4-7}
&  &  & $T$ & $C$ & \multicolumn{1}{|c|}{$E$} & $A$ & $%
+V_{tb}^{*}V_{td}P^{t} $ & $+V_{tb}^{*}V_{td}PA^{t}$ \\ \hline\hline
& \multicolumn{1}{|c|}{$\overline{K^{0}}\eta _{8}$} & $\frac{1}{\sqrt{6}}$ & 
$0$ & $1$ & \multicolumn{1}{|c|}{$0$} & $0$ & $-1$ & $0$ \\ \cline{2-9}
$B_{S}$ & \multicolumn{1}{|c|}{$\overline{K^{0}}\pi ^{0}$} & $\frac{1}{\sqrt{%
2}}$ & $0$ & $1$ & \multicolumn{1}{|c|}{$0$} & $0$ & $-1$ & $0$ \\ 
\cline{2-9}
& \multicolumn{1}{|c|}{$K^{-}\pi ^{+}$} & $1$ & $1$ & $0$ & 
\multicolumn{1}{|c|}{$0$} & $0$ & $1$ & $0$ \\ \hline\hline
& \multicolumn{1}{|c|}{$\pi ^{+}\eta _{8}$} & $\frac{1}{\sqrt{6}}$ & $1$ & $%
1 $ & \multicolumn{1}{|c|}{$0$} & $2$ & $2$ & $0$ \\ \cline{2-9}
$B^{+}$ & \multicolumn{1}{|c|}{$\overline{K^{0}}K^{+}$} & $1$ & $0$ & $0$ & 
\multicolumn{1}{|c|}{$0$} & $1$ & $1$ & $0$ \\ \cline{2-9}
& \multicolumn{1}{|c|}{$\pi ^{+}\pi ^{0}$} & $\frac{1}{\sqrt{2}}$ & $1$ & $1$
& \multicolumn{1}{|c|}{$0$} & $0$ & $0$ & $0$ \\ \hline\hline
& \multicolumn{1}{|c|}{$K^{+}K^{-}$} & $1$ & $0$ & $0$ & 
\multicolumn{1}{|c|}{$1$} & $0$ & $0$ & $1$ \\ \cline{2-9}
& \multicolumn{1}{|c|}{$K^{0}\overline{K^{0}}$} & $1$ & $0$ & $0$ & 
\multicolumn{1}{|c|}{$0$} & $0$ & $1$ & $1$ \\ \cline{2-9}
$B^{0}$ & \multicolumn{1}{|c|}{$\eta _{8}\eta _{8}$} & $\frac{1}{\sqrt{2}}$
& $0$ & $1/3$ & \multicolumn{1}{|c|}{$1/3$} & $0$ & $1/3$ & $1$ \\ 
\cline{2-9}
& \multicolumn{1}{|c|}{$\pi ^{0}\eta _{8}$} & $\frac{1}{\sqrt{3}}$ & $0$ & $%
0 $ & \multicolumn{1}{|c|}{$1$} & $0$ & $-1$ & $0$ \\ \cline{2-9}
& \multicolumn{1}{|c|}{$\pi ^{+}\pi ^{-}$} & $1$ & $1$ & $0$ & 
\multicolumn{1}{|c|}{$1$} & $0$ & $1$ & $1$ \\ \cline{2-9}
& \multicolumn{1}{|c|}{$\pi ^{0}\pi ^{0}$} & $\frac{1}{\sqrt{2}}$ & $0$ & $%
-1 $ & \multicolumn{1}{|c|}{$1$} & $0$ & $1$ & $1$ \\ \hline\hline
\end{tabular}
\]

\[
\begin{tabular}{|cc|c|c|c|c|c|c|c|}
\hline
$\Delta S=1$ &  &  & \multicolumn{4}{|c|}{$V_{ub}^{*}V_{us}$} & $%
V_{ub}^{*}V_{us}P$ & $V_{ub}^{*}V_{us}PA$ \\ 
\multicolumn{2}{|c|}{$B\rightarrow PP$} & $Prefactors$ & 
\multicolumn{4}{|c|}{} & $+V_{ub}^{*}V_{us}P^{c}$ & $+V_{cb}^{*}V_{cs}PA^{c}$
\\ \cline{4-7}\cline{4-7}
&  &  & $T$ & $C$ & \multicolumn{1}{|c|}{$E$} & $A$ & $%
+V_{tb}^{*}V_{ts}P^{t} $ & $+V_{tb}^{*}V_{ts}PA^{t}$ \\ \hline\hline
& \multicolumn{1}{|c|}{$K^{0}\eta _{8}$} & $\frac{1}{\sqrt{6}}$ & $0$ & $1$
& \multicolumn{1}{|c|}{$0$} & $0$ & $-1$ & $0$ \\ \cline{2-9}
$B^{0}$ & \multicolumn{1}{|c|}{$K^{0}\pi ^{0}$} & $\frac{1}{\sqrt{2}}$ & $0$
& $1$ & \multicolumn{1}{|c|}{$0$} & $0$ & $-1$ & $0$ \\ \cline{2-9}
& \multicolumn{1}{|c|}{$K^{+}\pi ^{-}$} & $1$ & $1$ & $0$ & 
\multicolumn{1}{|c|}{$0$} & $0$ & $1$ & $0$ \\ \hline\hline
& \multicolumn{1}{|c|}{$K^{+}\eta _{8}$} & $\frac{1}{\sqrt{6}}$ & $1$ & $1$
& \multicolumn{1}{|c|}{$0$} & $-1$ & $-1$ & $0$ \\ \cline{2-9}
$B^{+}$ & \multicolumn{1}{|c|}{$K^{+}\pi ^{0}$} & $\frac{1}{\sqrt{2}}$ & $1$
& $1$ & \multicolumn{1}{|c|}{$0$} & $1$ & $1$ & $0$ \\ \cline{2-9}
& \multicolumn{1}{|c|}{$K^{0}\pi ^{+}$} & $1$ & $0$ & $0$ & 
\multicolumn{1}{|c|}{$0$} & $1$ & $1$ & $0$ \\ \hline\hline
& \multicolumn{1}{|c|}{$K^{+}K^{-}$} & $1$ & $1$ & $0$ & 
\multicolumn{1}{|c|}{$1$} & $0$ & $1$ & $1$ \\ \cline{2-9}
& \multicolumn{1}{|c|}{$K^{0}\overline{K^{0}}$} & $1$ & $0$ & $0$ & 
\multicolumn{1}{|c|}{$0$} & $0$ & $1$ & $1$ \\ \cline{2-9}
$B_{S}$ & \multicolumn{1}{|c|}{$\eta _{8}\eta _{8}$} & $\frac{1}{\sqrt{2}}$
& $0$ & $-2/3$ & \multicolumn{1}{|c|}{$1/3$} & $0$ & $4/3$ & $1$ \\ 
\cline{2-9}
& \multicolumn{1}{|c|}{$\pi ^{0}\eta _{8}$} & $\frac{1}{\sqrt{3}}$ & $0$ & $%
-1$ & \multicolumn{1}{|c|}{$1$} & $0$ & $0$ & $0$ \\ \cline{2-9}
& \multicolumn{1}{|c|}{$\pi ^{+}\pi ^{-}$} & $1$ & $0$ & $0$ & 
\multicolumn{1}{|c|}{$1$} & $0$ & $0$ & $1$ \\ \cline{2-9}
& \multicolumn{1}{|c|}{$\pi ^{0}\pi ^{0}$} & $\frac{1}{\sqrt{2}}$ & $0$ & $0$
& \multicolumn{1}{|c|}{$1$} & $0$ & $0$ & $1$ \\ \hline\hline
\end{tabular}
\]

\newpage

\section{D decays decompositions}

{\small (Cabibbo Approximation)}

\[
\begin{tabular}{|c|c|c|c|c|c||c|c|c|c|}
\hline
\multicolumn{2}{|c|}{$\Delta S=0$} & $Prefactors$ & \multicolumn{3}{|c||}{$%
\lambda $} & \multicolumn{4}{c|}{$\lambda $} \\ \cline{4-10}\cline{4-10}
\multicolumn{2}{|c|}{$\overline{D}\rightarrow PP$} &  & $A^{27}$ & $A^{8}$ & 
$B^{8}$ & $T$ & $C$ & $E$ & $A$ \\ \hline\hline
& \multicolumn{1}{|c|}{$K^{0}K^{-}$} & $1$ & $6$ & $4$ & $2$ & $-1$ & $0$ & $%
0$ & $1$ \\ \cline{2-10}
$D^{-}$ & \multicolumn{1}{|c|}{$\pi ^{-}\eta _{8}$} & $\frac{1}{\sqrt{6}}$ & 
$-18$ & $8$ & $4$ & $1$ & $3$ & $0$ & $2$ \\ \cline{2-10}
& \multicolumn{1}{|c|}{$\pi ^{-}\pi ^{0}$} & $\frac{1}{\sqrt{2}}$ & $10$ & $%
0 $ & $0$ & $-1$ & $-1$ & $0$ & $0$ \\ \hline\hline
& \multicolumn{1}{|c|}{$K^{-}\eta _{8}$} & $\frac{1}{\sqrt{6}}$ & $-24$ & $4$
& $2$ & $2$ & $3$ & $0$ & $1$ \\ \cline{2-10}
$D_{S}^{-}$ & \multicolumn{1}{|c|}{$K^{-}\pi ^{0}$} & $\frac{1}{\sqrt{2}}$ & 
$4$ & $-4$ & $-2$ & $0$ & $-1$ & $0$ & $-1$ \\ \cline{2-10}
& \multicolumn{1}{|c|}{$\overline{K^{0}}\pi ^{-}$} & $1$ & $-6$ & $-4$ & $-2$
& $1$ & $0$ & $0$ & $-1$ \\ \hline\hline
& \multicolumn{1}{|c|}{$K^{+}K^{-}$} & $1$ & $4$ & $-4$ & $2$ & $-1$ & $0$ & 
$-1$ & $0$ \\ \cline{2-10}
& \multicolumn{1}{|c|}{$K^{0}\overline{K^{0}}$} & $1$ & $0$ & $0$ & $0$ & $0$
& $0$ & $0$ & $0$ \\ \cline{2-10}
$\overline{D^{0}}$ & \multicolumn{1}{|c|}{$\eta _{8}\eta _{8}$} & $\frac{1}{%
\sqrt{2}}$ & $-6$ & $-4$ & $2$ & $0$ & $1$ & $-1$ & $0$ \\ \cline{2-10}
& \multicolumn{1}{|c|}{$\pi ^{0}\eta _{8}$} & $\frac{1}{\sqrt{3}}$ & $-6$ & $%
-4$ & $2$ & $0$ & $1$ & $-1$ & $0$ \\ \cline{2-10}
& \multicolumn{1}{|c|}{$\pi ^{+}\pi ^{-}$} & $1$ & $-4$ & $4$ & $-2$ & $1$ & 
$0$ & $1$ & $0$ \\ \cline{2-10}
& \multicolumn{1}{|c|}{$\pi ^{0}\pi ^{0}$} & $\frac{1}{\sqrt{2}}$ & $6$ & $4$
& $-2$ & $0$ & $-1$ & $1$ & $0$ \\ \hline\hline
\end{tabular}
\]

\[
\begin{tabular}{|c|c|c|c|c|c||c|c|c|c|}
\hline
\multicolumn{2}{|c|}{$\Delta S=-1$} & $Prefactors$ & \multicolumn{3}{|c||}{$%
V_{cd}^{*}V_{us}$} & \multicolumn{4}{c|}{$V_{cd}^{*}V_{us}$} \\ 
\cline{4-10}\cline{4-10}
\multicolumn{2}{|c|}{$\overline{D}\rightarrow PP$} &  & $A^{27}$ & $A^{8}$ & 
$B^{8}$ & $T$ & $C$ & $E$ & $A$ \\ \hline\hline
& \multicolumn{1}{|c|}{$K^{-}\eta _{8}$} & $\frac{1}{\sqrt{6}}$ & $-6$ & $-4$
& $-2$ & $1$ & $0$ & $0$ & $-1$ \\ \cline{2-10}
$D^{-}$ & \multicolumn{1}{|c|}{$K^{-}\pi ^{0}$} & $\frac{1}{\sqrt{2}}$ & $6$
& $4$ & $2$ & $-1$ & $0$ & $0$ & $1$ \\ \cline{2-10}
& \multicolumn{1}{|c|}{$\overline{K^{0}}\pi ^{-}$} & $1$ & $-4$ & $4$ & $2$
& $0$ & $1$ & $0$ & $1$ \\ \hline\hline
$D_{S}^{-}$ & \multicolumn{1}{|c|}{$\overline{K^{0}}K^{-}$} & $1$ & $-10$ & $%
0$ & $0$ & $1$ & $1$ & $0$ & $0$ \\ \hline\hline
& \multicolumn{1}{|c|}{$\overline{K^{0}}\eta _{8}$} & $\frac{1}{\sqrt{6}}$ & 
$-6$ & $-4$ & $2$ & $0$ & $1$ & $-1$ & $0$ \\ \cline{2-10}
$\overline{D^{0}}$ & \multicolumn{1}{|c|}{$\overline{K^{0}}\pi ^{0}$} & $%
\frac{1}{\sqrt{2}}$ & $-6$ & $-4$ & $2$ & $0$ & $1$ & $-1$ & $0$ \\ 
\cline{2-10}
& \multicolumn{1}{|c|}{$K^{-}\pi ^{+}$} & $1$ & $-4$ & $4$ & $-2$ & $1$ & $0$
& $1$ & $0$ \\ \hline\hline
\end{tabular}
\]

\[
\begin{tabular}{|c|c|c|c|c|c||c|c|c|c|}
\hline
\multicolumn{2}{|c|}{$\Delta S=+1$} & $Prefactors$ & \multicolumn{3}{|c||}{$%
V_{cs}^{*}V_{ud}$} & \multicolumn{4}{c|}{$V_{cs}^{*}V_{ud}$} \\ 
\cline{4-10}\cline{4-10}
\multicolumn{2}{|c|}{$\overline{D}\rightarrow PP$} &  & $A^{27}$ & $A^{8}$ & 
$B^{8}$ & $T$ & $C$ & $E$ & $A$ \\ \hline\hline
& $K^{0}K^{-}$ & $1$ & $-4$ & $4$ & $2$ & $0$ & $1$ & $0$ & $1$ \\ 
\cline{2-10}
$D_{S}^{-}$ & $\pi ^{-}\eta _{8}$ & $\frac{1}{\sqrt{6}}$ & $12$ & $8$ & $4$
& $-2$ & $0$ & $0$ & $2$ \\ \cline{2-10}
& $\pi ^{-}\pi ^{0}$ & $\frac{1}{\sqrt{2}}$ & $0$ & $0$ & $0$ & $0$ & $0$ & $%
0$ & $0$ \\ \hline\hline
$D^{-}$ & $K^{0}\pi ^{-}$ & $1$ & $-10$ & $0$ & $0$ & $1$ & $1$ & $0$ & $0$
\\ \hline\hline
& $K^{0}\eta _{8}$ & $\frac{1}{\sqrt{6}}$ & $-6$ & $-4$ & $2$ & $0$ & $1$ & $%
-1$ & $0$ \\ \cline{2-10}
$\overline{D^{0}}$ & $K^{0}\pi ^{0}$ & $\frac{1}{\sqrt{2}}$ & $-6$ & $-4$ & $%
2$ & $0$ & $1$ & $-1$ & $0$ \\ \cline{2-10}
& $K^{+}\pi ^{-}$ & $1$ & $-4$ & $4$ & $-2$ & $1$ & $0$ & $1$ & $0$ \\ 
\hline\hline
\end{tabular}
\]

\pagebreak

\section{Demonstration of the Generalized Watson's theorem}

The \textbf{S }matrix is given by $:$%
\begin{equation}
\left( 
\begin{array}{cc}
1 & iW_{1}^{t} \\ 
iCP(W_{1})\equiv iW_{2} & S
\end{array}
\right)  \label{cpltSmatrix1}
\end{equation}

Unitarity implies, in the lowest order in electroweak interactions that :

\begin{equation}
\mathbf{S}^{\dagger }\mathbf{S=SS}^{\dagger }=1\Longleftrightarrow \left\{ 
\begin{array}{l}
S^{\dagger }S=SS^{\dagger }=1 \\ 
W_{1}=SW_{2}^{*} \\ 
W_{2}=SW_{1}^{*}
\end{array}
\right. \Longleftrightarrow \left\{ 
\begin{array}{l}
S\quad unitary \\ 
\left( W_{1}+W_{2}\right) =S\ \left( W_{1}+W_{2}\right) ^{*} \\ 
\left( W_{1}-W_{2}\right) =-S\ \left( W_{1}-W_{2}\right) ^{*}
\end{array}
\right.  \label{unitarityimp}
\end{equation}

Since S is symmetric and unitary, there is a real orthogonal transformation
which diagonalizes it :

\begin{equation}
S=O^{t}S_{diag}O  \label{diago3}
\end{equation}

with, since S is unitary, a diagonal form like : 
\begin{equation}
S_{diag}=\left( 
\begin{array}{llll}
e^{2i\delta _{1}} & 0 & \cdots & 0 \\ 
0 & e^{2i\delta _{2}} & \cdots & 0 \\ 
\vdots & \vdots & \ddots & \vdots \\ 
0 & 0 & \cdots & e^{2i\delta _{n}}
\end{array}
\right)  \label{diago2}
\end{equation}

Multiplying $\left( \text{\ref{unitarityimp}}\right) \ $by O :

\begin{equation}
\left\{ 
\begin{array}{l}
O\left( W_{1}+W_{2}\right) =S_{diag}\ \left( O\left( W_{1}+W_{2}\right)
\right) ^{*} \\ 
O\left( W_{1}-W_{2}\right) =-S_{diag}\ \left( O\left( W_{1}-W_{2}\right)
\right) ^{*}
\end{array}
\right.  \label{resol1}
\end{equation}

or explicitly, in terms of components :

\begin{equation}
\left\{ 
\begin{array}{l}
\left[ O\left( W_{1}+W_{2}\right) \right] _{\alpha }=e^{2i\delta _{\alpha
}}\ \left[ \left( O\left( W_{1}+W_{2}\right) \right) \right] _{\alpha }^{*}
\\ 
\left[ O\left( W_{1}-W_{2}\right) \right] _{\alpha }=-e^{2i\delta _{\alpha
}}\ \left[ \left( O\left( W_{1}-W_{2}\right) \right) \right] _{\alpha }^{*}
\end{array}
\right.  \label{resol2}
\end{equation}

This implies that each $\left[ O\left( W_{1}+W_{2}\right) \right] _{\alpha }$
is a complex number, with phase $\delta _{\alpha }$ (and similarly for $%
\left[ O\left( W_{1}-W_{2}\right) \right] _{\alpha }$ ):

\begin{equation}
\left\{ 
\begin{array}{l}
\left[ O\left( W_{1}+W_{2}\right) \right] _{\alpha }\equiv 2e^{i\delta
_{\alpha }}\ \left[ R\right] _{\alpha } \\ 
\left[ O\left( W_{1}-W_{2}\right) \right] _{\alpha }\equiv 2ie^{i\delta
_{\alpha }}\ \left[ I\right] _{\alpha }
\end{array}
\right.  \label{resol3}
\end{equation}

with R and I real (the factor 2 is unimportant). Solving for OW :

\begin{equation}
\left\{ 
\begin{array}{l}
\left[ OW_{1}\right] _{\alpha }=e^{i\delta _{\alpha }}\ \left[ R+iI\right]
_{\alpha } \\ 
\left[ OW_{2}\right] _{\alpha }=e^{i\delta _{\alpha }}\ \left[ R-iI\right]
_{\alpha }
\end{array}
\right.  \label{resol4}
\end{equation}

\smallskip This is the generalized Watson's theorem.

\paragraph{Bare amplitudes :}

The main point is to consider $\left[ R\pm iI\right] $ as weak amplitudes
without final state interactions, i.e. as bare amplitudes :

\begin{equation}
\left\{ 
\begin{array}{l}
\left[ R+iI\right] _{\alpha }=\left[ OW_{1,b}\right] _{\alpha } \\ 
\left[ R-iI\right] _{\alpha }=\left[ OW_{2,b}\right] _{\alpha }
\end{array}
\right.  \label{resol5}
\end{equation}

So we see that W$_{1,b}$ and its CP conjugate W$_{2,b}$ are complex
conjugate, as they should for weak amplitudes without FSI. In doing this, we
put all the FSI effects in the phases, and therefore the norms are not
modified. Expressing the full amplitudes in terms of these bare ones (eq. (%
\ref{resol5}) and (\ref{resol4})), we get :

\begin{equation}
\left\{ 
\begin{array}{l}
\left[ OW_{1}\right] _{\alpha }=e^{i\delta _{\alpha }}\ \left[
OW_{1,b}\right] _{\alpha } \\ 
\left[ OW_{2}\right] _{\alpha }=e^{i\delta _{\alpha }}\ \left[
OW_{2,b}\right] _{\alpha }
\end{array}
\right. \Leftrightarrow \left\{ 
\begin{array}{l}
OW_{1}=\sqrt{S_{diag}}OW_{1,b} \\ 
OW_{2}=\sqrt{S_{diag}}OW_{2,b}
\end{array}
\right.  \label{resol6}
\end{equation}

\begin{equation}
W_{i}=\sqrt{S}W_{i,b}  \label{solutwatson}
\end{equation}

where we identify the square root as :

\begin{equation}
\sqrt{S}=O^{t}\sqrt{S_{diag}}O  \label{solutwatson2}
\end{equation}

We have thus demonstrated the form (\ref{watsontheo}), which is the
generalized Watson's theorem rewritten using bare amplitude identifications.

In summary, Watson's theorem allows us to extract from the full weak
amplitudes the hadronic FSI part, leaving real bare amplitudes. We can say
that we have unitarized weak bare decay amplitudes, since with the
adjunction of the strong phases, the full \textbf{S} matrix is unitary. We
can also say that we have renormalized the weak bare amplitudes by $\sqrt{S}$%
, i.e. the effect of FSI factorize.

\subsection{Restriction on mixings}

Neglecting some mixings, we can impose a block-diagonal form for S :

\begin{equation}
\left( 
\begin{array}{ccc}
1 & iW_{1}^{t} & iZ_{1}^{t} \\ 
iW_{2} & S_{1} & 0 \\ 
iZ_{2} & 0 & S_{2}
\end{array}
\right)  \label{twoblockwatson}
\end{equation}

For each decoupled part, we can repeat the whole analysis. Indeed, unitarity
implies

\begin{equation}
\left\{ 
\begin{array}{l}
S_{1}^{\dagger }S_{1}=S_{1}S_{1}^{\dagger }=1 \\ 
W_{1}=S_{1}W_{2}^{*} \\ 
W_{2}=S_{1}W_{1}^{*}
\end{array}
\right. \ \quad \text{and}\quad \left\{ 
\begin{array}{l}
S_{2}^{\dagger }S_{2}=S_{2}S_{2}^{\dagger }=1 \\ 
Z_{1}=S_{2}Z_{2}^{*} \\ 
Z_{2}=S_{2}Z_{1}^{*}
\end{array}
\right.  \label{twoblockwatson2}
\end{equation}

and the remaining discussion is straightforward.

The probability conservation allows a characterization of this
approximation. It is now expressed, if S$_{1}$ is n$_{1}$ x n$_{1}$ and S$%
_{2}$ is n$_{2}$ x n$_{2}$ with n = n$_{1}$+n$_{2}$:

\begin{equation}
\stackunder{i=1}{\stackrel{n}{\dsum }}\left\| \left( B\rightarrow
x_{i}\right) \right\| ^{2}=\stackunder{i=1}{\stackrel{n}{\dsum }}\left\|
\left( B\rightarrow \left\{ x_{i}\right\} \right) \right\| ^{2}\rightarrow
\left\{ 
\begin{array}{c}
\stackunder{i=1}{\stackrel{n_{_{1}}}{\dsum }}\left\| \left( B\rightarrow
x_{i}\right) \right\| ^{2}=\stackunder{i=1}{\stackrel{n_{_{1}}}{\dsum }}%
\left\| \left( B\rightarrow \left\{ x_{i}\right\} \right) \right\| ^{2} \\ 
\stackunder{i=n_{_{1}}}{\stackrel{n}{\dsum }}\left\| \left( B\rightarrow
x_{i}\right) \right\| ^{2}=\stackunder{i=n_{_{1}}}{\stackrel{n}{\dsum }}%
\left\| \left( B\rightarrow \left\{ x_{i}\right\} \right) \right\| ^{2}
\end{array}
\right.  \label{probconstwoblocwat}
\end{equation}

It remains to be seen in each case wether this is an appropriate restriction
or not.

\subsection{Mixing parameters}

The most general mixings will be specified by a general orthogonal
transformation O on M$_{diag}$. The determinant of this matrix can be $\pm $%
1, but we can restrict our attention to orthogonal matrices of determinant
+1, and introduce a diagonal matrix P with diagonal element $\pm $1. This P
matrix will always disappear when calculating $M$ since $%
M=O^{t}PM_{diag}PO=O^{t}M_{diag}O.$

For example, a general two-channel mixing can be described from

\begin{equation}
O=\left( 
\begin{array}{cc}
\cos \alpha & -\sin \alpha \\ 
\sin \alpha & \cos \alpha
\end{array}
\right) \text{\quad and\qquad }M=O^{t}M_{diag}O  \label{generalmix}
\end{equation}

where $\alpha $ is a mixing parameter. For a three-channel mixing, we will
need tree mixing parameters (Euler's angles) and so on.

\subsection{M matrix properties}

From their building as real orthogonal transformations on diagonal unitary
matrices, M matrices have a number of properties.

\begin{quote}
(i) They are symmetric and unitary.

(ii) The n$^{th}$ power is trivial : just multiply all phases by n. This is
also valid for n rational (see (\ref{solutwatson2})).

(iii) When all the phases are equal to $\delta $, M is simply $e^{i\delta }1$
since $O^{t}1O=$ $1$. In other words, when all the eigenphases are equal,
mixings disappear.

(iv) Finally, write M$_{diag}$ as :

\begin{equation}
M_{diag}=e^{i\delta _{1}}{\large 1+}\left( e^{i\delta _{2}}-e^{i\delta
_{1}}\right) \left( 
\begin{array}{cccc}
0 & 0 & 0 & \cdots \\ 
0 & 1 & 0 & \cdots \\ 
0 & 0 & 0 & \cdots \\ 
\vdots & \vdots & \vdots & \ddots
\end{array}
\right) +\left( e^{i\delta _{3}}-e^{i\delta _{1}}\right) \left( 
\begin{array}{cccc}
0 & 0 & 0 & \cdots \\ 
0 & 0 & 0 & \cdots \\ 
0 & 0 & 1 & \cdots \\ 
\vdots & \vdots & \vdots & \ddots
\end{array}
\right) +\ldots  \label{termdecomp1}
\end{equation}

and apply the orthogonal transformation. In this way, we have decomposed M
into a sum of numerical matrices, with phase differences as coefficients :

\begin{equation}
M=e^{i\delta _{1}}{\large 1+}\stackunder{i=2}{\stackrel{n}{\sum }}\left(
e^{i\delta _{i}}-e^{i\delta _{1}}\right) A_{i}=e^{i\delta _{1}}\left( 
{\large 1+}\stackunder{i=2}{\stackrel{n}{\sum }}\left[ \left( e^{i\left(
\delta _{i}-\delta _{1}\right) }-1\right) A_{i}\right] \right)
\label{termdecomp2}
\end{equation}

with$\quad A_{i}A_{j}=\delta _{ij}A_{i}$. We can, of course, factor another
phase than $e^{i\delta _{1}}$. These forms can be useful
phenomenologically.{}
\end{quote}

\section{Passage from a mixing parameter formulation towards an elasticity
parameter formulation.}

The form for the S matrix in terms of elasticity parameters are build in the
following way.

- Define the base transformations as :

\begin{equation}
\left| \overrightarrow{C}\right\rangle =O(\alpha )\left| \overrightarrow{T}%
\right\rangle =O(\alpha )O_{SU(2)}\left| \overrightarrow{X}\right\rangle
=O(\beta )\left| \overrightarrow{X}\right\rangle  \label{3bases}
\end{equation}

with $O(\gamma =\alpha ,\beta )=$ $\left( 
\begin{array}{cc}
\cos \gamma & -\sin \gamma \\ 
\sin \gamma & \cos \gamma
\end{array}
\right) $.

- From this, we can directly write the link between the $%
S_{eigen},S_{isospin}$ and $S_{physical}$ :

\begin{equation}
\left\{ 
\begin{array}{c}
S_{isospin}(\delta _{1},\delta _{2},\alpha )=O^{t}(\alpha )S_{eigen}O(\alpha
) \\ 
S_{physical}(\delta _{1},\delta _{2},\beta )=O^{t}(\beta )S_{eigen}O(\beta )
\end{array}
\right.  \label{Sin3bases}
\end{equation}

and $S_{physical}=O_{SU(2)}^{t}S_{isospin}O_{SU(2)},$ where we have
explicitly written the parameters : the eigenphases and the mixing
parameters.

- In these last expressions, we will change the parameters from :

\begin{equation}
\left\{ 
\begin{array}{c}
\delta _{1},\delta _{2},\alpha \rightarrow w_{1},w_{2},\eta _{w} \\ 
\delta _{1},\delta _{2},\beta \rightarrow \alpha _{1},\alpha _{2},\eta
\end{array}
\right.  \label{mixtoelast}
\end{equation}

And the elasticity parameters are defined in terms of eigenphases and mixing
parameters as

\begin{equation}
\left\{ 
\begin{array}{c}
\eta _{w}=\sqrt{1-4\varepsilon _{w}^{2}\sin ^{2}(\delta _{2}-\delta _{1})}%
;\varepsilon _{w}=\cos \alpha \sin \alpha \\ 
\eta =\sqrt{1-4\varepsilon ^{2}\sin ^{2}(\delta _{2}-\delta _{1})}%
;\varepsilon =\cos \beta \sin \beta
\end{array}
\right.  \label{mixtoelast2}
\end{equation}

For the phases, we have expressions like :

\begin{equation}
\left\{ 
\begin{array}{c}
2\alpha _{1}=\arg \left( \cos ^{2}\beta e^{2i\delta _{1}}+\sin ^{2}\beta
e^{2i\delta _{2}}\right) \\ 
2\alpha _{2}=\arg \left( \sin ^{2}\beta e^{2i\delta _{1}}+\cos ^{2}\beta
e^{2i\delta _{2}}\right)
\end{array}
\right.  \label{mixtoelast3}
\end{equation}

and similarly for $w_{1},w_{2}$ in $S_{isospin}$.

- Finally, the different limits for the passage $\delta _{1},\delta
_{2},\beta \longleftrightarrow \alpha _{1},\alpha _{2},\eta $ are :

\begin{equation}
\left\{ 
\begin{array}{l}
\beta =0\Rightarrow \eta =1,\alpha _{1}=\delta _{1},\alpha _{2}=\delta _{2}
\\ 
\delta _{2}=\delta _{1}\Rightarrow \alpha _{1}=\alpha _{2},\eta =0
\end{array}
\right.  \label{elaslimits}
\end{equation}

We can also characterize the $maximal$\ $mixing$ : The limit $\alpha
_{1}=\alpha _{2},\eta \neq 0$ can be obtained with $\beta =45{{}^{\circ }}$;
and this gives the smallest value for $\eta $ for a given $\delta
_{2}-\delta _{1},$ i.e. $\eta =\cos (\delta _{2}-\delta _{1})$.

\begin{quote}
\pagebreak
\end{quote}

{\LARGE Figures\bigskip \medskip }

\textbf{Figure 1} : The two basic topologies and Quark Diagrams extractions.

\bigskip

\textbf{Figure 2 }: Illustration of the Watson's theorem using quark
diagrams : FSI are viewed as mixings of basic quark diagram topologies.


\begin{thebibliography}{99}
\bibitem{su(3)fond}  J. de Swart, Rev.Mod.Phys. 35, 916 (1963).

\bibitem{su(3)bdec1}  D. Zeppenfeld, Z.Phys. C8, 77 (1981).

\bibitem{su(3)bdec2}  M. Savage and M. Wise, Phys.Rev. D39, 3346 (1989).

\bibitem{su(3)bdec3}  L.-L. Chau, H.-Y. Cheng, W.K. Sze, H. Yao, B. Tseng,
Phys.Rev. D43, 2176 (1991).

\bibitem{su(3)gronau1}  M. Gronau, O. Hernandez, D. London, J. Rosner,
Phys.Lett. B333, 500 (1994) (hep-ph/9404281),Phys.Rev.D50, 4529 (1994)
(hep-ph/9404283), Phys.Rev. D52, 6374 (1995), Phys.Rev D52, 6356 (1995).

\bibitem{su(3)gronau2}  M. Gronau, D. London, J. Rosner, Phys.Rev.Lett.73,
21 (1994) (hep-ph/9404282).

\bibitem{su(3)gronau3}  M. Gronau, A. Dighe, J. Rosner, Phys.Rev.Lett. 79,
4333, (1997).

\bibitem{su(3)gronau4}  M. Gronau, J. Rosner, Phys.Rev. D57, 6843 (1998)
(hep-ph/9711246).

\bibitem{su(3)ddec1}  R. Kingsley, S. Treiman, F. Wilzcek, A. Zee, Phys.Rev.
D11, 1919 (1975).

\bibitem{su(3)ddec2}  L. Chau, H. Cheng, Phys.Rev.Lett. 56, 1655 (1986),
Phys.Rev. D36, 137 (1987), Phys.Lett. B222, 285 (1989).

\bibitem{su(3)ddec3}  L. Chau, F. Wilzcek, Phys.Rev.Lett. 43, 816 (1979).

\bibitem{critic1}  M. Suzuki, hep-ph/9807414.

\bibitem{fsikamal1}  C. Sorensen, Phys.Rev. D23, 2618 (1981).

\bibitem{dsikamal2}  A. Kamal, R. Verma, Alberta thy-13-86 (aug. 1986).

\bibitem{fsikamal3}  A. Kamal, Int.J.Mod. Phys. A7, 3515 (1992).

\bibitem{fsikamal4}  A. Kamal, C.W. Luo, Phys.Rev. D57, 4275 (1998)
(hep-ph/9710275), hep-ph/9705396, hep-ph/9702289.

\bibitem{fsikamal5}  A. Kamal, N. Sinha, R. Sinha, Z.Phys. C41, 207 (1988).

\bibitem{fsikamal6}  P. Zenczykowski, Acta Phys. Polon. B28, 1605 (1997)
(hep-ph/9601265).

\bibitem{roma1}  F. Buccella, M. Lusignoli, A. Pugliese, Phys. Lett. B379,
249 (1996) (hep-ph/9601343).

\bibitem{roma2}  F. Buccella, M. Lusignoli, G. Miele, A. Pugliese, P.
Santorelli, Phys. Rev. D51, 3478 (1995) (hep-ph/9411286).

\bibitem{roma3}  F. Buccella, M. Forte, G. Miele, G. Ricciardi, Z. Phys.
C48, 47 (1990).

\bibitem{fsiother1}  O. Babelon, J.-L. Basdevant, D. Caillerie, G.
Mennessier, Nucl.Phys. B113, 445 (1976).

\bibitem{fsiother2}  J. Donoghue, hep-ph/9607351, hep-ph/9607352.

\bibitem{fsiother3}  M. Neubert, Phys. Lett. B424, 152 (1998)
(hep-ph/9712224).

\bibitem{fsiother4}  R. Cahn, M. Suzuki, hep-ph/9708208.

\bibitem{fsiother5}  H. Zheng, Phys.Lett. B356, 107 (1995) (hep-ph/9504360).

\bibitem{fsiother6}  R. Fleischer, hep-ph/9802433, hep-ph/9804319.

\bibitem{fsiother}  L. Wolfenstein, Phys.Rev. D43, 151 (1991).

\bibitem{lipkin1}  H. Lipkin, Phys.Rev.Lett. 44, 710 (1980), Phys.Rev.Lett.
46, 1307 (1981).

\bibitem{weinberg}  S. Weinberg, Quantum Field Theory, Cambridge University
Press, 1995.

\bibitem{gpw1}  J.-M. G\'{e}rard, J. Weyers, hep-ph/9711469.

\bibitem{gpw2}  D. Del\'{e}pine, J.-M. G\'{e}rard, J. Pestieau, J. Weyers,
Phys.Lett. B429, 106 (1998) (hep-ph/9802361).

\bibitem{gpw3}  J.-M. G\'{e}rard, J. Pestieau, J. Weyers, hep-ph/9803328.

\bibitem{expe1}  Cleo collaboration, Cleo-conf 97-22.

\bibitem{expe2}  T. Browder, CLNS 93/1226.

\bibitem{expe3}  A. Weinstein, Cleo Beauty '97, UCLA.

\bibitem{expe4}  F. W\"{u}rthwein, hep-ph/9706010.

\bibitem{expe5}  Cleo collaboration, Phys.Rev.Lett. 71, 1973 (1993),
Phys.Rev.Lett. 71, 3070 (1993), Phys.Rev. D48, 4007 (1993), Phys.Rev. D54,
4211 (1996), Phys.Rev.Lett. 78, 3261 (1997).
\end{thebibliography}
\end{document}